\documentclass{emulateapj}
\usepackage{lscape}

\shorttitle{Simulations of Winds of Weak-Lined T Tauri Stars. II.}
\shortauthors{Vidotto et al.}

\begin{document}

\title{Simulations of Winds of Weak-Lined T Tauri Stars. II.: The Effects of a Tilted Magnetosphere and Planetary Interactions}

\author{A. A. Vidotto}
\affil{University of S\~ao Paulo, Rua do Mat\~ao 1226, S\~ao Paulo, SP, 05508-090, Brazil} 
\affil{School of Physics and Astronomy, University of St Andrews, North Haugh, St Andrews, KY16 9SS, UK}
\email{Aline.Vidotto@st-andrews.ac.uk}              

\author{M. Opher}
\affil{George Mason University, 4400 University Drive, Fairfax, VA, 22030-4444, USA}

\author{V. Jatenco-Pereira}
\affil{University of S\~ao Paulo, Rua do Mat\~ao 1226, S\~ao Paulo, SP, 05508-090, Brazil}

\and

\author{T. I. Gombosi}
\affil{University of Michigan, 1517 Space Research Building, Ann Arbor, MI, 48109-2143, USA}

\begin{abstract}
Based on our previous work \citep{paper2}, we investigate here the effects on the wind and magnetospheric structures of weak-lined T Tauri stars due to a misalignment between the axis of rotation of the star and its magnetic dipole moment vector. In such configuration, the system loses the axisymmetry presented in the aligned case, requiring a fully three-dimensional approach. We perform three-dimensional numerical magnetohydrodynamic simulations of stellar winds and study the effects caused by different model parameters, namely the misalignment angle $\theta_t$, the stellar period of rotation, the plasma-$\beta$, and the heating index $\gamma$. Our simulations take into account the interplay between the wind and the stellar magnetic field during the time evolution. The system reaches a periodic behavior with the same rotational period of the star. We show that the magnetic field lines present an oscillatory pattern. Furthermore, we obtain that by increasing $\theta_t$, the wind velocity increases, especially in the case of strong magnetic field and relatively rapid stellar rotation. Our three-dimensional, time-dependent wind models allow us to study the interaction of a magnetized wind with a magnetized extra-solar planet. Such interaction gives rise to reconnection, generating electrons that propagate along the planet's magnetic field lines and produce electron cyclotron radiation at  radio wavelengths. The power released in the interaction depends on the planet's magnetic field intensity, its orbital radius, and on the stellar wind local characteristics. We find that a close-in Jupiter-like planet orbiting at $0.05$~AU presents a radio power that is $\sim 5$ orders of magnitude larger than the one observed in Jupiter, which suggests that the stellar wind from a young star has the potential to generate strong planetary radio emission that could be detected in the near future with LOFAR. This radio power varies according to the phase of rotation of the star. For three selected simulations, we find a variation of the radio power of a factor $1.3$ to $3.7$, depending on $\theta_t$. Moreover, we extend the investigation done in \citet{paper2} and analyze whether winds from misaligned stellar magnetospheres could cause a significant effect on planetary migration. Compared to the aligned case, we show that the time-scale $\tau_w$ for an appreciable radial motion of the planet is shorter for larger misalignment angles. While for the aligned case $\tau_w\simeq 100$~Myr, for a stellar magnetosphere tilted by $\theta_t = 30^{\rm o}$, $\tau_w$ ranges from $\sim 40$ to $70$~Myr for a planet located at a radius of $0.05$~AU. Further reduction on $\tau_w$ might occur for even larger misalignment angles and/or different wind parameters.
\end{abstract}


\section{INTRODUCTION}
T Tauri stars are pre-main sequence low-mass stars ($0.5 \lesssim M/M_\odot \lesssim 2$), with a range of spectral types from F to M, and radius  $\lesssim 3-4~R_\odot$. They are usually classified in two categories, depending on their evolutionary stage. In an earlier stage, they are known as classical T Tauri stars (CTTS), surrounded by circumstellar disks. In a later stage, with the dissipation of the accretion disk, they are known as weak-lined T Tauri stars (WTTSs). Thanks to spectropolarimetric measurements, the number of young low-mass stars with detected magnetic fields has significantly increased in the past decade \citep{1999ApJ...516..900J, 2004Ap&SS.292..619V, 2007ApJ...664..975J, 2007MNRAS.380.1297D, 2008MNRAS.386.1234D, 2010MNRAS.402.1426D, 2008AJ....136.2286Y, 2009MNRAS.398..189H,  2010MNRAS.403..159S}. These detections have suggested that T Tauri stars present mean surface field strengths of the order of kG. Surface magnetic maps, derived from spectropolarimetric data, indicate that the surface fields on T Tauri stars are more complex than that of a simple dipole and are often misaligned with the rotational axis of the star \citep{2007MNRAS.380.1297D, 2008MNRAS.386.1234D}. CTTSs such as BP Tau \citep{2008MNRAS.386.1234D}, V2129 Oph \citep{2007MNRAS.380.1297D}, CR Cha and CV Cha  \citep{2009MNRAS.398..189H}, and V2247 Oph \citep{2010MNRAS.402.1426D}  present dipolar and octupolar components of the surface magnetic field moment that are asymmetric with respect to the rotational axis of the star. More recently, the first determination of surface magnetic maps for a WTTS, V410 Tau, has been acquired \citep{2010MNRAS.403..159S}, showing that, similar to the less evolved CTTSs, V410 Tau also presents a non-axisymmetric poloidal field.

Despite the existing knowledge of the surface magnetic fields in young stars, the global structure of the stellar magnetic field is unknown. Magnetic field extrapolations from surface magnetograms using the potential field source surface (PFSS) method has been used to help us elucidate the geometry of the large-scale field around T Tauri stars \citep{2008MNRAS.386..688J, 2008MNRAS.389.1839G}. Such extrapolations, however, neglect the interaction of the field with the stellar wind and the temporal evolution of the system. Full magnetohydrodynamics (MHD) numerical simulations \citep{2003ApJ...595L..57R,paper2} allow us to study the interplay between the stellar magnetic field and the wind. In this method, the dynamical interaction of the stellar wind and the magnetic field lines is a result of the action of magnetic, thermal, gravitational, and inertial forces. MHD simulations can be, however, computationally expensive and time consuming. A comparison between PFSS and MHD models that used observed surface magnetic maps as a boundary condition can be found in \citet{2006ApJ...653.1510R} in the context of the solar wind, showing that PFSS models are able to reconstruct the large-scale structure of the solar corona when time-dependent changes in the photospheric flux can be neglected, although nonpotential effects can have a significant effect on the magnetic structure of the corona.

The accurate determination of the wind properties and topology of the magnetic field of a star is necessary to solve a series of open questions. The rotational evolution of the star, for example, requires the knowledge of the topology of the field, as the shape of the magnetic field lines may enhance rotational braking caused by the stellar magnetized winds \citep{1967ApJ...150..551K, 1988ApJ...333..236K, 2005A&A...440..411H, 2009EAS....39..199B}. Furthermore, studies of the magnetic interaction between a CTTS and its disk require the knowledge of the structure of the magnetic field of the star \citep{1990RvMA....3..234C, 1991ApJ...370L..39K, 2008MNRAS.386.1274L}. Determining realistic magnetic field topologies and wind dynamics are also key to understand interactions between magnetized extra-solar planets and the star, such as interactions that lead to planetary migration \citep{2006ApJ...645L..73R, 2008MNRAS.389.1233L, paper2}, interactions between the stellar magnetic field and the planetary magnetosphere \citep{2004A&A...425..753G, 2004ApJ...602L..53I, 2006A&A...460..317P, 2007P&SS...55..589P, 2005MNRAS.356.1053S, 2009ApJ...704L..85C, 2010MNRAS.tmp..735F} and also with the planetary atmosphere \citep{2004A&A...425..753G}.

As a next step towards a more realistic wind and magnetic field modeling of WTTSs, in this work we extend the study performed in \citet{paper2}, where the stellar rotation and magnetic moment vectors were assumed to be parallel. We now consider cases where these vectors are not aligned. Some numerical and theoretical models exist considering the case of an oblique magnetic geometry, mainly applied to the study of pulsars \citep[e.g.,][]{1999A&A...349.1017B, 2006ApJ...648L..51S, 2009A&A...496..495K}, with a few applications to other astrophysical objects \citep[e.g.,][]{1998MNRAS.300..718L, 2003ApJ...595.1009R, 2004ApJ...610..920R, 2007MNRAS.382..139T}. As a consequence of the oblique magnetic geometry, the system loses the axisymmetry present in the aligned case \citep{paper2}, thus requiring a fully three-dimensional (3D) approach. We perform here 3D MHD numerical simulations of magnetized stellar winds of WTTSs, by considering at the base of the coronal wind a dipolar magnetic field that is tilted with respect to the rotational axis of the star. Complex, high-order multipoles magnetic field configuration at the surface may exist, but a dipolar component should dominate at larger distances \citep[e.g.,][]{2007ApJ...664..975J,2007MNRAS.380.1297D}. As the simulation evolves in time, the initial field configuration is modified by the interaction with the stellar wind, which in turn also is modified by the magnetic field geometry.

The stellar wind of a host star is expected to directly influence an orbiting planet and its atmosphere. The interaction, for example, of a magnetized wind with a magnetized planet can give rise to reconnection of magnetic field lines. Reconnection processes appear in several places in the Solar System.  E.g., the magnetic field lines of the Earth magnetospheric day-side (i.e., the side of the Earth that is facing the Sun) are compressed due to the interaction with the solar wind, while in the opposite side of the Earth's magnetosphere (the night-side), a tear-dropped-shaped tail is formed \citep[e.g.,][]{1930Natur.126..129C}. The solar wind interaction with the magnetic planets of the Solar System (Earth, Jupiter, Saturn, Uranus, and Neptune) accelerates electrons that propagate along the planets magnetic field lines, producing electron cyclotron radiation at radio wavelengths \citep{1998JGR...10320159Z}. By analogy to the magnetic planets in the Solar System, predictions have been made that extra-solar planets should produce cyclotron maser emission \citep[e.g.,][]{1999JGR...10414025F, 2001Ap&SS.277..293Z, 2004ApJ...612..511L, 2004P&SS...52.1469F, 2004A&A...425..753G, 2005A&A...437..717G, 2007A&A...475..359G, 2007P&SS...55..618G, 2005MNRAS.356.1053S, 2007P&SS...55..598Z, 2008A&A...490..843J}, if they harbor intrinsic magnetic fields, although predictions also exist for the case of non-magnetized planets \citep{2001Ap&SS.277..293Z, 2007P&SS...55..598Z, 2007A&A...475..359G}. Evidence that extrasolar giant planets can be magnetized was found by \citet{2005ApJ...622.1075S, 2008ApJ...676..628S}, who observed modulations of the Ca II H\&K lines in phase with planetary orbital periods on extrasolar planetary systems. Such modulations were interpreted as induced activity on the stellar chromosphere caused by the interaction between the stellar and planetary magnetic fields.\footnote{\citet{2006A&A...460..317P} suggest that a conductor planet moving relatively to the stellar wind is also able to generate perturbations that can trigger the chromospheric modulations observed by \citet{2005ApJ...622.1075S, 2008ApJ...676..628S} without the requirement of a magnetized planet.}

The consideration of a realistic wind is crucial to determine how the interaction between the stellar wind and the magnetosphere of an extrasolar planet occur. Using the 3D, time-dependent MHD wind models developed in this paper, we investigate the planet-wind interaction. Such interaction can give rise to reconnection processes, which result in transfer of energy from the stellar wind to the planet's magnetosphere. Analogously to the interaction of Jupiter's magnetosphere with the solar wind, we estimate the radio power released from the interaction of a close-in giant planet with the wind of its host star.

We have organized this paper as follows. \S\ref{sec.numerics} presents the MHD numerical model adopted to describe a magnetized stellar wind of WTTSs. In \S\ref{sec.results}, we present the simulations made and the results we achieved, along with a comparison between wind models with different parameters. In \S\ref{sec.discussion}, we discuss our stellar wind results, performing comparisons with other simpler stellar wind models. The interaction between a close-in magnetized giant planet and the stellar wind and an estimate of the radio power released from this interaction are presented in \S\ref{sec.reconnection}. In \S\ref{sec.migration}, we investigate whether the action of magnetic torques from the stellar wind acting on a close-in giant planet is able to cause planetary migration. This investigation extends the one presented in \citet{paper2}, where the winds analyzed in that case assumed that the axis of rotation of the star and the surface magnetic dipole moment were aligned. In \S\ref{sec.conclusions}, we present the conclusions.

\section{THE NUMERICAL MODEL} \label{sec.numerics}
To perform the simulations, we use the Block Adaptive Tree Solar-wind Roe Upwind Scheme (BATS-R-US), a 3D ideal MHD numerical code developed at the Center for Space Environment Modeling at University of Michigan \citep{1999JCoPh.154..284P}. BATS-R-US has a block-based computational domain, consisting of Cartesian blocks of cells that can be adaptively refined for the region of interest. It has been used to simulate the heliosphere \citep{2003ApJ...595L..57R, 2007ApJ...654L.163C}, the outer-heliosphere \citep{1998JGR...103.1889L, 2003ApJ...591L..61O, 2006ApJ...640L..71O, 2007Sci...316..875O}, coronal mass ejections \citep{2004JGRA..10901102M,2005ApJ...627.1019L}, the Earth's magnetosphere \citep{2006AdSpR..38..263R} and the magnetosphere of Saturn \citep{2005GeoRL..3220S06H} and Uranus \citep{2004JGRA..10911210T}, among others. In this work, we extend the model developed in \citet{paper2} to study the wind structure of WTTSs, specifically when the magnetic moment of the star and the stellar rotational axis are non-parallel. 

BATS-R-US solves the ideal MHD equations, that in the conservative form are given by (in cgs units)
\begin{equation}
\label{eq:continuity_conserve}
\frac{\partial \rho}{\partial t} + \nabla\cdot \left(\rho {\bf u}\right) = 0
\end{equation}
\begin{equation}
\label{eq:momentum_conserve}
\frac{\partial \left(\rho {\bf u}\right)}{\partial t} + \nabla\cdot\left[ \rho{\bf u\,u}+ \left(p + \frac{B^2}{8\pi}\right)I - \frac{{\bf B\,B}}{4\pi}\right] = \rho {\bf g}
\end{equation}
\begin{equation}
\label{eq:bfield_conserve}
\frac{\partial {\bf B}}{\partial t} + \nabla\cdot\left({\bf u\,B} - {\bf B\,u}\right) = 0
\end{equation}
\begin{equation}
\label{eq:energy_conserve}
\frac{\partial\varepsilon}{\partial t} +  \nabla \cdot \left[ {\bf u} \left( \varepsilon + p + \frac{B^2}{8\pi} \right) - \frac{\left({\bf u}\cdot{\bf B}\right) {\bf B}}{4\pi}\right] = \rho {\bf g}\cdot {\bf u} \, ,
\end{equation}
where $\rho$ is the mass density, ${\bf u}$ the plasma velocity, ${\bf B}$ the magnetic field, $p$ the gas pressure, ${\bf g}$ the gravitational acceleration due to the central body, and  $\varepsilon$ is the total energy density given by 
\begin{equation}\label{eq:energy_density}
\varepsilon=\frac{\rho u^2}{2}+\frac{p}{\gamma-1}+\frac{B^2}{8\pi} \, .
\end{equation}
We consider an ideal gas, so $p=\rho k_B T/(\mu m_p)$, where $k_B$ is the Boltzmann constant, $T$ is the temperature, $m_p$ is the proton mass, $\mu =0.5 $ is the mean molecular weight of a totally ionized hydrogen gas, and $\gamma$ is the ratio of the specific heats (or heating parameter). In our simulations, we adopt either $\gamma=1.1$ or $\gamma=1.2$.

\begin{figure}
  \includegraphics[height=7cm]{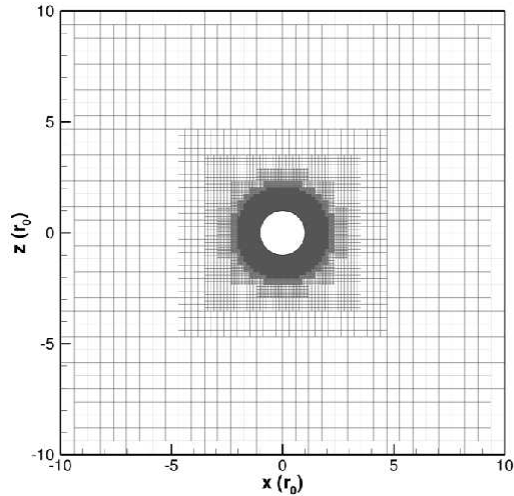}
  \caption{Meridional cut ($xz$-plane) of the adopted 3D grid in the simulations of misaligned magnetospheres, illustrating the refinement at the inner portion of the grid. Immediately around the star, the cell resolution is  $0.018~r_0$ ($r_0$ is the stellar radius), and with distance from the star, the grid gets coarser. The coarsest resolution is in the outer corners of the grid (not shown above) and is $3.75~r_0$.  \label{fig.grid}}
\end{figure}

The adopted grid is Cartesian and the star is placed at the origin. The axes $x$, $y$, and $z$ extend from $-75~r_0$ to $75~r_0$, where $r_0$ is the stellar radius. For all the cases studied, we apply $11$ levels of refinement in the simulation domain. Figure~\ref{fig.grid} presents a cut along the meridional $xz$-plane illustrating the refinement at the inner portion of the grid. We note that a higher resolution is used around the central star. This configuration has a total of $\sim 2.5 \times 10^7$ cells in the simulation domain. The smallest cells (closest to the star) have a size of $0.018~r_0$ and the maximum cell size is $3.75~r_0$.

The star has $M_\star = 0.8~M_\odot$ and $r_0=2~R_\odot$. The grid is initialized with a 1D hydrodynamical wind for a fully ionized plasma of hydrogen. The solution for $u_r (r)$ depends on the choice of the temperature at the base of the wind and on $\gamma$, and the only physical possible solution is the one that becomes supersonic when passing through the critical radius \citep{1958ApJ...128..664P}.  Due to conservation of mass of a steady wind (i.e., $\rho u_r r^2={\rm constant}$), we obtain the density profile from the radial velocity profile $u_r (r)$. 

The star is considered to be rotating as a solid body with a period of rotation $P_0=2\pi/\Omega$, where $\Omega$ is the angular velocity of the star. Its axis of rotation lies in the $z$-direction 
\begin{equation}
{\bf \Omega} = \Omega \hat{\bf z} \, .
\end{equation}
The surface magnetic moment vector ${\bf m}$ is tilted with respect to ${\bf \Omega}$ at an angle $\theta_t$
\begin{equation}
|{\bf m \cdot \Omega}| = m \Omega \cos\theta_t \, .
\end{equation}

\begin{figure}
  \includegraphics[height=7cm]{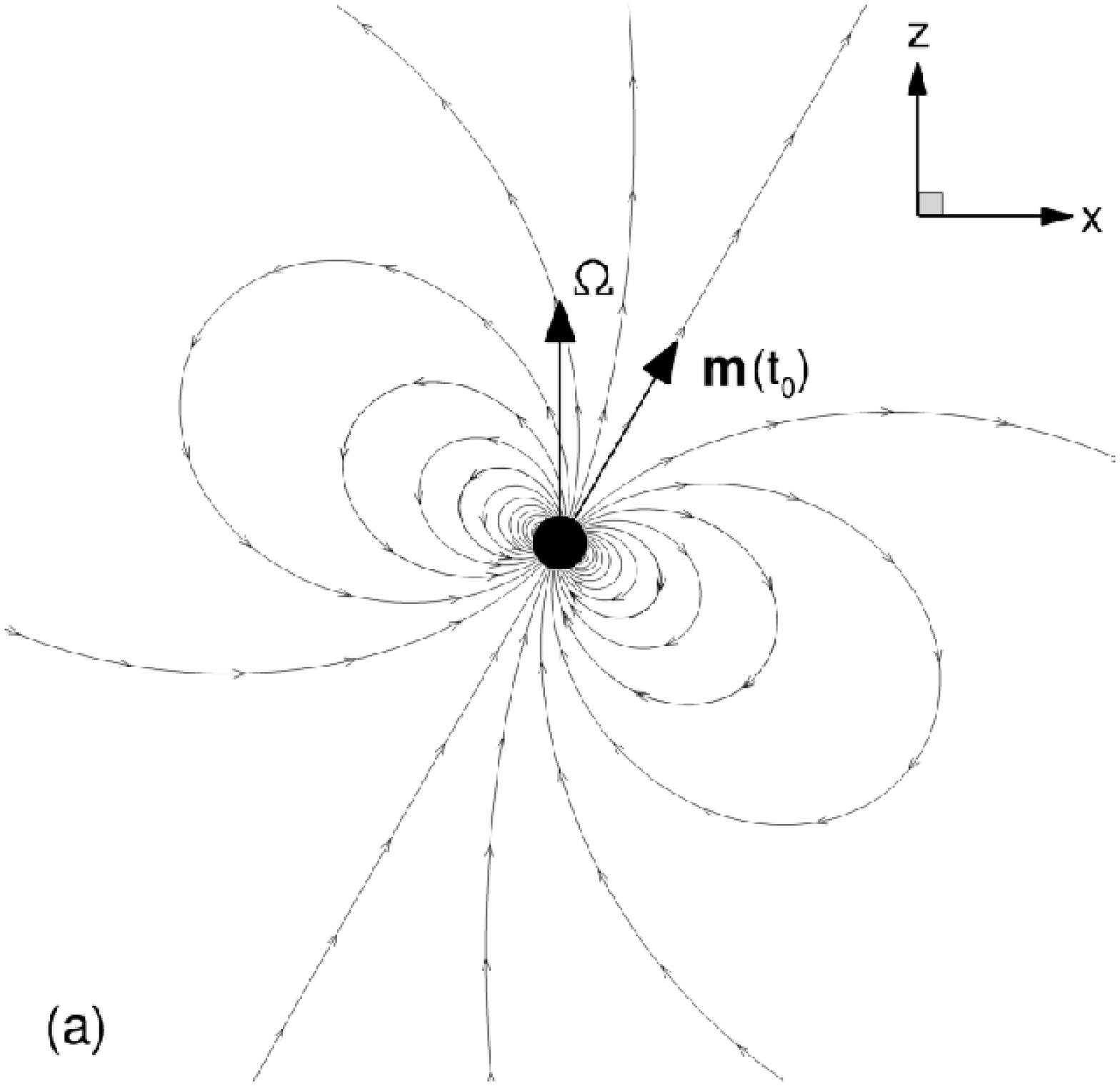}\\
  \includegraphics[height=7cm]{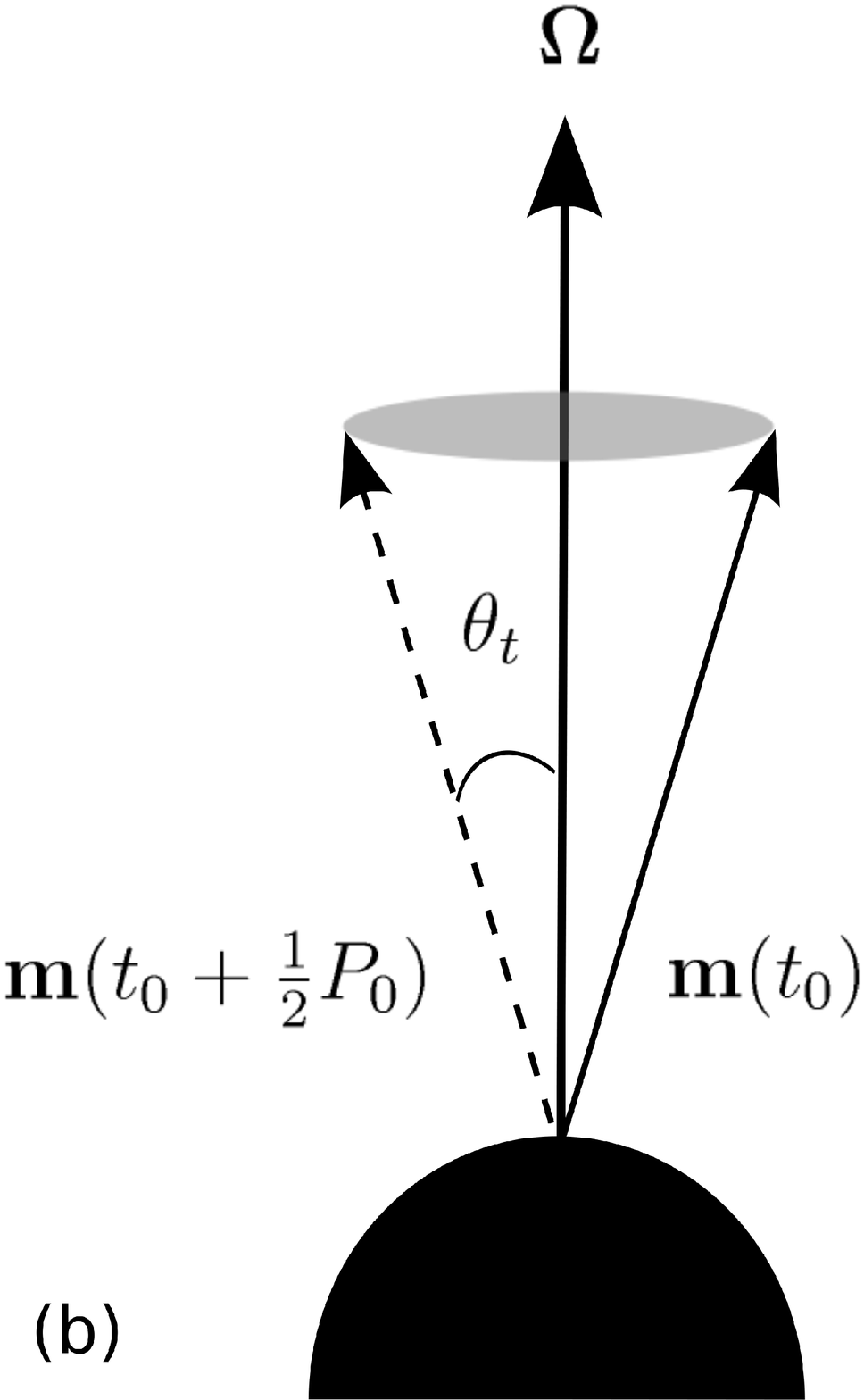}
  \caption{The magnetic field configuration in the simulations of misaligned magnetospheres. (a) Orientation of the magnetic field lines in the $xz$-plane at the initial instant $t_0$. (b) In one stellar rotational period, the magnetic moment vector ${\bf m}$ draws a cone, whose central axis is the $z$-axis. Shown above are the magnetic moment vector for the initial instant $t_0$ and half-period later. $\theta_t$ is the angle between the vectors ${\bf m}$ and $\Omega$. \label{fig.tilted-initial} }
\end{figure}

The simulations are initialized with a dipolar magnetic field described in spherical coordinates $\{ r, \theta, \varphi \}$ by
\begin{equation}\label{eq:dipoler}
B_r (t_0)=  \frac{B_0 r_0 ^3}{r ^3} (\cos \theta \cos \theta_t  + \sin \theta \cos \varphi (t_0) \sin \theta_t ) \, ,
\end{equation}
\begin{equation}\label{eq:dipoletheta}
B_\theta (t_0) = \frac{B_0 r_0 ^3}{r ^3} \left(\frac12 \sin \theta \cos \theta_t  - \frac12\cos \theta \cos \varphi (t_0) \sin \theta_t  \right) \, ,
\end{equation}
\begin{equation}\label{eq:dipolephi}
B_\varphi (t_0) = \frac{B_0 r_0 ^3}{r ^3} \frac{\sin \varphi (t_0) \sin \theta_t}{2} \, ,
\end{equation}
where $B_0$ is the magnetic field intensity at the {\it magnetic poles} (where $\theta = \theta_t$ and $r=r_0$), $r$ is the radial coordinate, $\theta$ is the co-latitude, and $\varphi$ is the azimuthal angle measured in the equatorial plane. At the initial instant $t_0$, the vector {\bf m} is in the $xz$-plane, tilted by an angle $\theta_t$ in the clockwise direction around the $y$-axis (Fig.~\ref{fig.tilted-initial}a). 

The inner boundary of the system is the base of the wind at $r=r_0$, where fixed boundary conditions were adopted. The outer boundary has outflow conditions, i.e., a zero gradient is set to all the primary variables (${\bf u}$, ${\bf B}$, $p$, and $\rho$). As the magnetic field is anchored on the star, in one stellar rotational period, the surface magnetic moment vector ${\bf m}$ draws a cone, whose central axis is the $z$-axis (Fig.~\ref{fig.tilted-initial}b). Because of that, in the simulations with oblique magnetic geometries, the boundary conditions are time-dependent and the simulations reach a periodic configuration (\S\ref{sec.temp.beh}).

The MHD solution is evolved in time from the initial dipolar configuration for the magnetic field to a fully self-consistent non-dipolar solution. The wind interacts with the magnetic field lines and deforms the initial dipolar configuration of the field. The stellar wind is also modified by the magnetic field, i.e., no fixed topologies for the magnetic field neither for the wind are assumed. 

\section{SIMULATION RESULTS}\label{sec.results}
Table~\ref{table} shows the parameters adopted in the simulations. Common to all simulations are the magnetic field intensity of $B_0=1$~kG at the magnetic poles of the star and the temperature $T_0=10^6$~K at the base of the wind. We varied the misalignment angle $\theta_t$, the period of rotation of the star $P_0$, the density at the base of the wind $\rho_0$ (and consequently the plasma-$\beta$ at the base of the coronal wind $\beta_0$), and the heating index $\gamma$. Observations of WTTSs show that they possess rotational periods ranging from $0.5$ to $13$~d \citep{2007A&A...463.1081M} with a distribution peaking at $P_0 \sim 2$~d \citep{2007ApJ...671..605C}. However, we did not consider $P_0>3$~d, as the dynamical effect of the misalignment is more significant in the cases where the star has a low period of rotation. This implies that we are in the lower range of observed rotational periods for WTTSs \citep[e.g.,][]{1993A&A...272..176B, 1996AJ....111..283C, 2002A&A...396..513H, 2006ApJ...646..297R}. A description of the choice of parameters used in the simulations (except for $\theta_t$) can be found in Section 3 of \citet{paper2}. The misalignment angle $\theta_t$ between the rotational axis of the star and the surface magnetic dipole moment vector was chosen to vary from $0^{\rm o}$ (aligned case) to $90^{\rm o}$, although surface magnetic maps of T Tauri stars have shown the existence of smaller angles \citep[$\theta_t \lesssim 30^{\rm o}$, ][]{2007MNRAS.380.1297D,2008MNRAS.386.1234D}. Simulation T01 considers the aligned case. Simulations T02 to T06 represent case with different $\theta_t$. With respect to our fiducial case T04, we varied $\gamma$ in case T07, $P_0$ in cases T08 and T09, and $\beta_0$ in case T10.

\begin{table}
\begin{center}
\caption{The set of simulations. The columns represent, respectively: the name of the simulation, the density $\rho_0$ at the base of the wind, the heating parameter $\gamma$, the period of rotation of the star $P_0$, the misalignment angle $\theta_t$, and the plasma-$\beta$ evaluated at the magnetic pole $\beta_0$. Case T01 represents the aligned case ($\theta_t=0^{\rm o}$) and case T04 the fiducial case ($\theta_t=30^{\rm o}$).\label{table}}   
\begin{tabular}{c c c c c c c}  
\tableline\tableline    
{Name} &  {$\rho_0$}  & {$\gamma$} &  {$P_0$} & {$\theta_t$}  & {$\beta_0$} \\
  & (g~cm$^{-3}$) &  &  (d) &  ($^{\rm o}$) & \\
\tableline    
T01 &    $1\times 10^{-11}$ & $1.2$ &  $1$    & $0$   &  $1/25$ \\
T02 &    $1\times 10^{-11}$ & $1.2$ &  $1$    & $10$  &  $1/25$ \\
T03 &    $1\times 10^{-11}$ & $1.2$ &  $1$    & $20$  &  $1/25$ \\
T04 &    $1\times 10^{-11}$ & $1.2$ &  $1$    & $30$  &  $1/25$ \\
T05 &    $1\times 10^{-11}$ & $1.2$ &  $1$    & $60$  &  $1/25$ \\
T06 &    $1\times 10^{-11}$ & $1.2$ &  $1$    & $90$  &  $1/25$ \\
T07 &    $1\times 10^{-11}$ & $1.1$ &  $1$    & $30$  &  $1/25$ \\ 
T08 &    $1\times 10^{-11}$ & $1.2$ &  $3$    & $30$  &  $1/25$ \\
T09 &    $1\times 10^{-11}$ & $1.2$ &  $0.5$  & $30$  &  $1/25$ \\
T10 &    $2.4\times 10^{-12}$ & $1.2$ &  $1$    & $30$  &  $1/100$ \\
\tableline                              
\end{tabular}
\end{center}
\end{table}

\subsection{Time-Dependent Behavior}\label{sec.temp.beh}
Because the star is rotating and the magnetic field is asymmetric with respect to the axis of rotation, the simulations with an oblique magnetosphere have a periodic behavior with the same period of rotation of the star. Depending on the physical conditions and grid size, after a certain number of stellar rotations, the system has relaxed and such periodic behavior is achieved. For example, for a wind expansion velocity of $\sim 200$~km~s$^{-1}$ propagating in a grid size of $75~r_0$, the time the solution will take to relax in the grid is of $\sim 75~r_0/(200$~km~s$^{-1}) \sim 6~$days. Cases T02, T03, and T04 were run for $10$~days, T05 for $9$~days, T06 for $8$~days, T07 for $5$~days, T08 and T09 for $6$~days, and T10 for $7$~days. These time intervals were sufficient for the solution to relax in the grid.

\begin{figure*}
  \includegraphics[height=5.5cm]{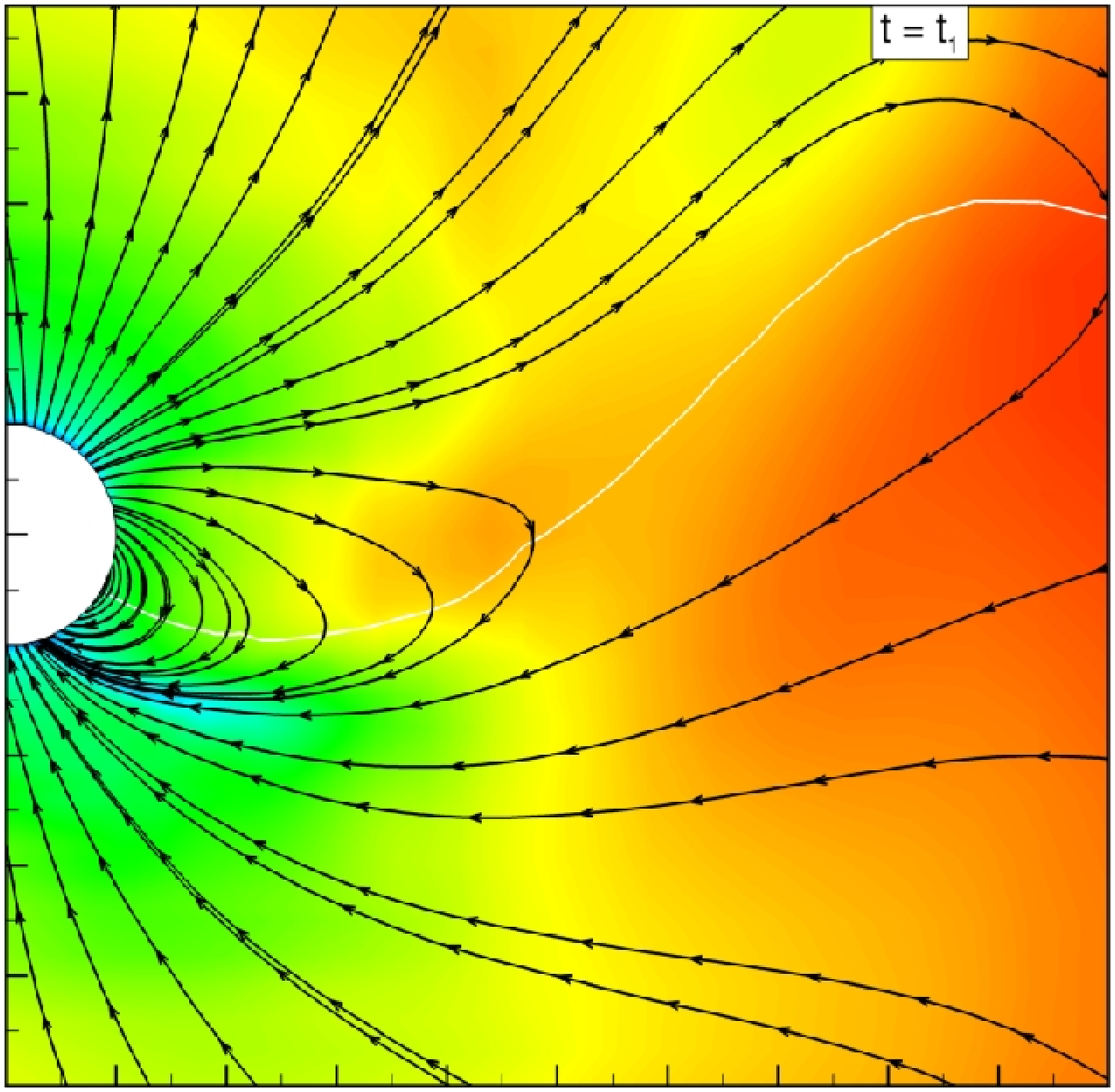}%
  \includegraphics[height=5.5cm]{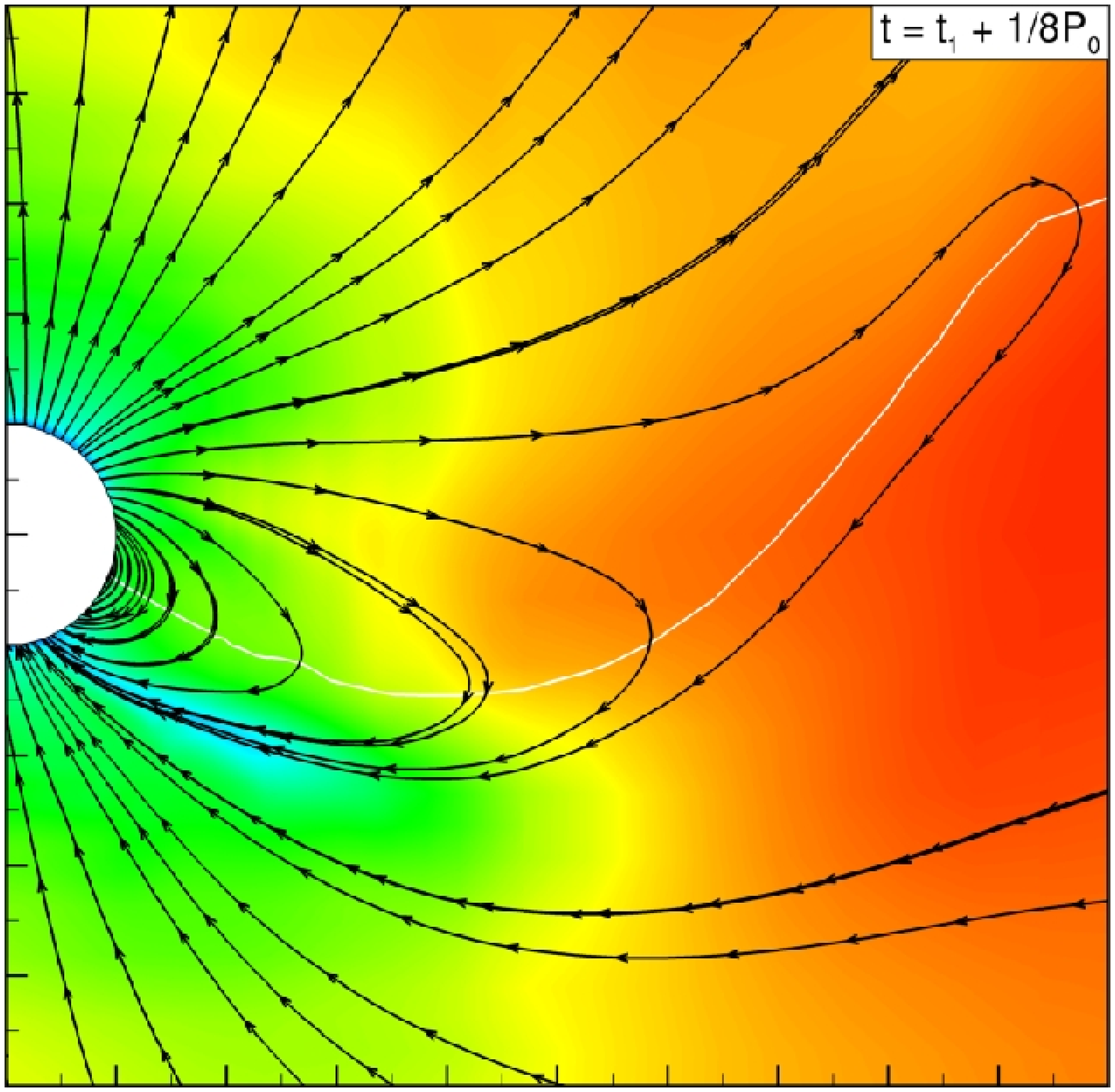}%
  \includegraphics[height=5.5cm]{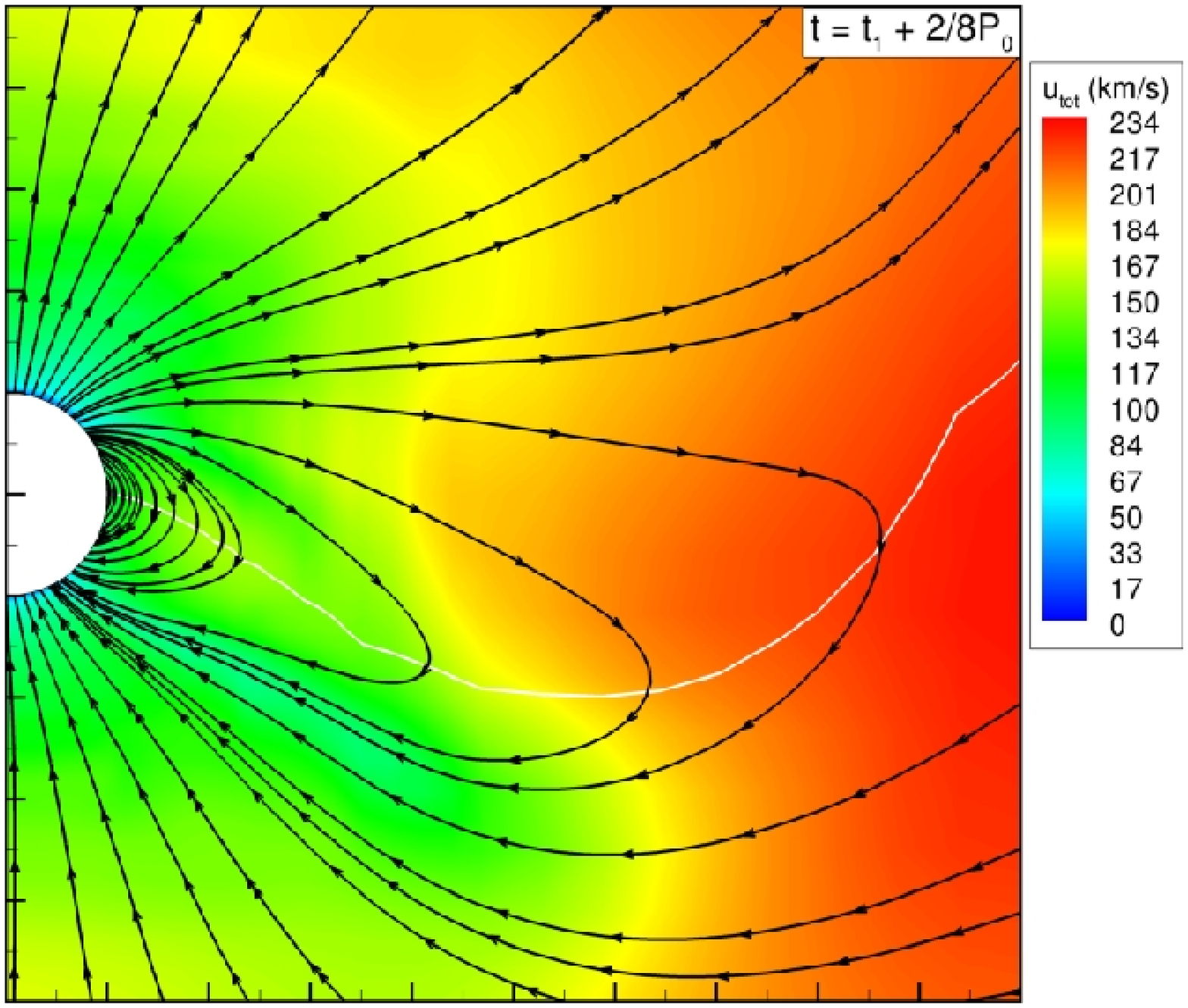}\\
  \includegraphics[height=5.5cm]{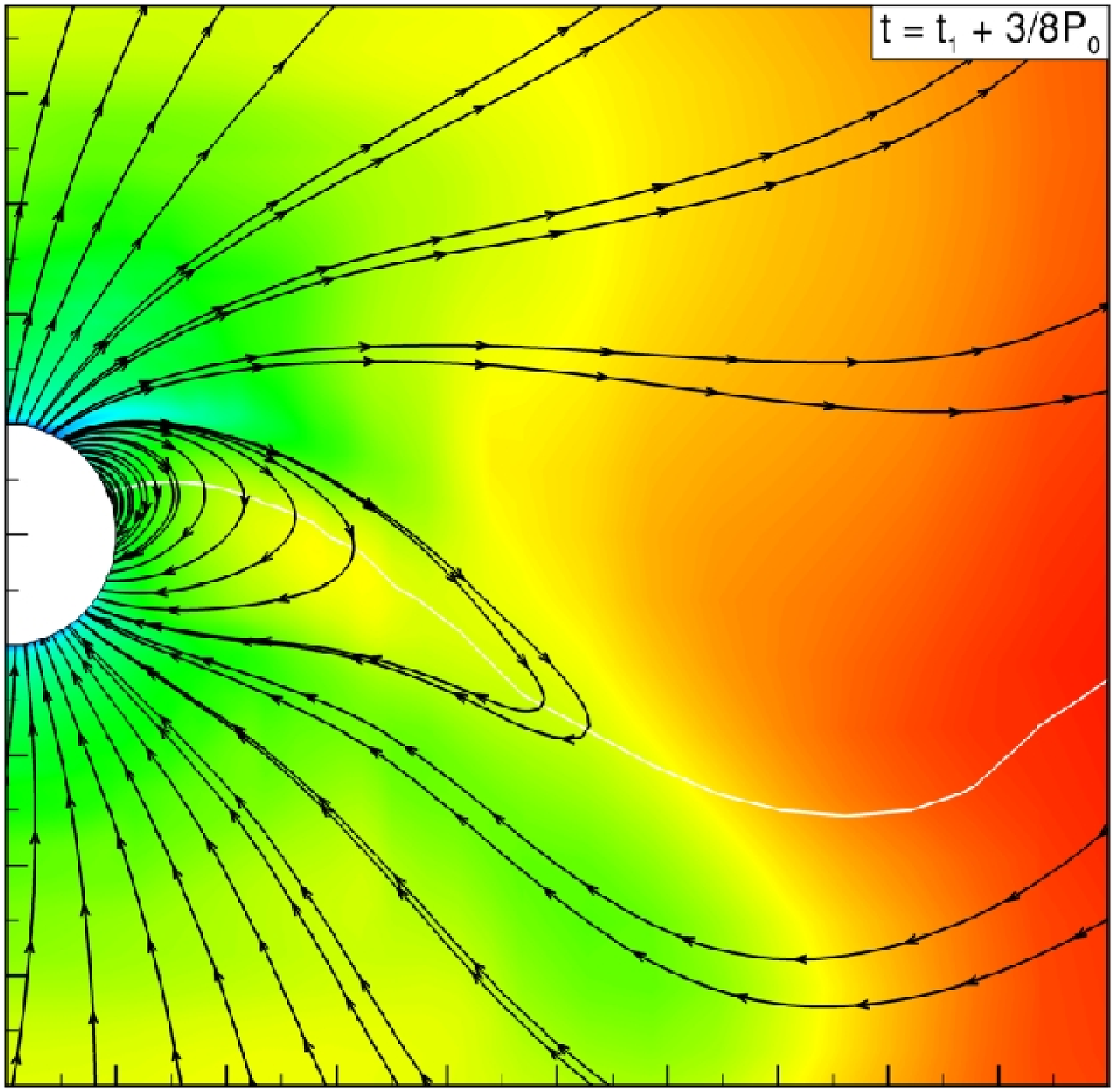}%
  \includegraphics[height=5.5cm]{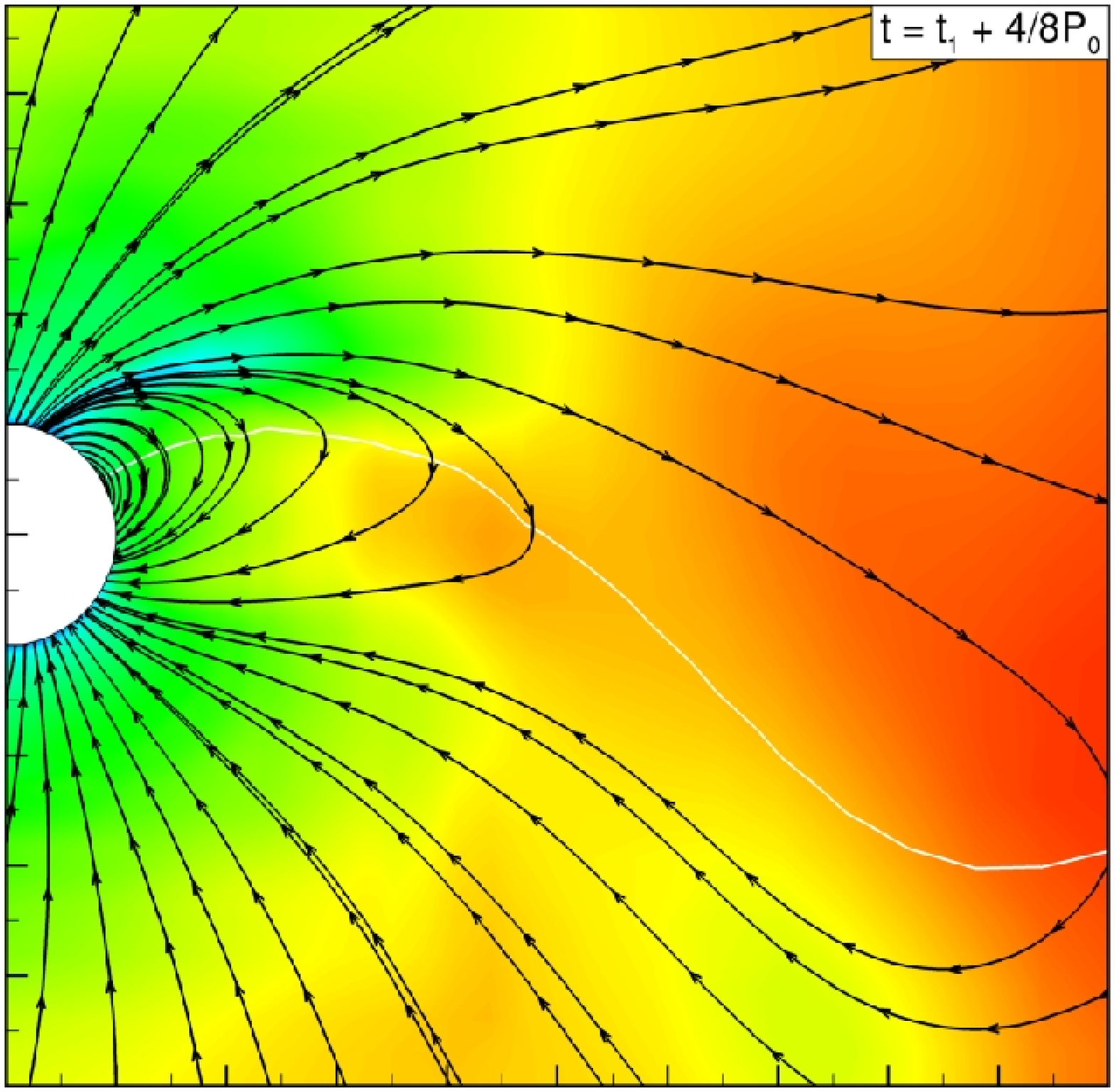}%
  \includegraphics[height=5.5cm]{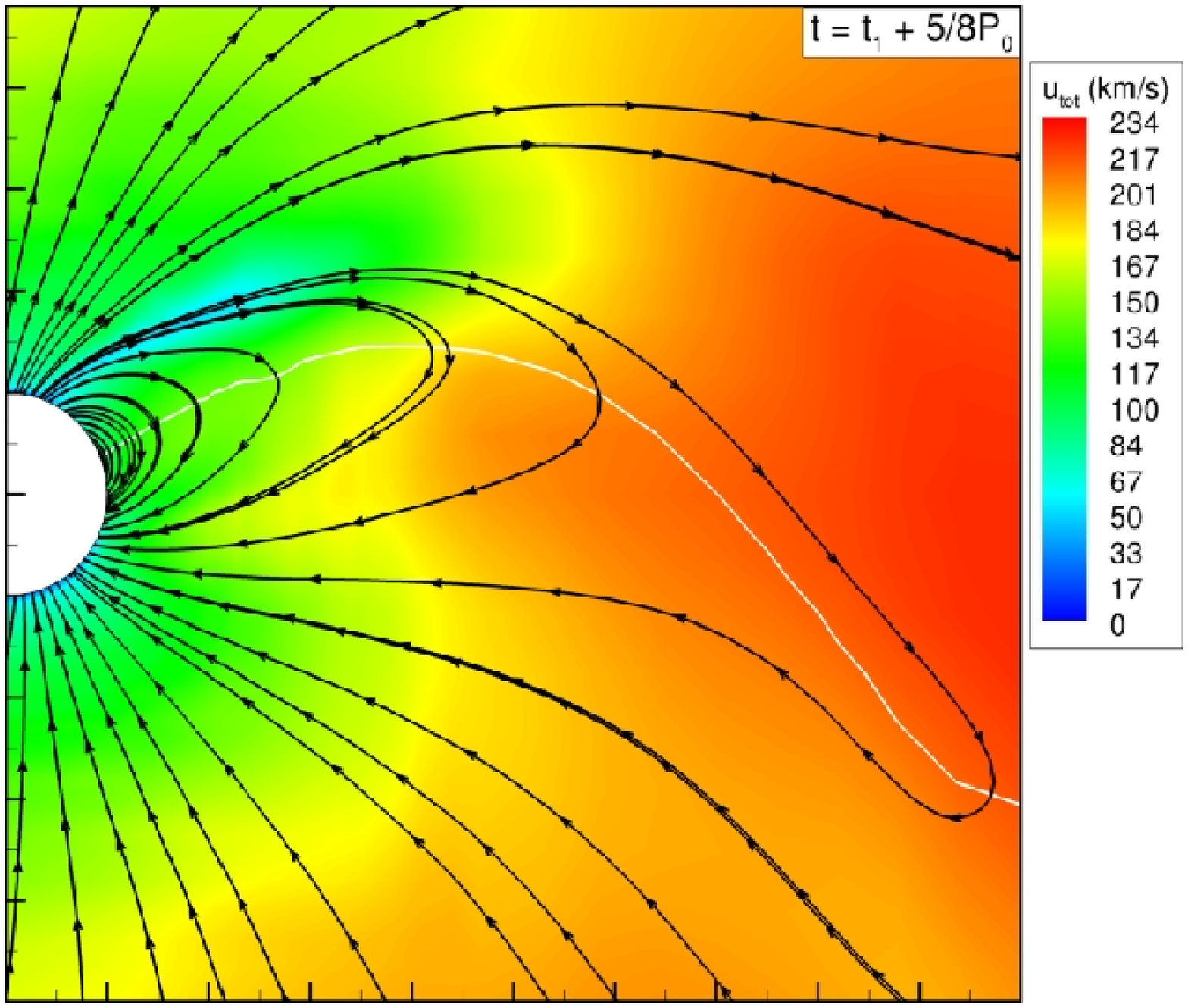}\\
  \includegraphics[height=5.5cm]{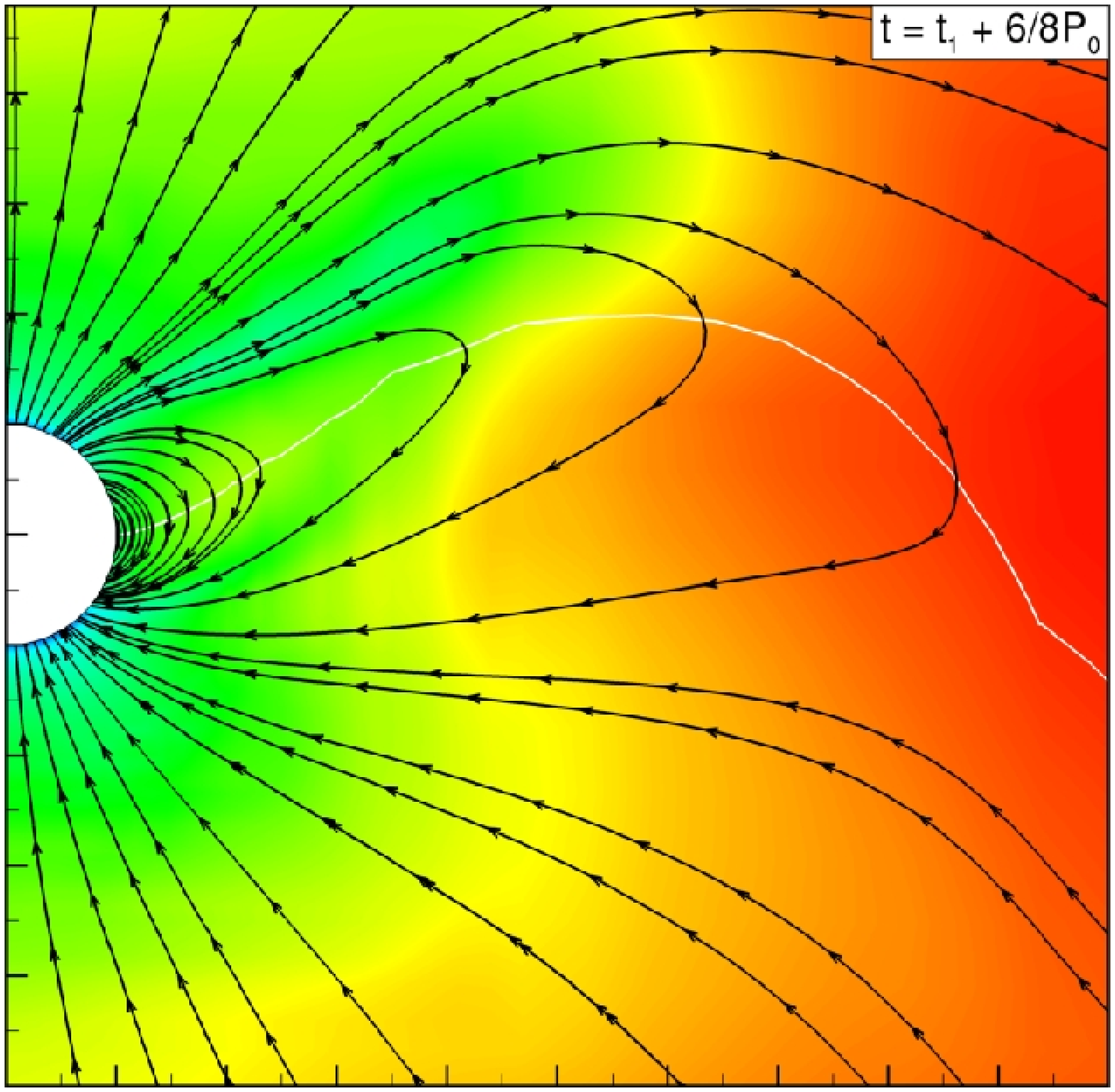}%
  \includegraphics[height=5.5cm]{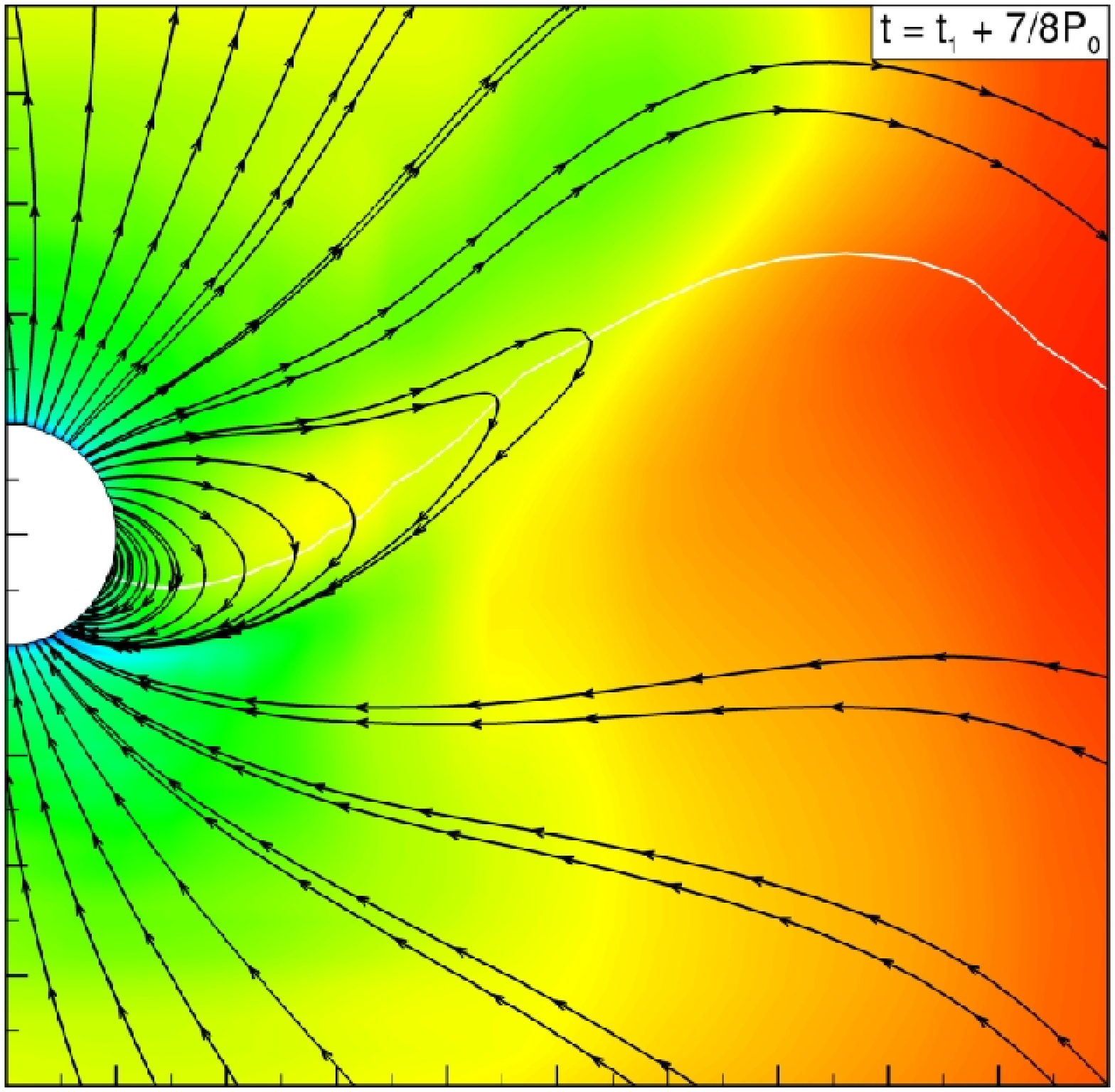}%
  \includegraphics[height=5.5cm]{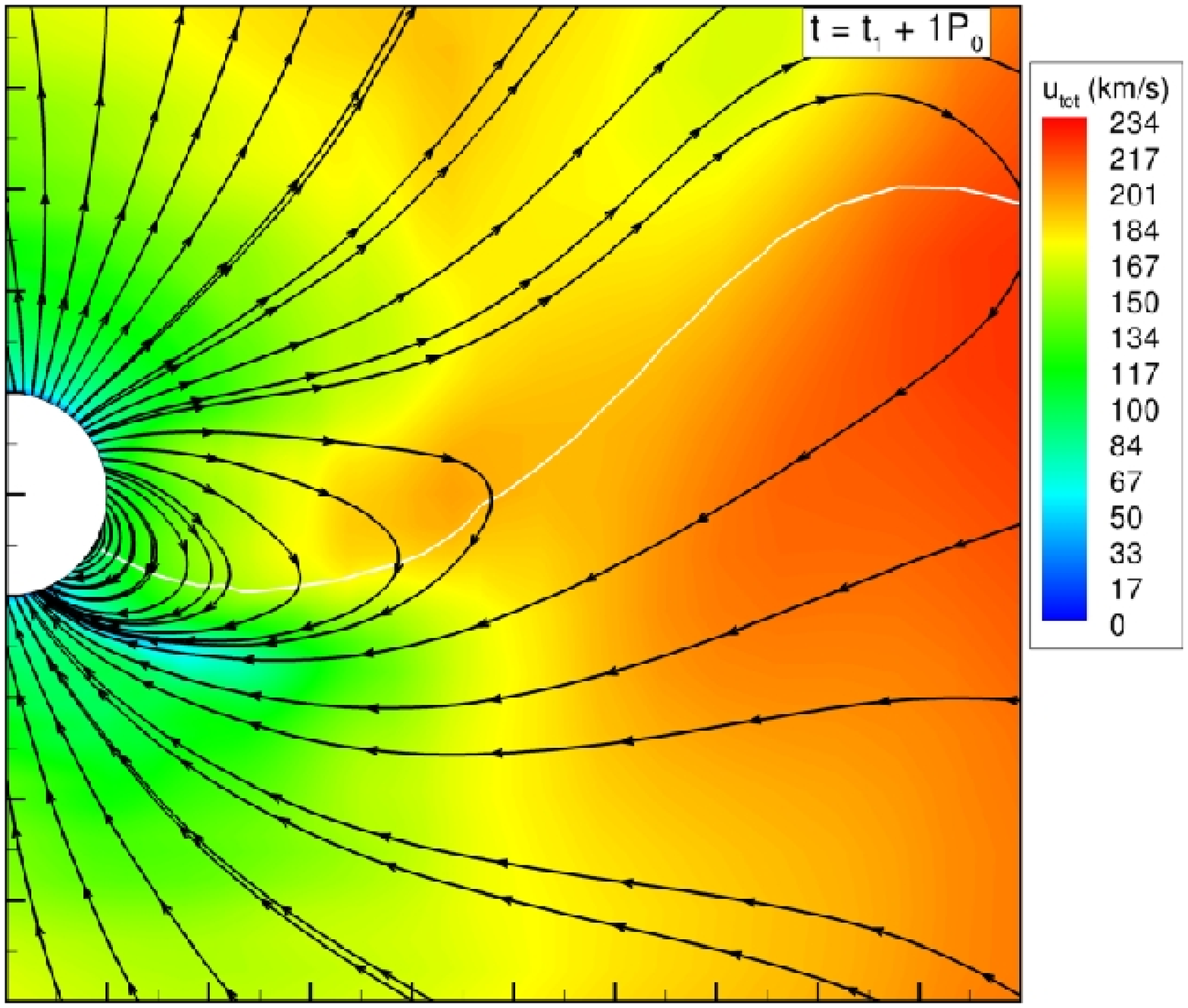}%
  \caption{Temporal evolution of the fiducial case T04 during one full stellar rotational period. The panels show $9$ different instants, starting from $t=t_1=216~$h, and increasing in multiples of $1/8~P_0 = 3~$h, until $t=t_1+P_0=240$~h. We plot meridional cuts of the total wind velocity $u_{\rm tot}$. The magnetic field lines are represented by black lines, and the white line is the contour-line of $B_r = 0$. The $x$-axis ranges from $0$ to $10~r_0$ and the $z$-axis ranges from $-5~r_0$ to $5~r_0$. \label{fig.evolution}}
\end{figure*}

Figure~\ref{fig.evolution} illustrates the time-dependent behavior of our simulations, where we show meridional cuts of the total velocity of the wind for nine instants during one full period of rotation of the star for case T04, our fiducial case. The first panel represents a given instant $t = t_1$; the subsequent panels increase in multiples of $1/8~P_0$, until the cycle completes at instant $t=t_1+P_0$. The magnetic field lines are represented by black lines and the white line denotes the contour where the magnetic field changes polarity, i.e., when $B_r = 0$. It can be seen that the initial instant $t = t_1$ and the final instant $t=t_1+P_0$ of a given stellar rotational period are identical, which exemplifies the periodic behavior of the simulation. The magnetic field has zones of open and closed field lines. The zone of closed field lines rotates around the stellar equatorial plane ($z=0$), as is readily seen if one uses the contour-line of $B_r=0$ (white line) as a guide.

\begin{figure} 
  \includegraphics[height=7cm]{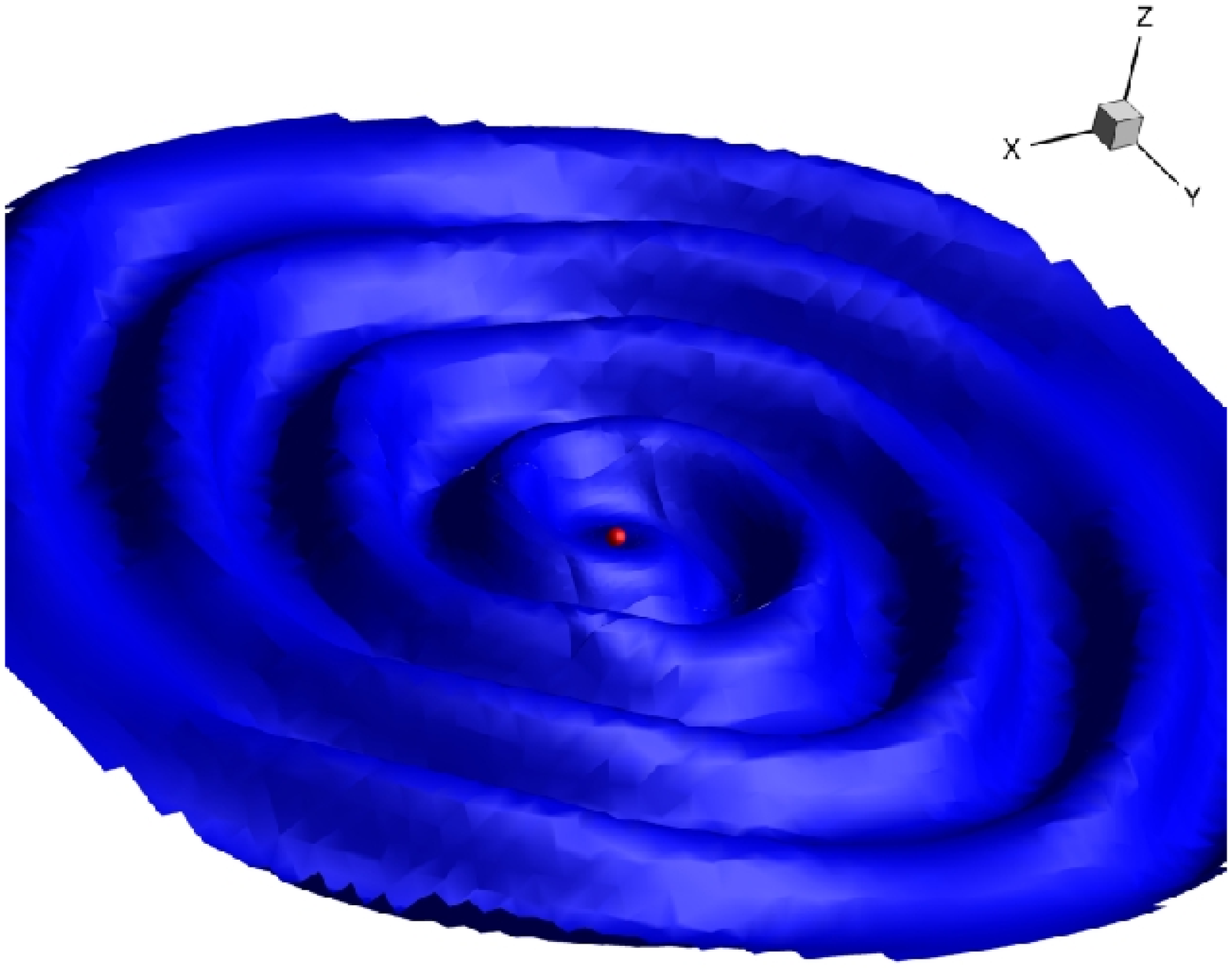}
  \caption{Three-dimensional view of the isosurface defined by $B_r=0$ for case T04 after $8$ rotations of the star ($t=192$~h). The star is shown in red. \label{fig.Br-isosurface}}
\end{figure}

It should be reminded that $B_r=0$ defines a bi-dimensional surface, but because Fig.~\ref{fig.evolution} (as well as several other figures we present later on) is a meridional cut, $B_r=0$ is shown as a contour-line. When there is no misalignment ($\theta_t=0$), the surface $B_r=0$ coincides with the equatorial plane ($z=0$). However, in the case where $\theta_t \ne 0$, $B_r=0$ defines a wavy, time-dependent surface for an observer that is in an inertial referential frame. In the corotating frame of the star, the surface would still be wavy, but it will appear as static. In both cases ($\theta_t = 0$ and $\theta_t \ne 0$), such a surface is the locus of points at the tip of the closed magnetic field lines. Therefore, we expect cusp-like structures (i.e., helmet-streamers) to also oscillate with the same rotational period of the star. The three-dimensional view of the surface $B_r=0$ can be seen in Fig.~\ref{fig.Br-isosurface}. 

\subsection{The effects of the misalignment angle $\theta_t$}
By comparing simulations T01 to T04, we can analyze the effects caused due to small misalignment between the stellar rotation axis and the stellar magnetic dipole moment vector in the wind structure and magnetic field configuration.

\begin{figure*} 
  \includegraphics[height=7cm]{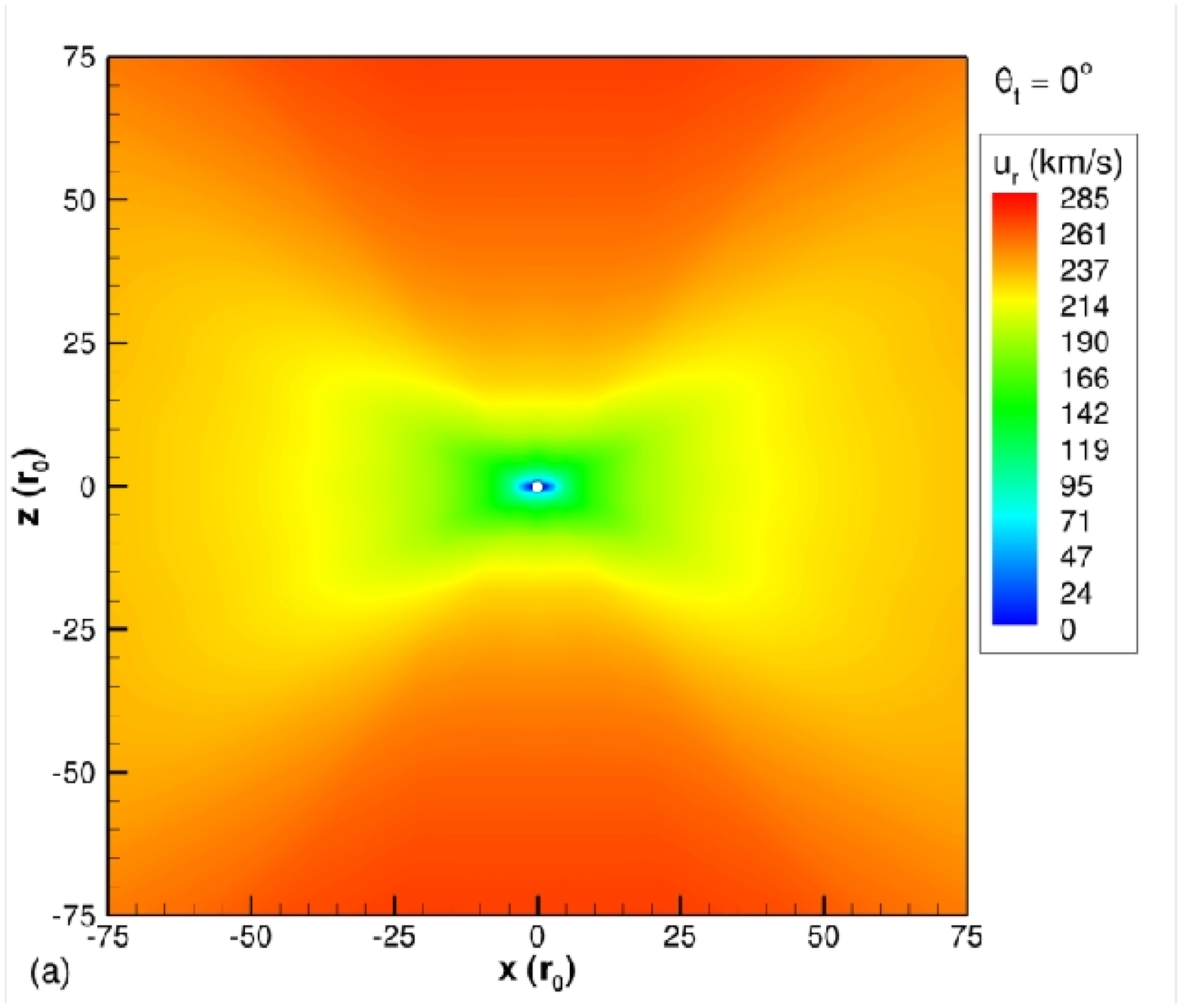}
  \includegraphics[height=7cm]{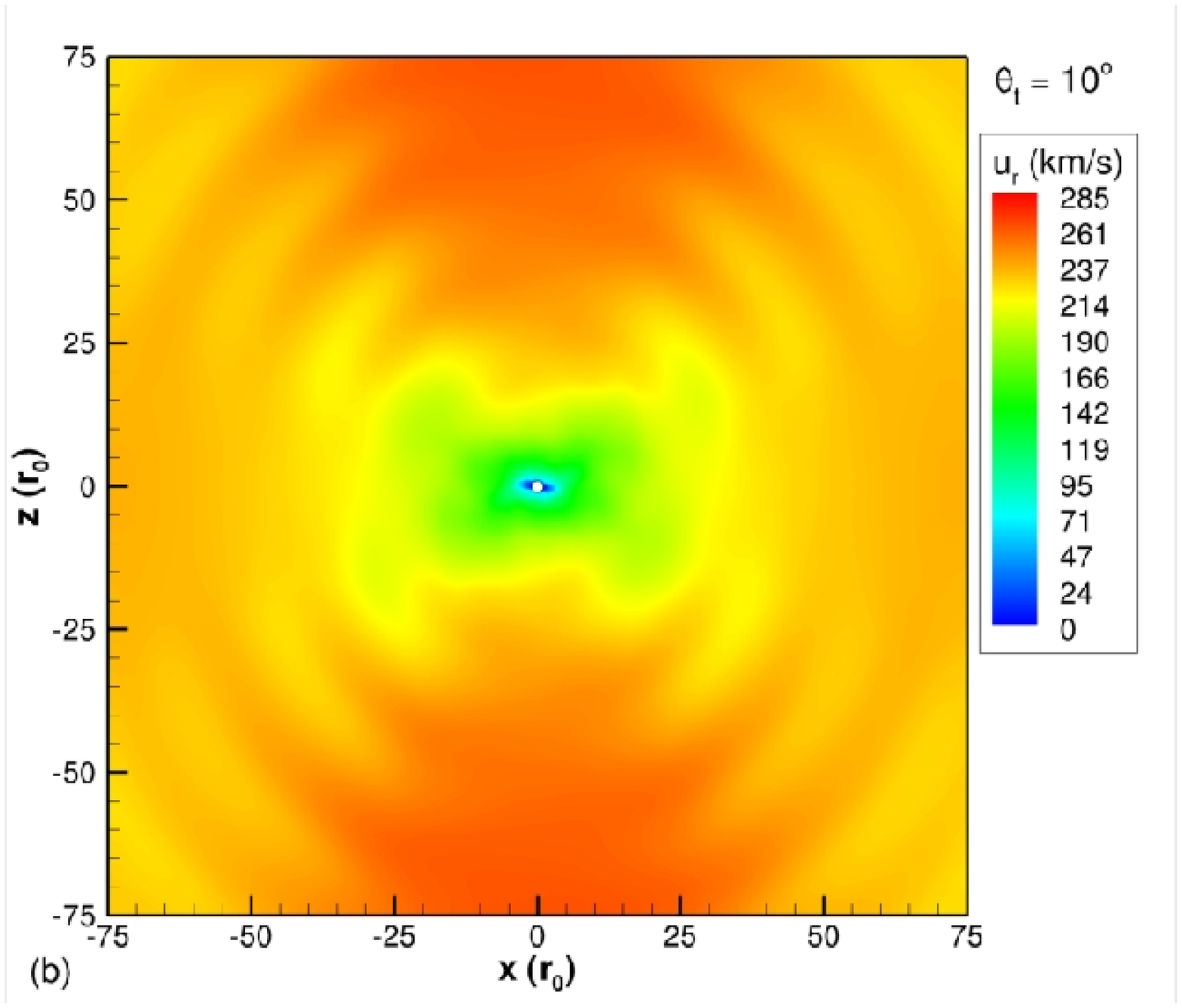}\\
  \includegraphics[height=7cm]{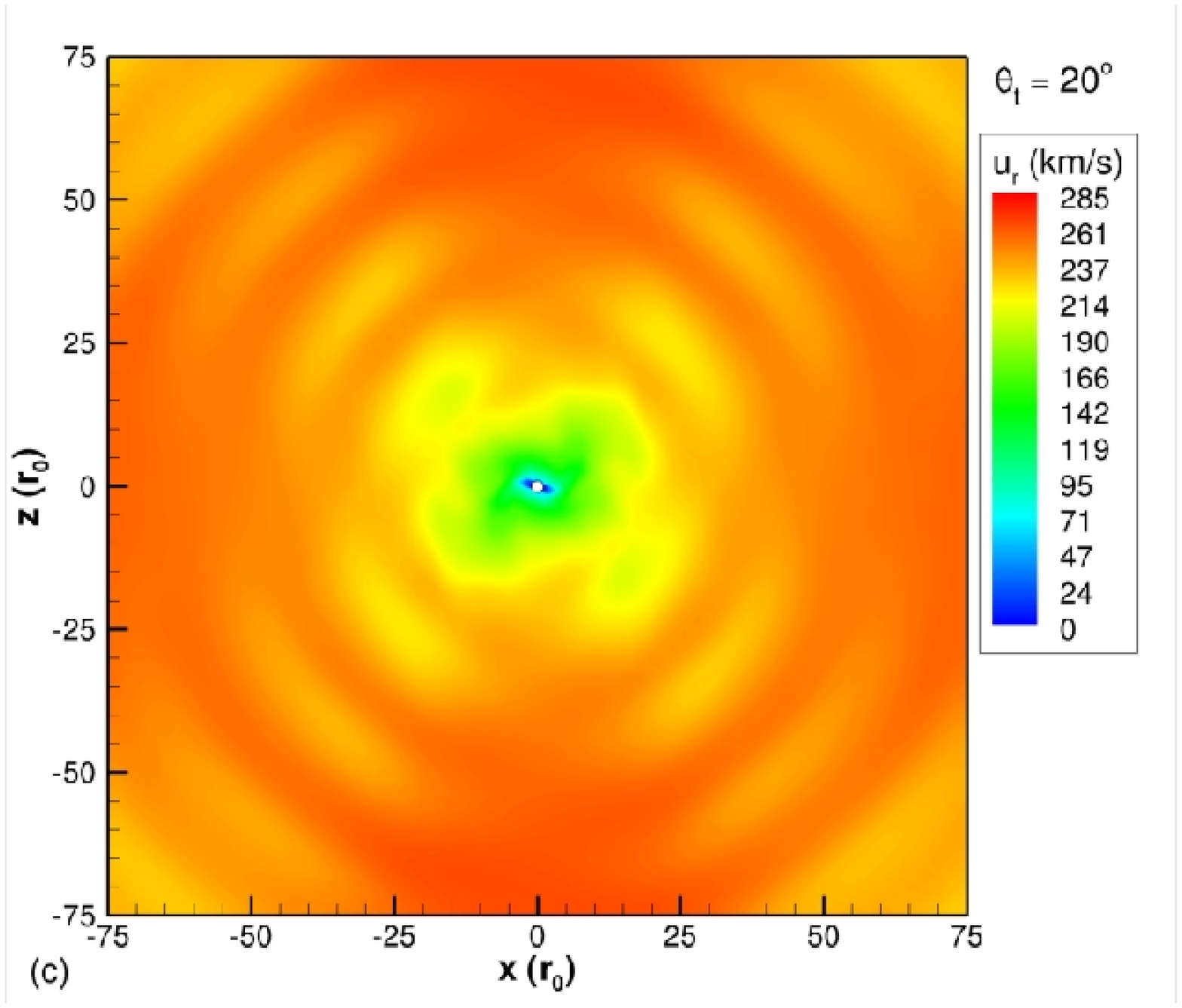}
  \includegraphics[height=7cm]{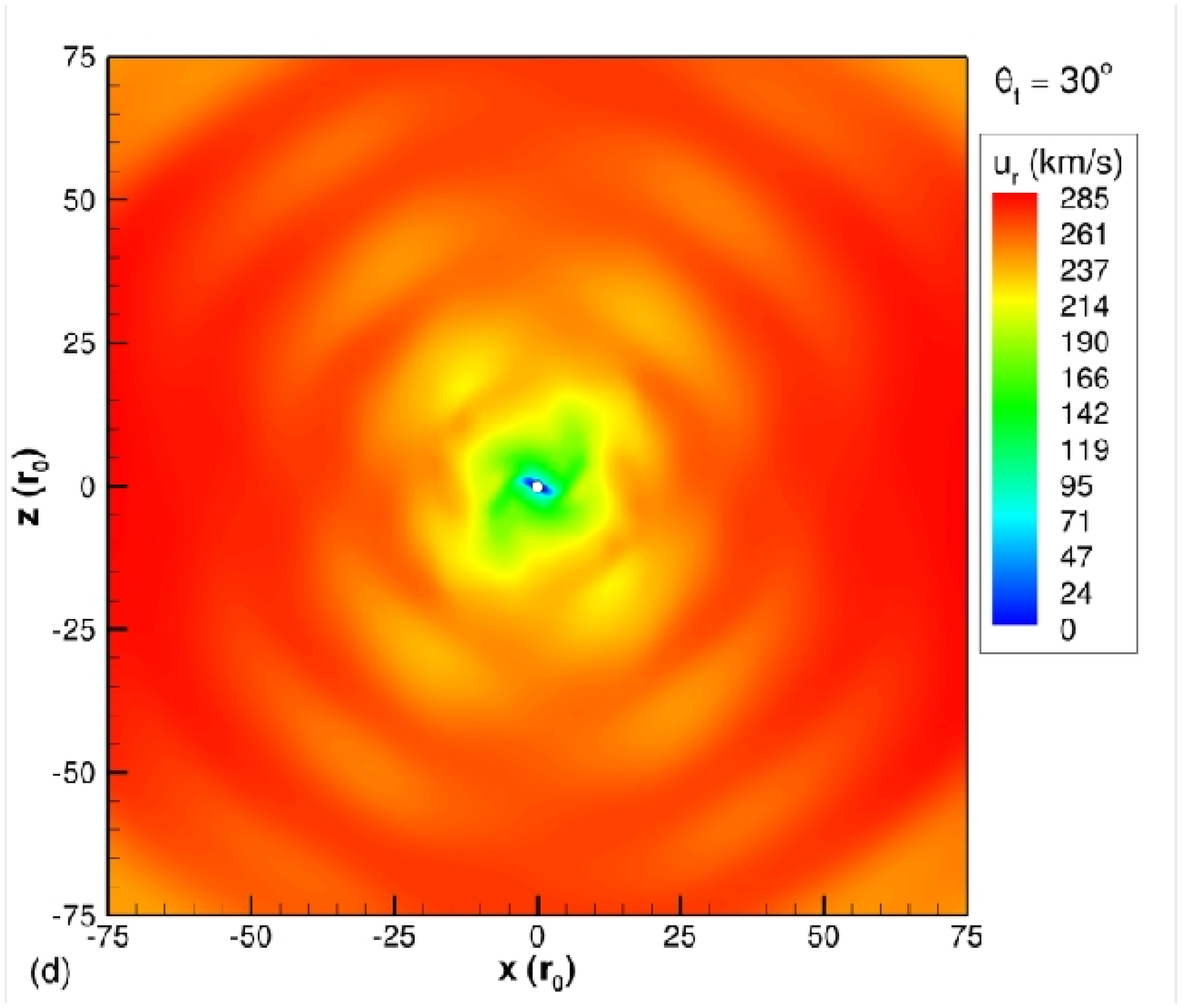}
  \caption{Meridional cuts of radial velocities $u_r$ for (a) the aligned case with $\theta_t=0^{\rm o}$ (T01) and cases with three different misalignment angles (b) $\theta_t=10^{\rm o}$ (T02), (c) $\theta_t=20^{\rm o}$ (T03), and (d) $\theta_t=30^{\rm o}$ (T04). The panels are snapshots of the simulations after $10$ stellar rotations ($t=240$~h). In this figure, we present the meridional cut of the entire simulation box. \label{fig.radial-velocity}}
\end{figure*}

Figure~\ref{fig.radial-velocity} presents radial velocity color maps for the cases T02 ($\theta_t=10^{\rm o}$), T03 ($\theta_t=20^{\rm o}$), T04 ($\theta_t=30^{\rm o}$), and the aligned case T01 ($\theta_t=0^{\rm o}$) for the entire simulation box (shown in the figure is a meridional cut in the $xz$-plane). The four panels are snapshots taken at $t=240$~h, which is a sufficient time to assure that the periodic behavior described in \S\ref{sec.temp.beh} has been achieved. For each of the misaligned cases, we note regions of higher velocities, surrounded by regions of smaller velocities. This oscillatory behavior is seen in all the variables of the wind and is caused by the precession of the stellar magnetic field around the polar axis of the star. We also note that the increase in $\theta_t$ leads to faster winds on average. This is caused by the azimuthal derivative ($\partial / \partial \varphi$) terms in Eqs.~(\ref{eq:continuity_conserve}) to (\ref{eq:energy_conserve}), which vanish in the aligned case. In the momentum equation [Eq.~\ref{eq:momentum_conserve}], for instance, these terms are: the inertial term 
\begin{equation}\label{eq.extrainertial}
\frac{u_\varphi}{r \sin \theta} \left( \frac{\partial {u_r}}{\partial \varphi}\hat{\bf r}  +  \frac{\partial {u_\theta}}{\partial \varphi}\hat{\bf \theta} +  \frac{\partial {u_\varphi}}{\partial \varphi} \hat{\bf \varphi}\right) \, ,
\end{equation}
the pressure gradient in the azimuthal direction 
\begin{equation}\label{eq.extrathermal}
- \frac{1}{r \sin \theta} \frac{\partial {p}}{\partial \varphi} \hat{\bf \varphi} \, ,
\end{equation}
 and the magnetic force $({\bf \nabla \times B}){\bf \times B}/4\pi$
\begin{equation}\label{eq.extramag}
\frac{1}{4\pi r \sin \theta} \left[ B_\varphi \frac{\partial}{\partial \varphi} (B_r \hat{\bf r} + B_\theta \hat{\bf \theta})  - \frac{\partial}{\partial \varphi}  \left( \frac{B_r^2}{2} + \frac{B_\theta^2}{2} \right) \hat{\bf \varphi}\right] \,,
\end{equation}
where the first term inside the brackets refers to a magnetic tension and the second term refers to a gradient of the magnetic pressure. The terms given by Eq.~(\ref{eq.extrainertial}) are negligible when compared to the total inertial force, as well as the terms in Eq.~(\ref{eq.extrathermal}) when compared to the magnitude of the total pressure gradient force. The magnetic tension and pressure presented in Eq.~(\ref{eq.extramag}) contribute more significantly to the acceleration of the wind under an oblique magnetic field configuration. Furthermore, the larger the misalignment angle $\theta_t$ is, the larger is the magnitude of this contribution and, consequently, the larger is the increase in the wind velocity. 

\begin{figure*}
  \includegraphics[height=7cm]{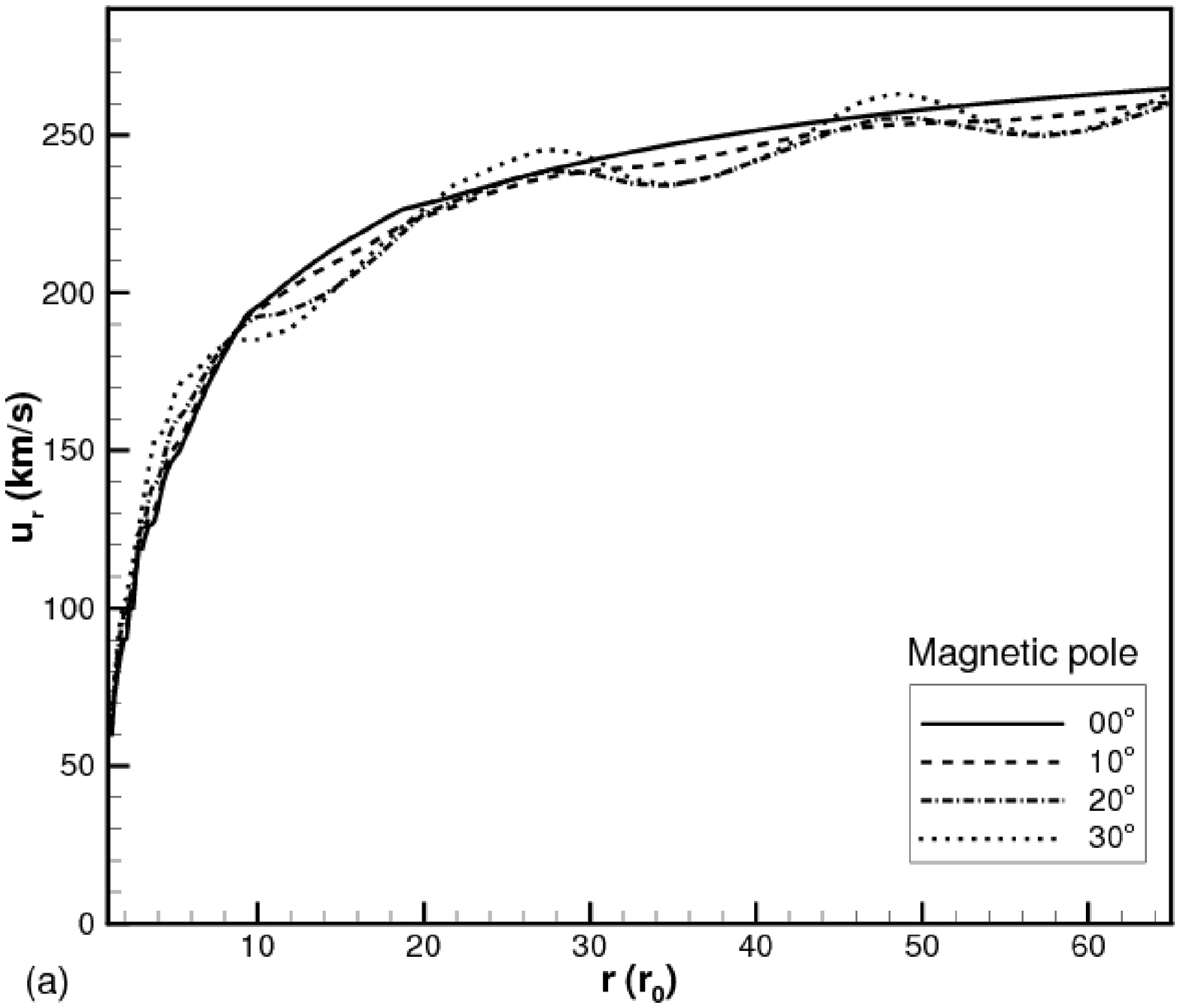}
  \includegraphics[height=7cm]{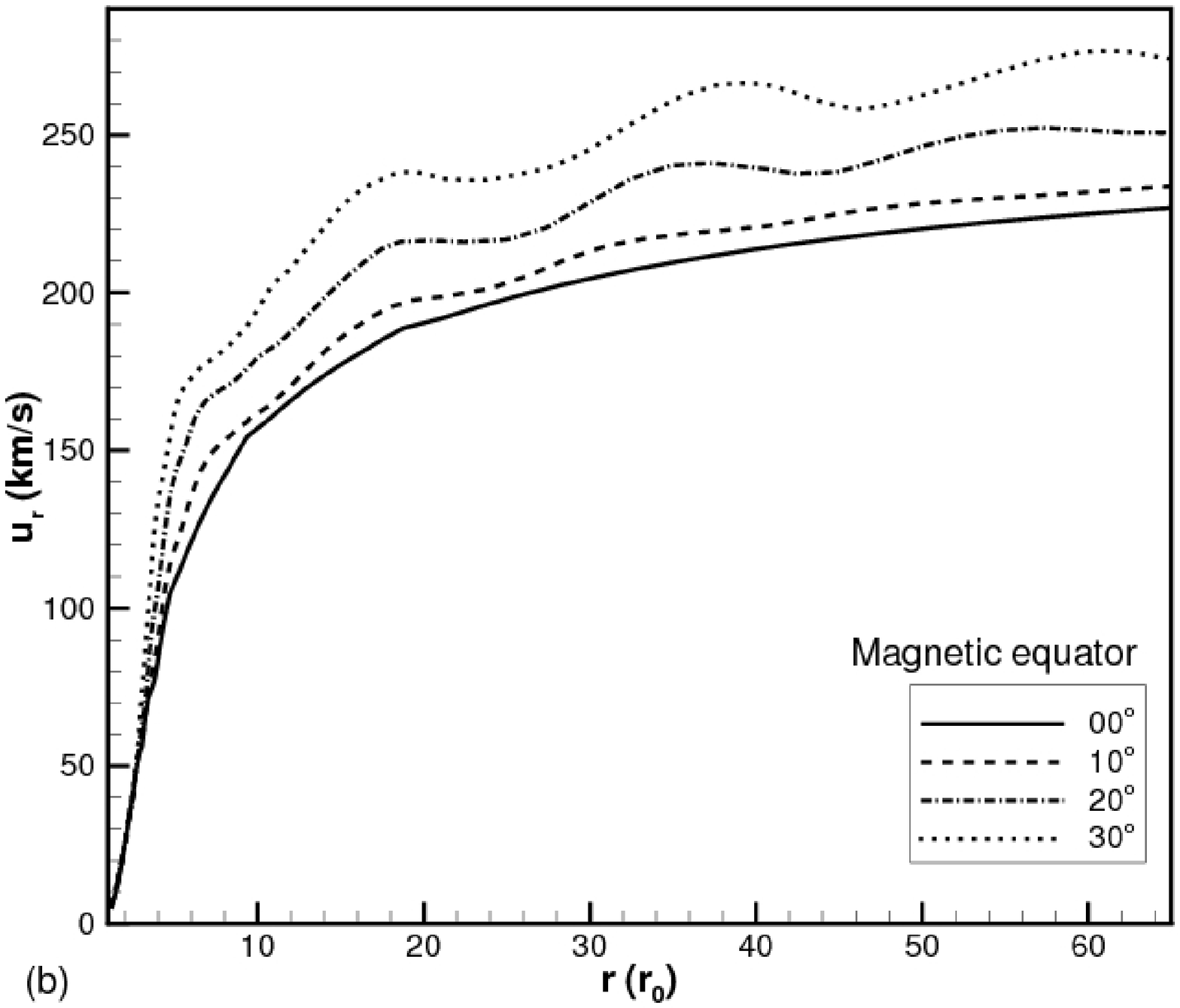}\\
  \includegraphics[height=7cm]{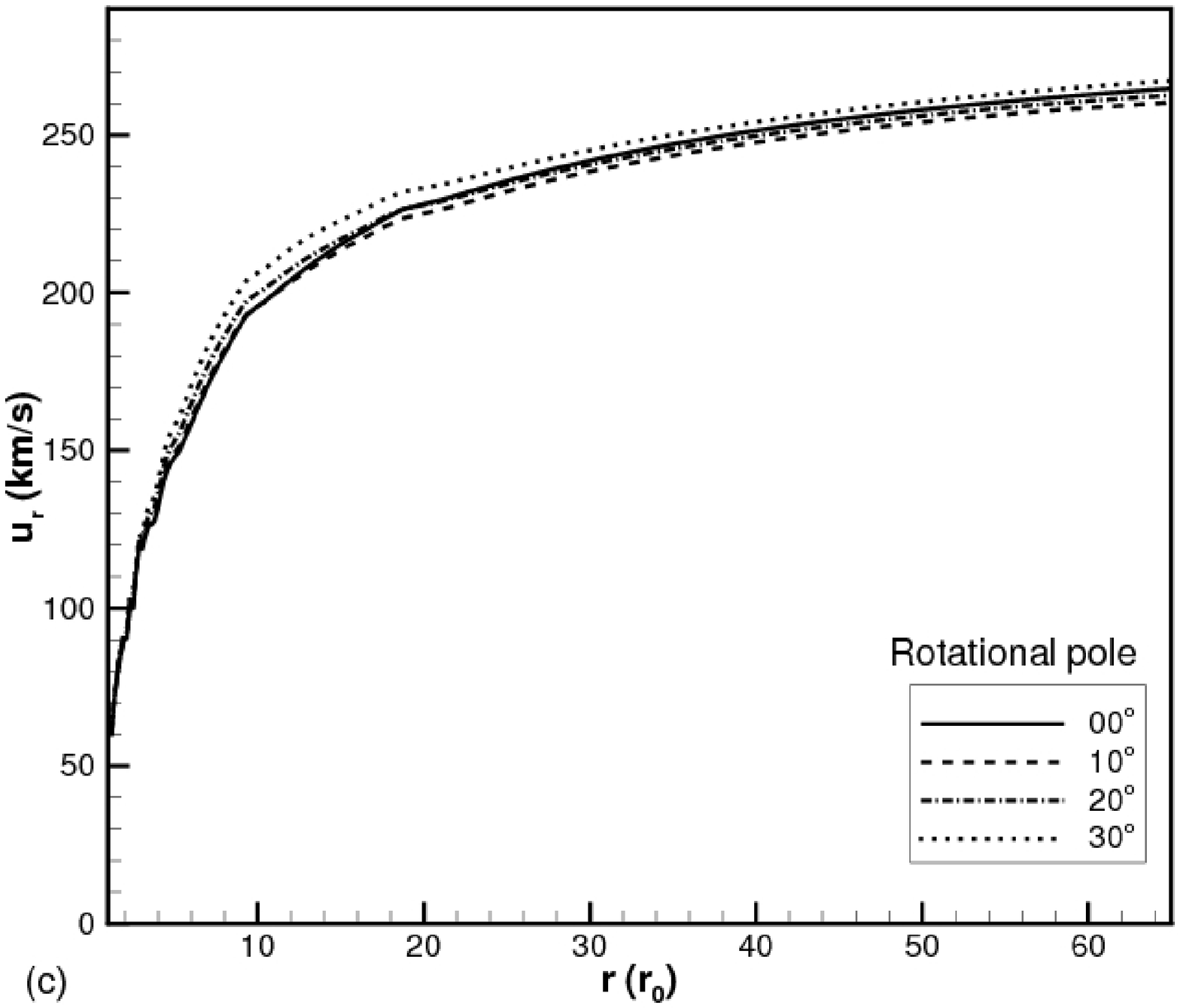}
  \includegraphics[height=7cm]{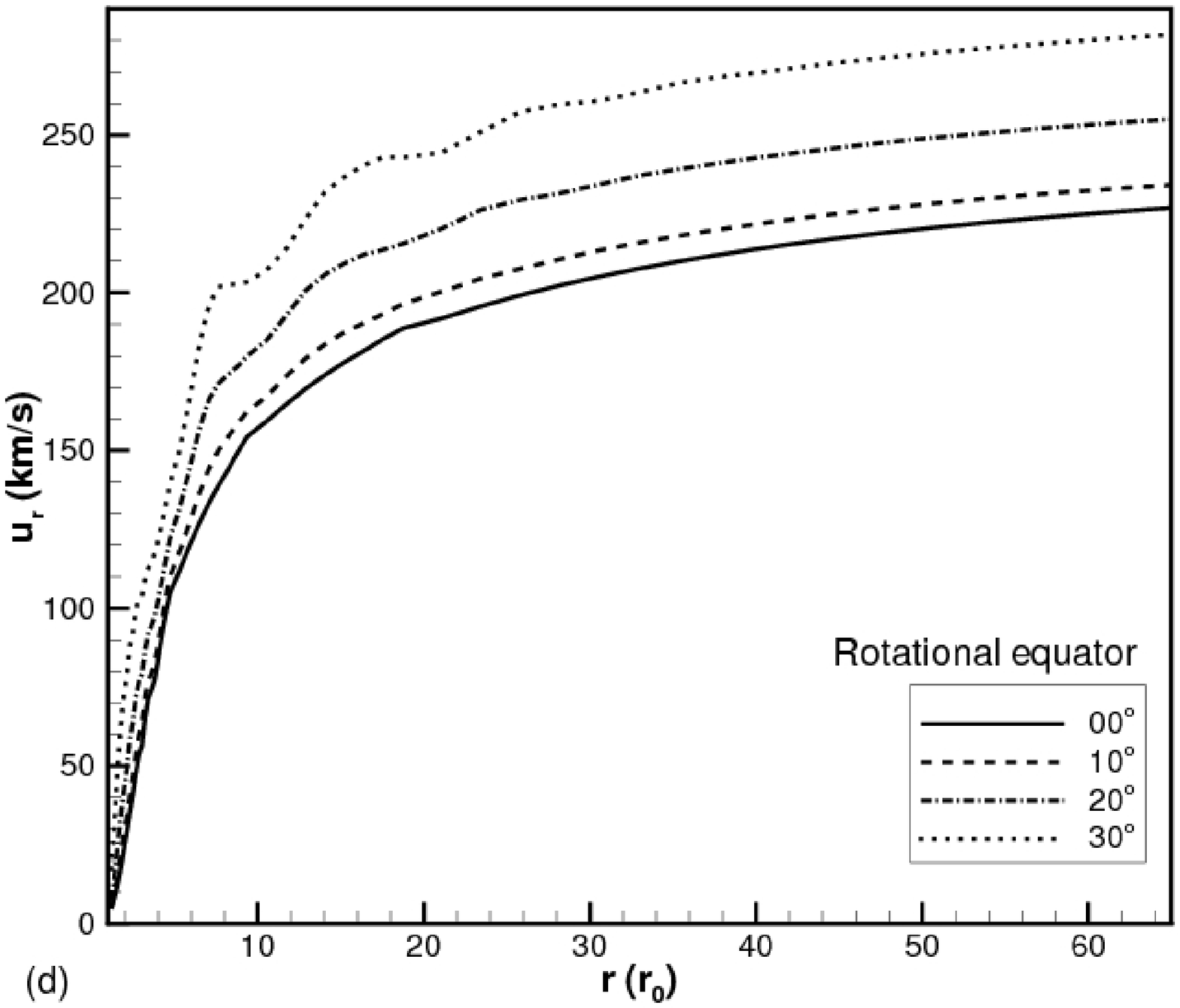}\\
  \caption{Line-cuts in the $xz$-plane of radial velocities $u_r$ for three misalignment angles $\theta_t=10^{\rm o}$ (T02, dashed line), $20^{\rm o}$ (T03, dot-dashed line), and $30^{\rm o}$ (T04, dotted line). For comparison purposes we also include the aligned case ($\theta_t=0^{\rm o}$, T01, solid line). The plots refer to $t=240$~h. Radial cuts along (a) the magnetic pole; (b) the magnetic equator; (c) the rotational pole; (d) the rotational equator.
\label{fig.radial-velocity-cuts}}
\end{figure*}

Figure~\ref{fig.radial-velocity-cuts} shows line-radial cuts in the $xz$-plane of radial velocity for the same instant shown in Fig.~\ref{fig.radial-velocity}: Fig.~\ref{fig.radial-velocity-cuts}a presents radial cuts along the {\it magnetic pole}, i.e., along co-latitude $\theta =10^{\rm o}$ for T02 (where $\theta_t=10^{\rm o}$), along co-latitude $\theta =20^{\rm o}$ for T03 (where $\theta_t=20^{\rm o}$) and so on; Fig.~\ref{fig.radial-velocity-cuts}b presents radial cuts along the {\it magnetic equator}, i.e., along co-latitude $\theta =100^{\rm o}$ for T02 (where $\theta_t=10^{\rm o}$), along co-latitude $\theta =110^{\rm o}$ for T03 (where $\theta_t=20^{\rm o}$) and so on. We note that along the magnetic pole, the curves of radial velocity for the tilted cases oscillate around the curve assigned for the aligned case (black solid line, $\theta_t=0^{\rm o}$). However, along the magnetic equator, the radial velocity increases as $\theta_t$ gets larger, as explained in the previous paragraph. The wind radial velocity at $\sim 75~r_0$ is $227~$km~s$^{-1}$ for case T01, $234~$km~s$^{-1}$ for T02, $251~$km~s$^{-1}$ for T03, and $275~$km~s$^{-1}$ for T04. Figures~\ref{fig.radial-velocity-cuts}c and \ref{fig.radial-velocity-cuts}d presents the same cuts along the rotational poles and equator, respectively.

\begin{figure*} 
  \includegraphics[height=7cm]{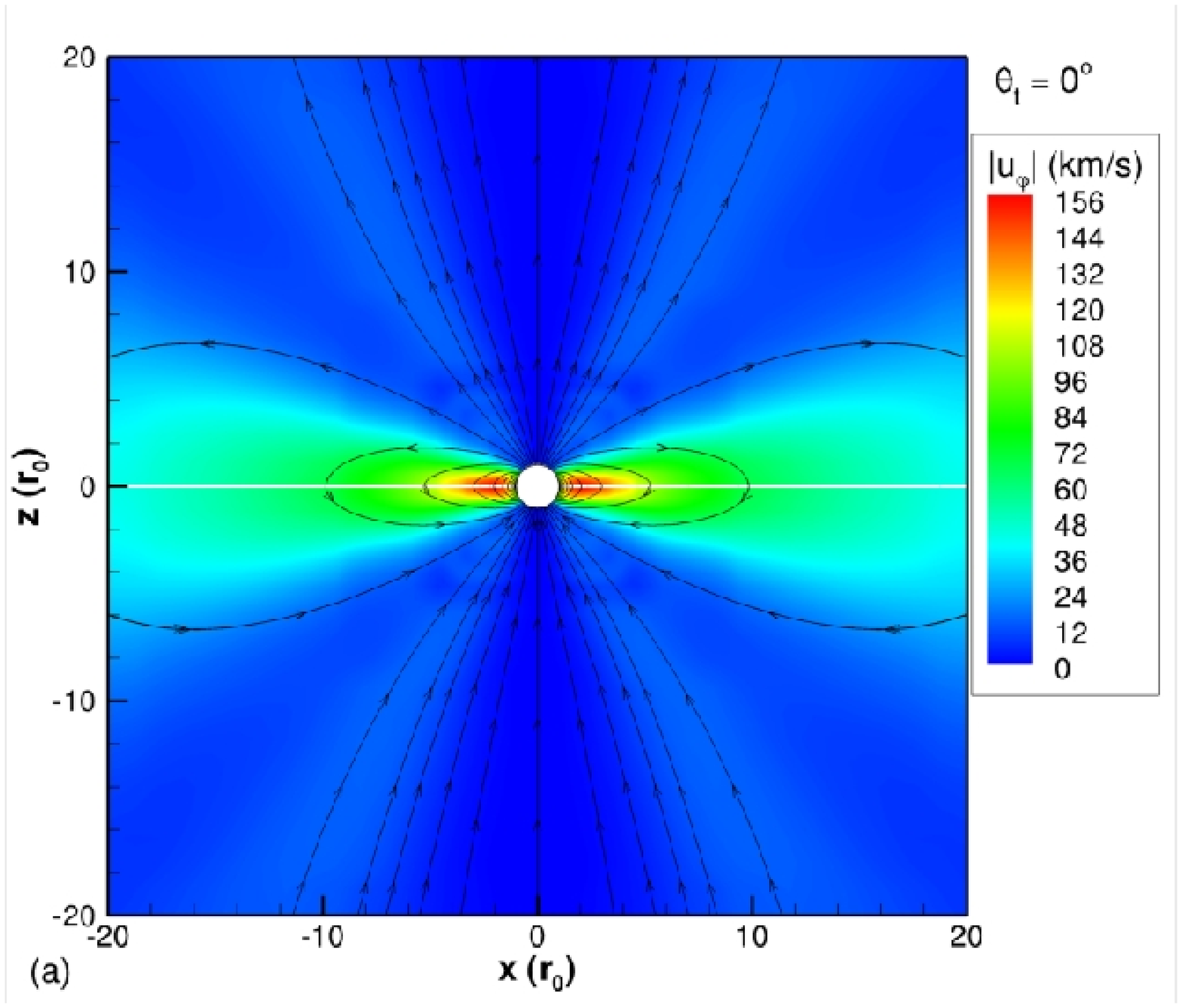}
  \includegraphics[height=7cm]{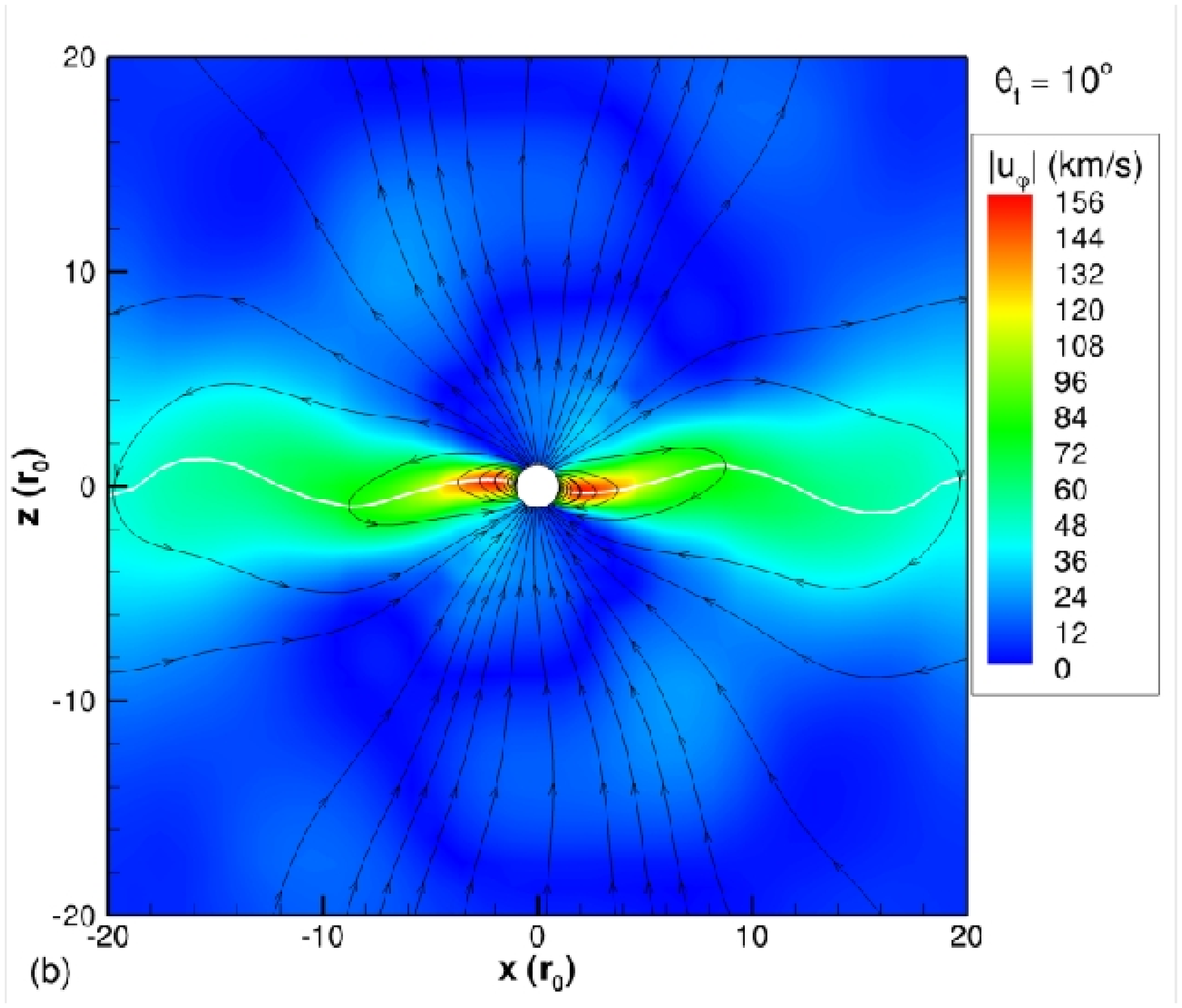}\\
  \includegraphics[height=7cm]{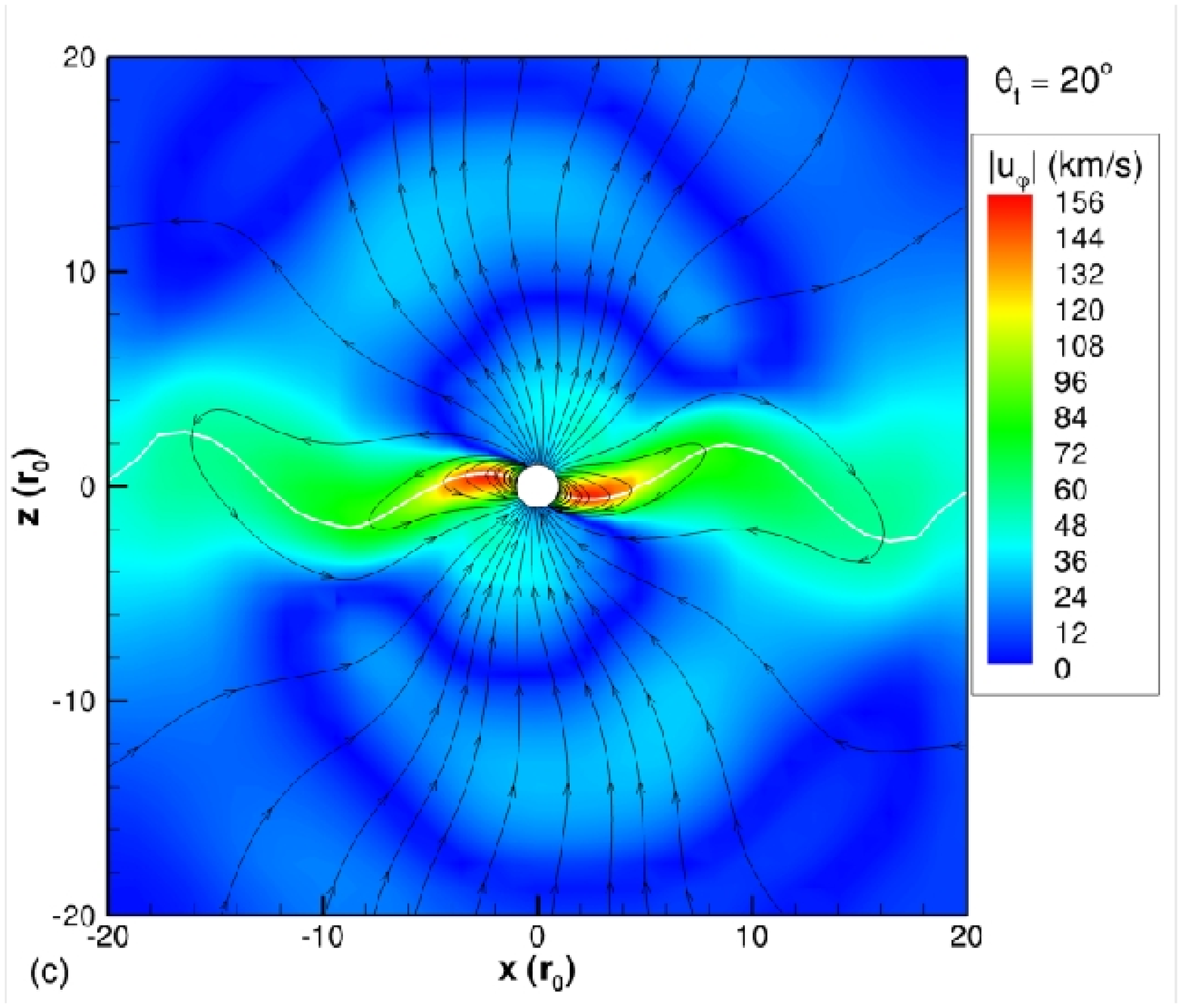}
  \includegraphics[height=7cm]{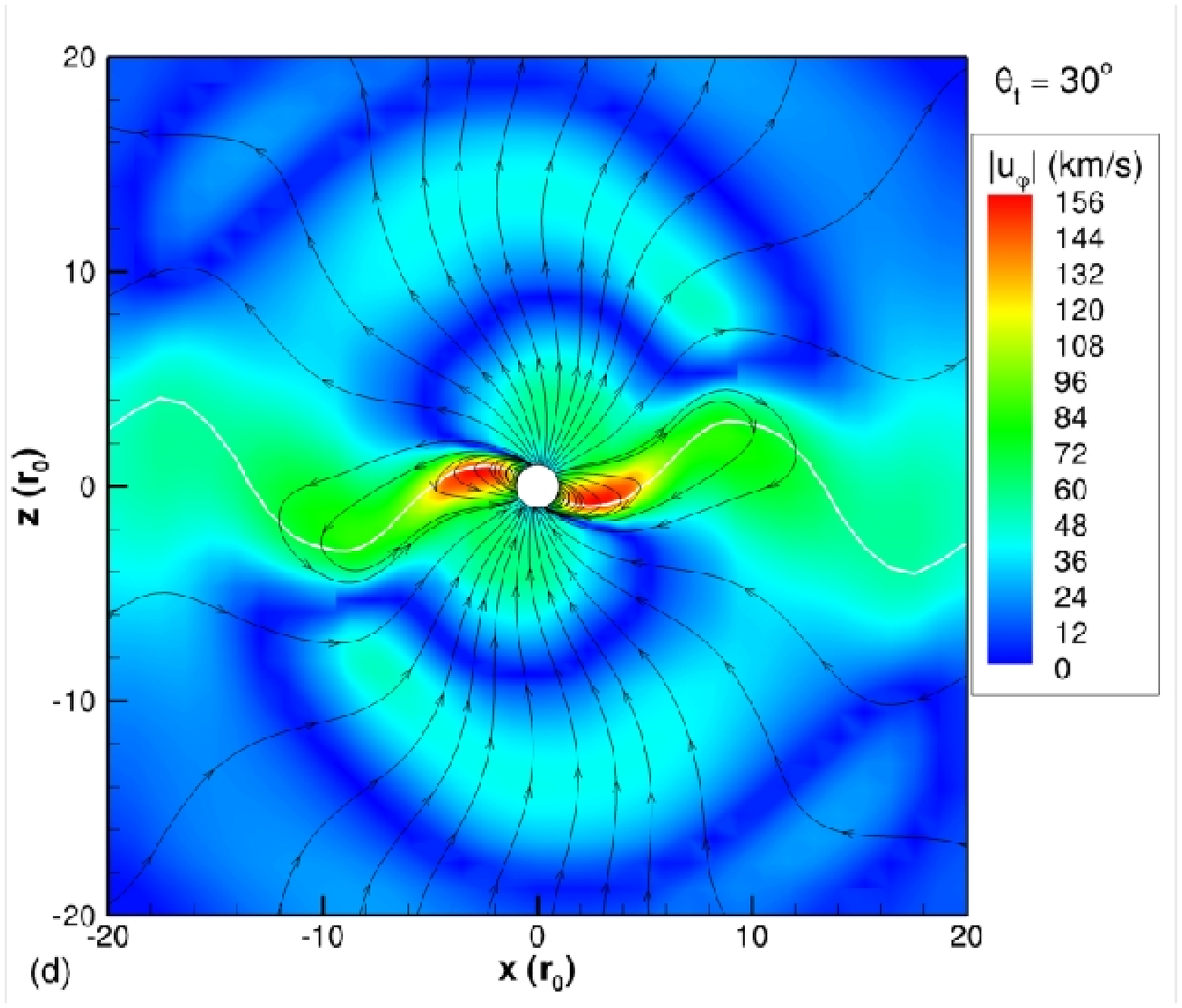}
  \caption{Meridional cuts of azimuthal velocities $|u_\varphi|$ for (a) the aligned case (T01), (b) $\theta_t=10^{\rm o}$ (T02), (c) $\theta_t=20^{\rm o}$ (T03), and (d) $\theta_t=30^{\rm o}$ (T04). The panels are snapshots of the simulations at $t=240$~h. Black and white lines have the same meaning as in Fig.~\ref{fig.evolution}.
\label{fig.Uphi-snapshot}}
\end{figure*}

Figure~\ref{fig.Uphi-snapshot} presents the inner portion of our simulation boxes at $t=240$~h for cases T01 to T04. The magnetic field lines are represented by black lines, and the white line represents the contour-line where $B_r = 0$. By following the white line in Fig.~\ref{fig.Uphi-snapshot}, we note that the closed magnetic field lines, as well as the open field lines, are not rigid, presenting a warped zone around the rotational equatorial plane of the star ($z= 0$). The amplitude of the oscillations gets larger as $\theta_t$ increases. The color maps show azimuthal velocity. There is not a significant variation in the magnitude of $u_\varphi$ between the simulations, but the spatial profile of $u_\varphi$ is highly dependent on the configuration of the magnetic field, and thus on $\theta_t$. As we can see, the highest values of $u_\varphi$ are achieved inside the closed magnetic field lines (close to the star) and the rotating wind is forced to follow the same oscillation pattern of the magnetic field lines. In case of perfect alignment, the maximum azimuthal velocity happens in the equatorial plane of the star (Fig.~\ref{fig.Uphi-snapshot}a).

\begin{figure} 
  \includegraphics[height=7cm]{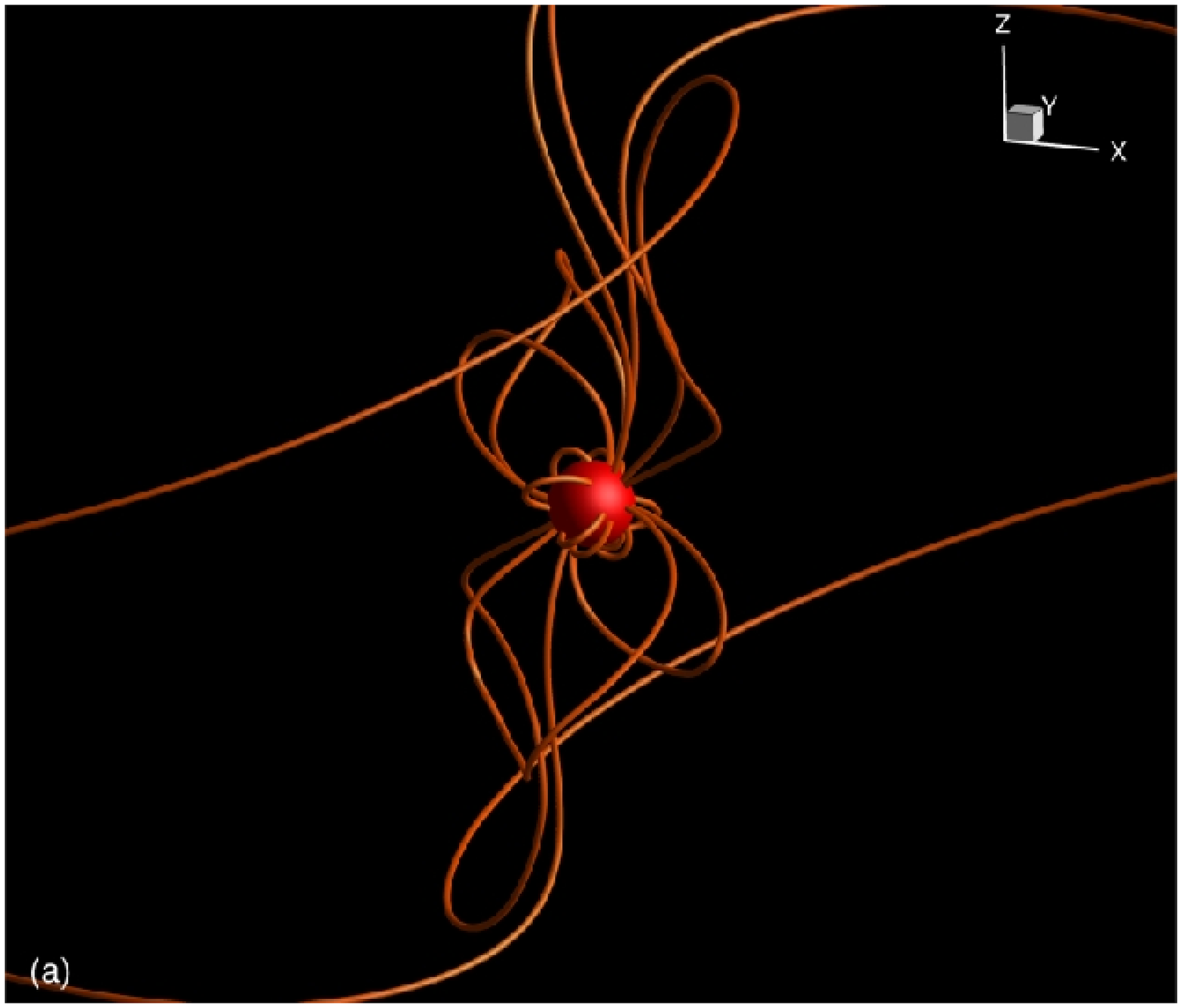}\\
  \includegraphics[height=7cm]{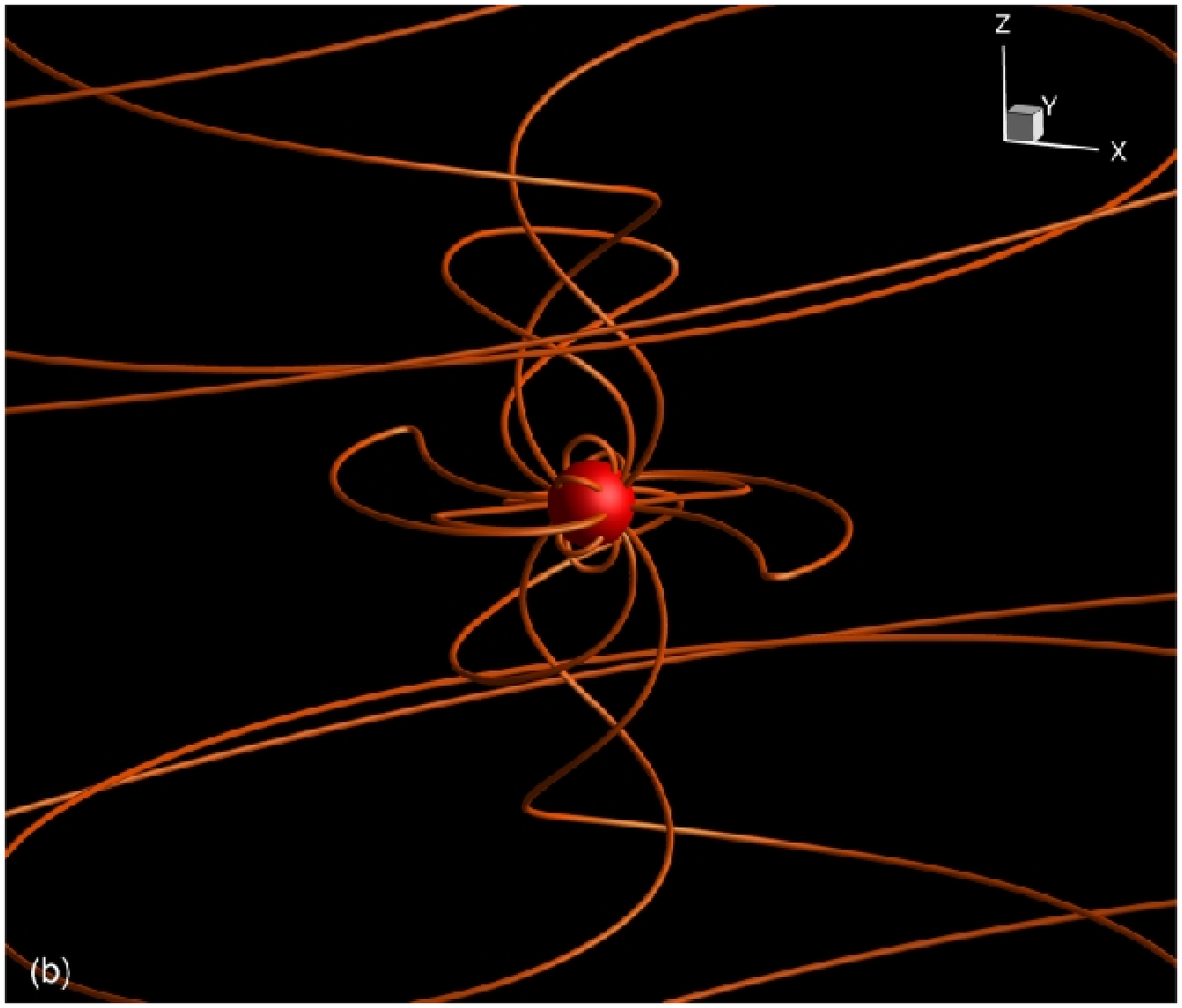}
  \caption{3D view of selected magnetic field lines for (a) $\theta_t=60^{\rm o}$ (T05) and (b) $\theta_t=90^{\rm o}$ (T06). \label{fig.3D.T23.T24}}
\end{figure}

Two further simulations for different misalignment angles were performed: T05 ($\theta_t=60^{\rm o}$) and T06 ($\theta_t=90^{\rm o}$). These simulations represent more extreme case of misalignment, being T06 the case where the axis of the magnetic moment at the base of the coronal wind is perpendicular to the rotational axis. Both of them presents similar characteristics as the cases presented so far for $\theta_t \leq 30^{\rm o}$, with an enhanced wind velocity though. For case T06, the region of lower radial velocity remains in the region of closed magnetic field lines, but this region is now around the rotational poles (at co-latitudes $\theta=0^{\rm o}$, $180^{\rm o}$). Figure~\ref{fig.3D.T23.T24} presents the 3D view of selected magnetic field lines for cases T05 (Fig.~\ref{fig.3D.T23.T24}a) and T06 (Fig.~\ref{fig.3D.T23.T24}b), where it illustrates the inherent three-dimensional nature of our simulations.

\begin{figure*}
  \includegraphics[height=7cm]{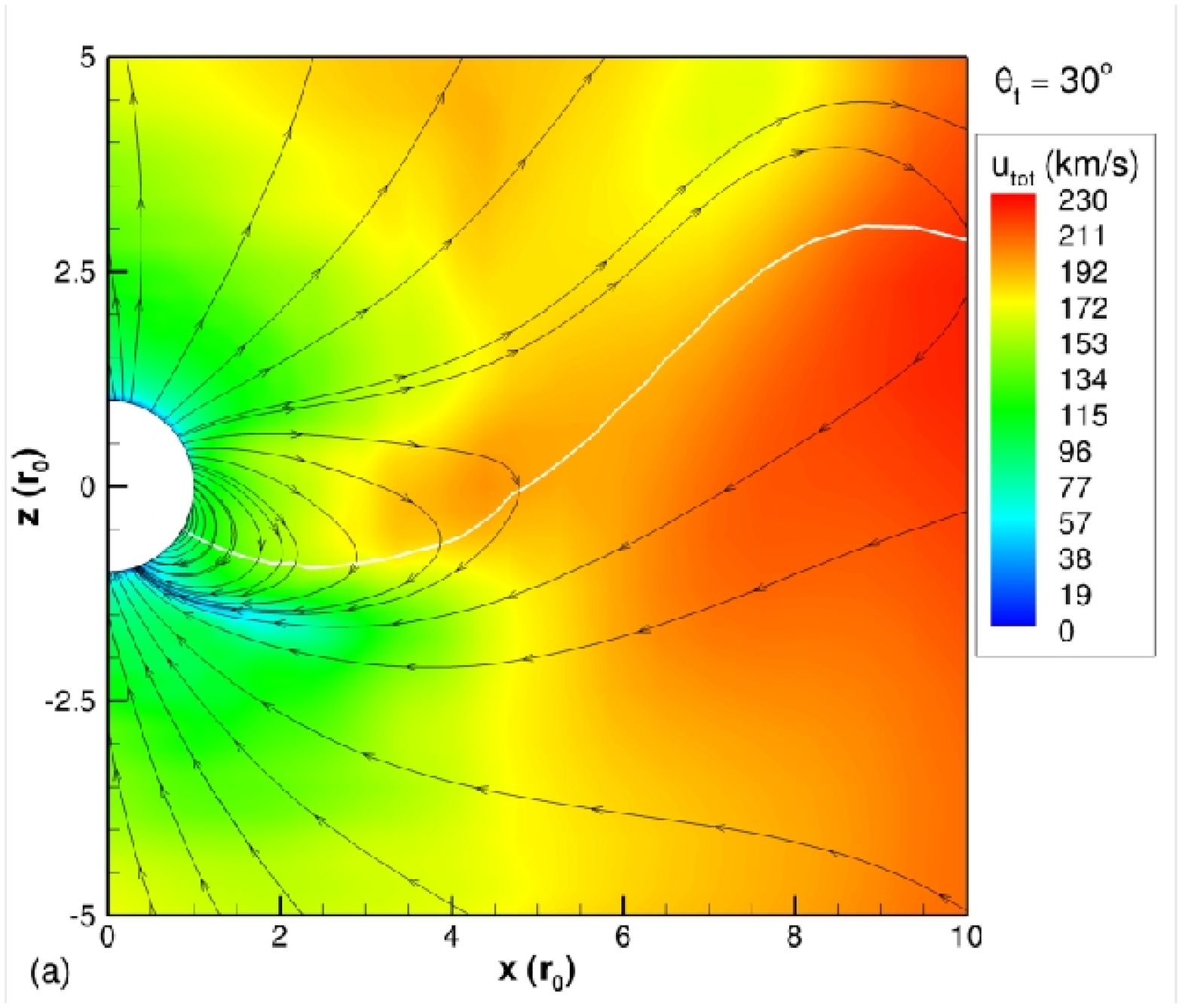}
  \includegraphics[height=7cm]{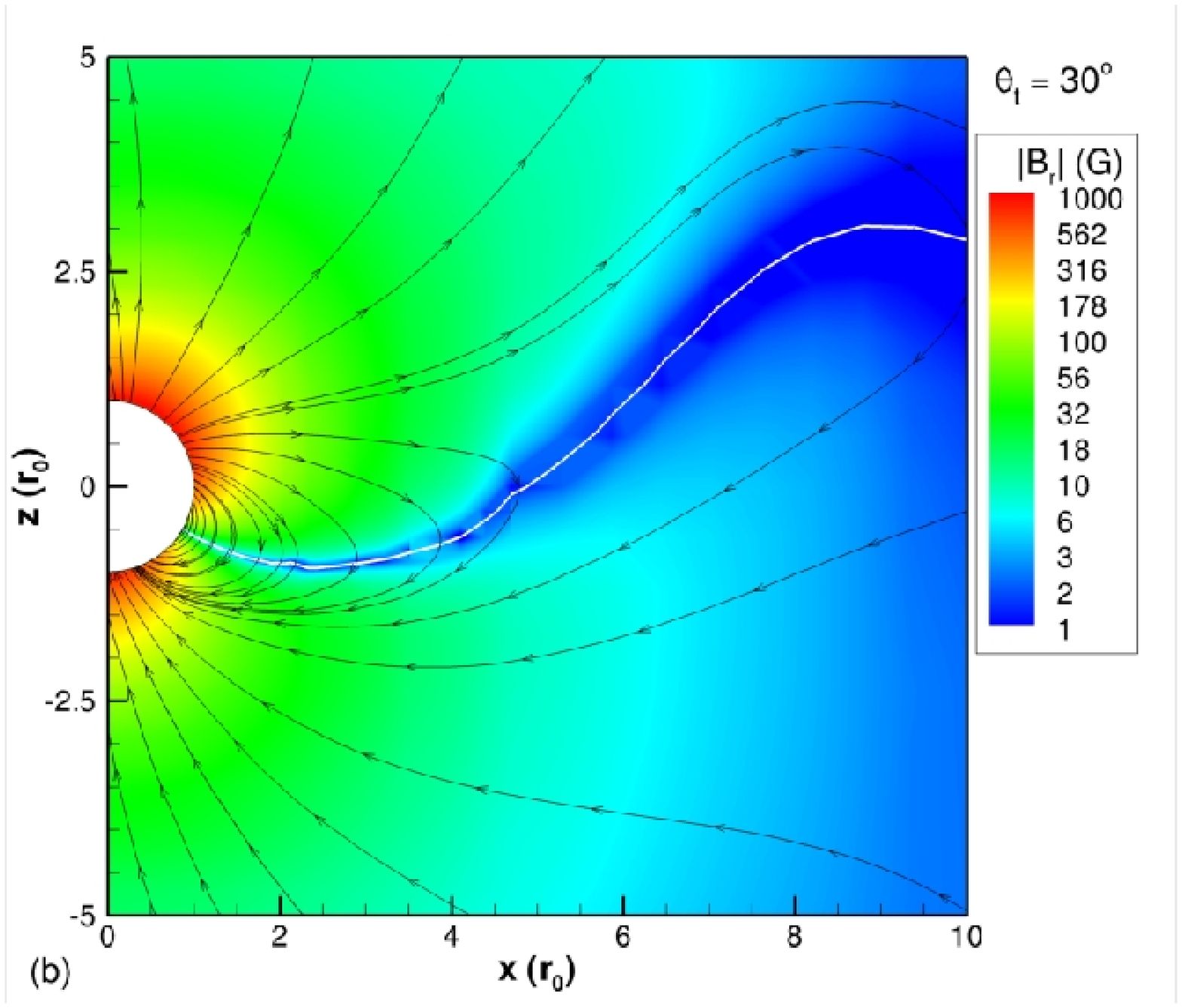}\\
  \includegraphics[height=7cm]{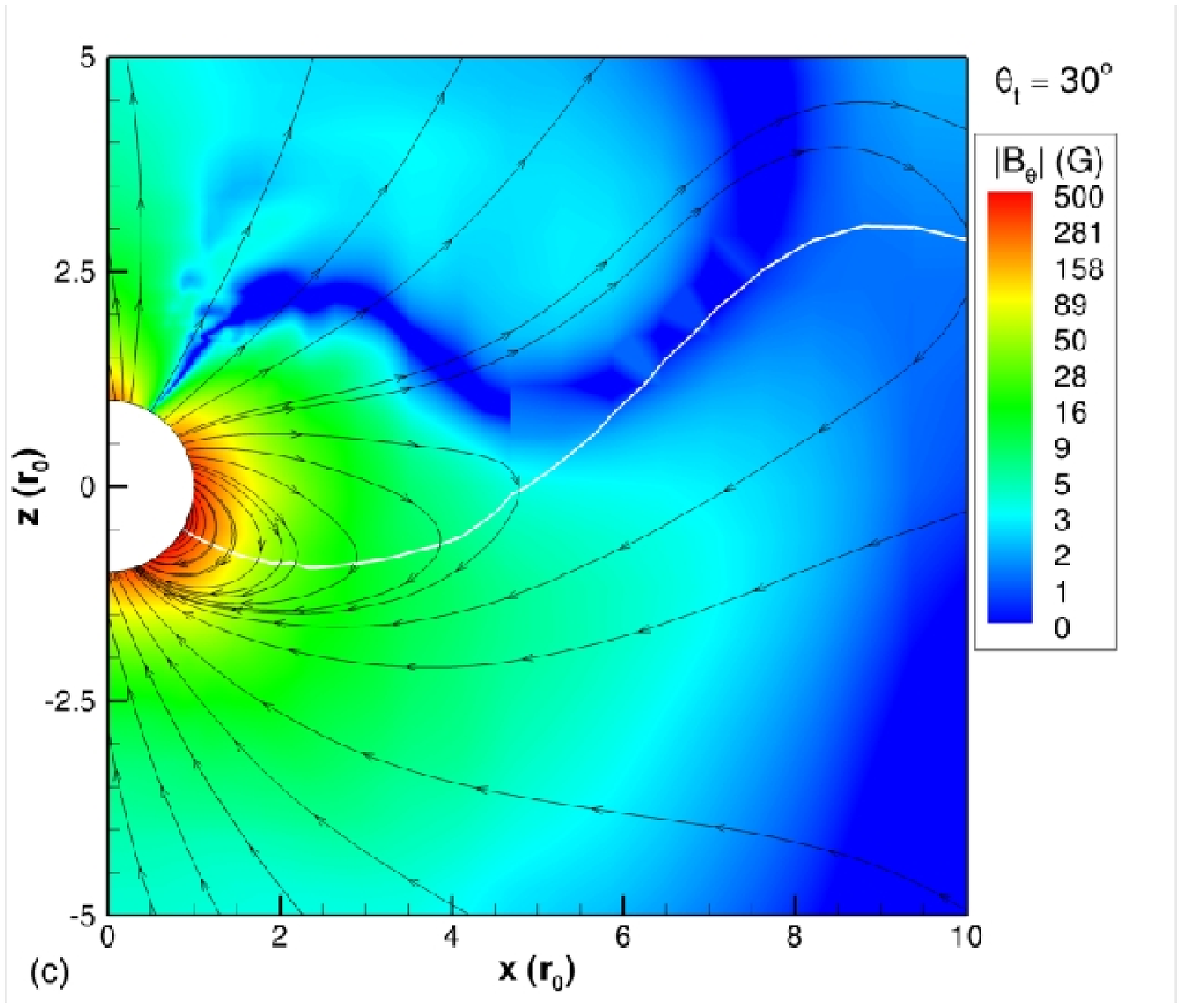}
  \includegraphics[height=7cm]{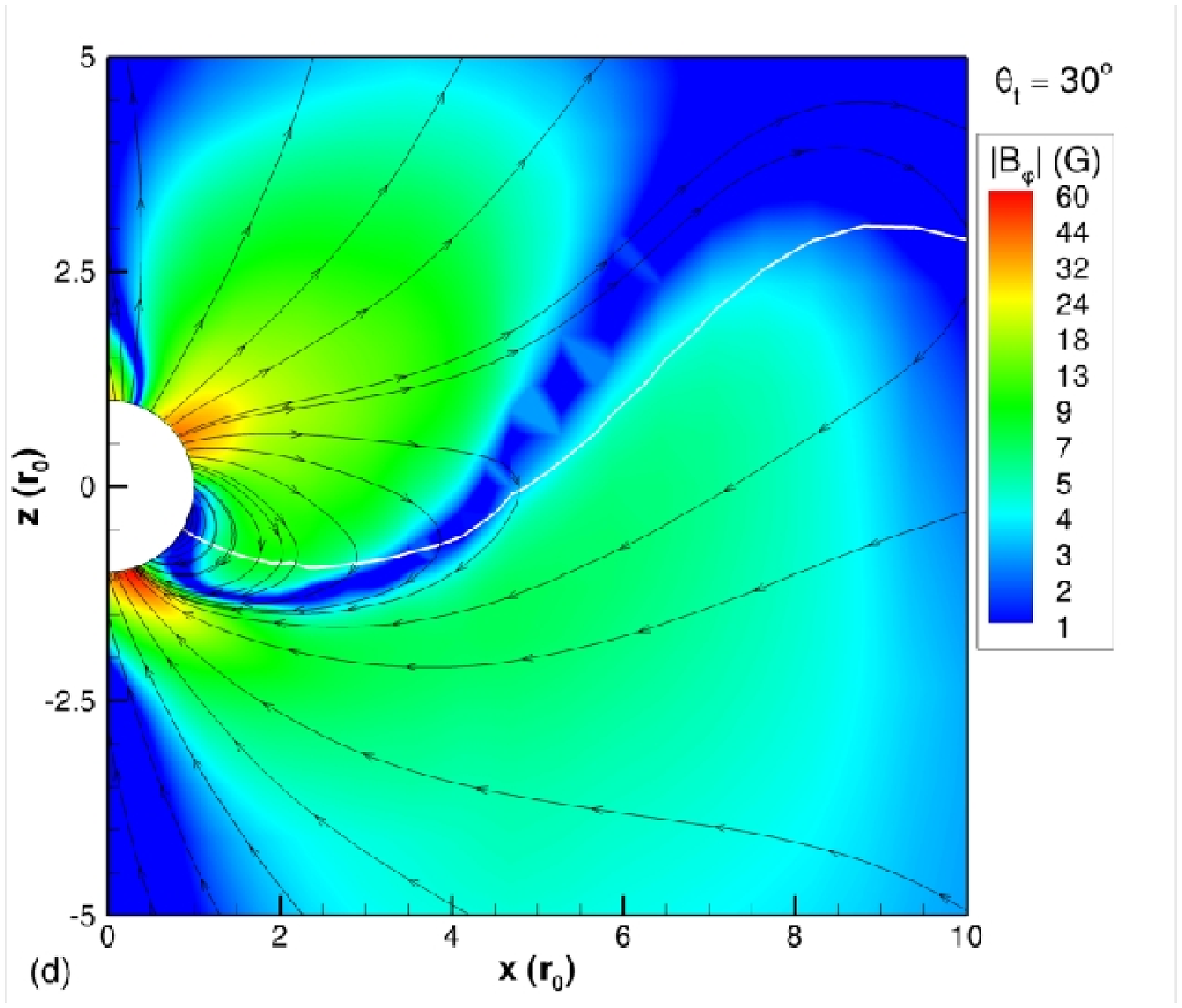}\\
  \includegraphics[height=7cm]{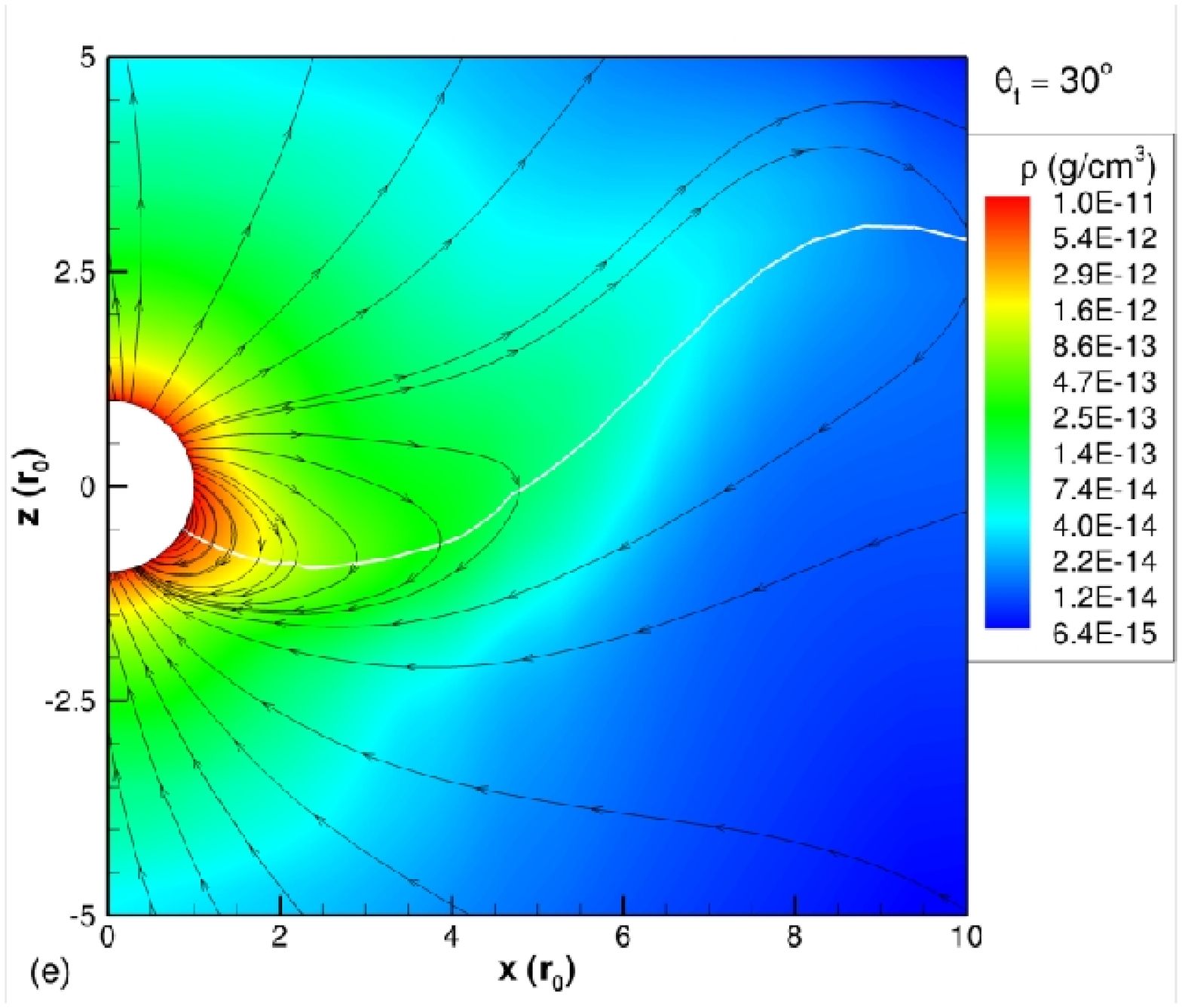}
  \includegraphics[height=7cm]{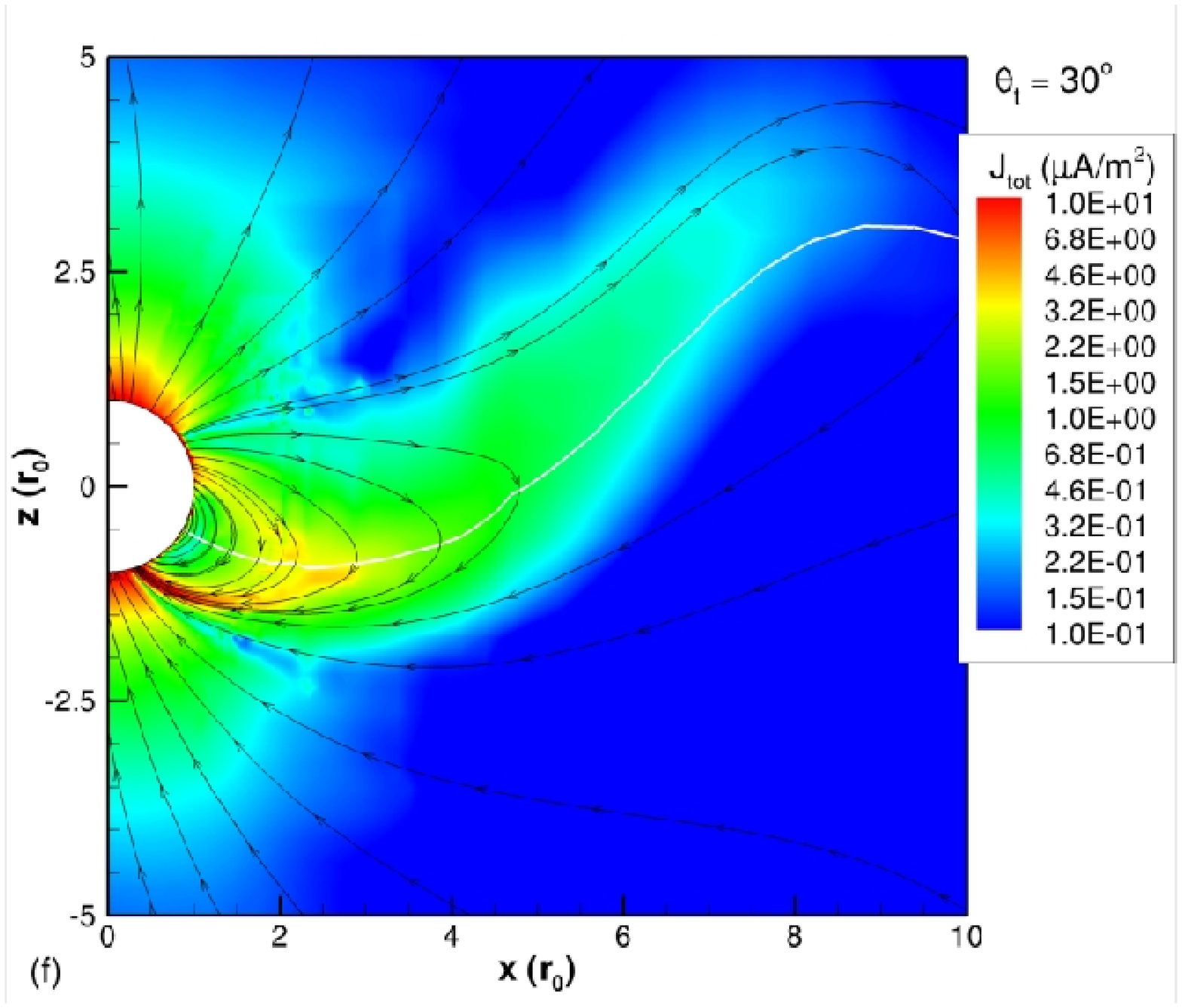}
  \caption{Meridional cuts of several wind variables for case T04 ($\theta_t = 30^{\rm o}$). The panels are snapshots of the simulation at $t=240$~h. Color-maps are for the (a) total wind velocity $u_{\rm tot}$; (b) radial component of the magnetic field $|B_r|$; (c) $\theta$-component of the magnetic field $|B_\theta|$; (d) azimuthal component of the magnetic field $|B_\varphi|$; (e) mass density of the wind $\rho$; (f) total current density $J_{\rm tot}$. All color-maps, except panel (a), are in logarithmic scale. \label{fig.t21.typical}}
\end{figure*}

We selected case T04 to describe the wind characteristics. The periodic movement of the stellar magnetosphere affects the entire wind structure, as can be seen in Fig.~\ref{fig.t21.typical}, where we present meridional cuts of the following wind variables: $u_{\rm tot}$, $|B_r|$, $|B_\theta|$, $|B_\varphi|$, $\rho$, and $J_{\rm tot}$. Meridional cuts of $u_r$ and $|u_\varphi|$ can be found in Figs.~\ref{fig.radial-velocity} and \ref{fig.Uphi-snapshot}, respectively. $|B_r|$, $|B_\theta|$ and $|B_\varphi|$ present dipolar configuration at the base of the coronal wind, but for other radii, their solution is dependent on the resultant interaction of the magnetic field with the wind. Because of this interaction, the stellar magnetosphere acquires an azimuthal component for the magnetic field, which can be seen in Fig.~\ref{fig.t21.typical}d, presenting maximum intensity in the interface between closed and open field lines. The density of the wind is not spherically symmetric, presenting higher densities around the $B_r=0$ surface (Fig.~\ref{fig.t21.typical}e). The total current density $J_{\rm tot} \propto |\nabla \times {\bf B}|$ is shown in Fig.~\ref{fig.t21.typical}f. 

\subsection{The effect of a different $\gamma$ on the wind}
We now compare the results from simulations T04 and T07, where different values of $\gamma$ were adopted. The value of $\gamma$ influences the thermal acceleration of the wind. It defines the input of thermal energy and $\gamma$ obeys the relation $p\propto \rho^{\gamma}$. Low values of $\gamma$ imply a proportionally large input of thermal energy in the wind, and consequently, high wind terminal velocities. It was shown in \citet{paper1} that when the wind is magnetized, the value chosen for $\gamma$ can alter the ratio between thermal and magnetic forces, and thus, accentuate the latitudinal dependence of the wind.

\begin{figure}
  \includegraphics[height=7cm]{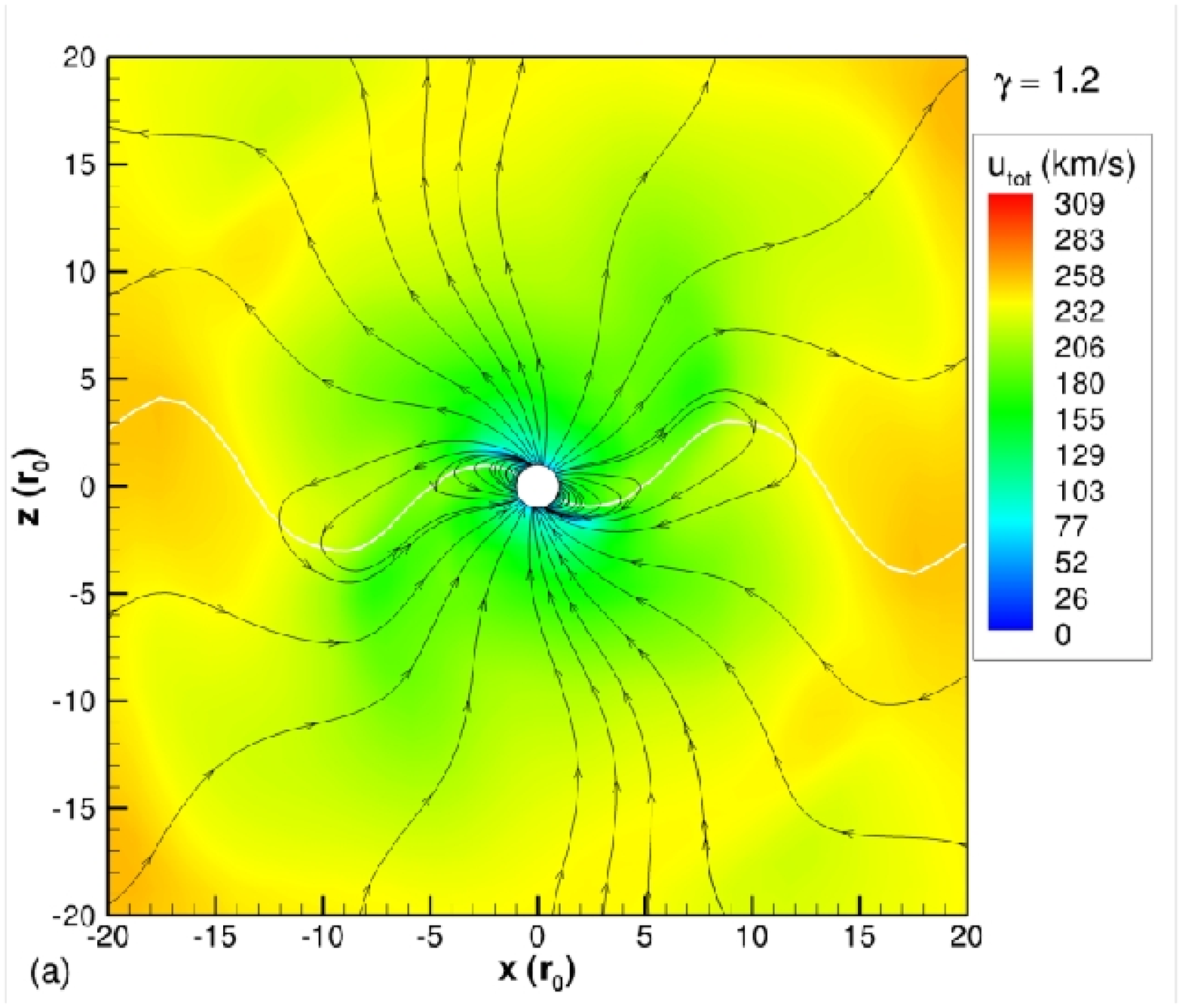}\\
  \includegraphics[height=7cm]{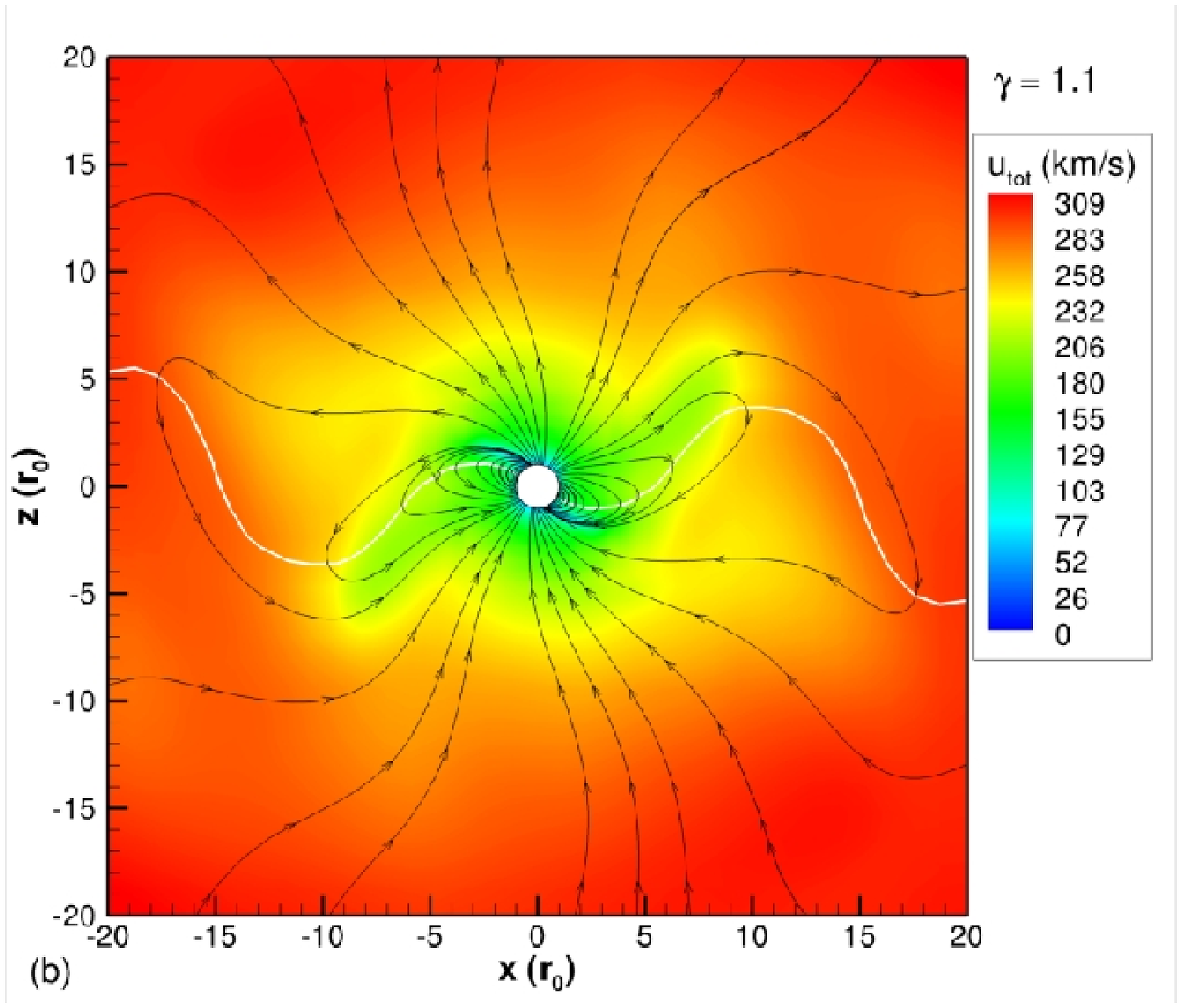}
  \caption{Meridional cuts of the total wind velocity $u_{\rm tot}$ for cases with (a) $\gamma=1.2$ (T04) and (b) $\gamma = 1.1$ (T07). Black and white lines have the same meaning as before. 
\label{fig.gamma}}
\end{figure}

In this work, we do not invoke the physical processes that may cause a larger input of energy in our models. Nevertheless, we study the effects of a smaller $\gamma$ in the wind of a magnetized star with oblique magnetic geometries. Figure~\ref{fig.gamma} presents the total velocity of the wind for both cases. By comparing simulations T04 ($\gamma=1.2$) and T07 ($\gamma=1.1$), we find that the terminal velocity achieved by simulation T07 is around $22\%$ larger than the one achieved in T04. The wind temperature profile for both cases is different, as expected: at the equatorial plane, near $10~r_0$, for instance, $T \simeq 3.4 \times 10^5~$K for case T04 and $T \simeq 5.5 \times 10^5~$K for case T07. Furthermore, the magnetic field configuration differs from both cases, with case T07 presenting a more compact zone of closed field lines. These results agree with our previous ones \citep{paper2,paper1}, which show that the heating parameter is a important in defining the acceleration of the wind and the magnetic configuration around the star. 

\subsection{The effect of different stellar rotational periods}
\begin{figure}
  \includegraphics[height=7cm]{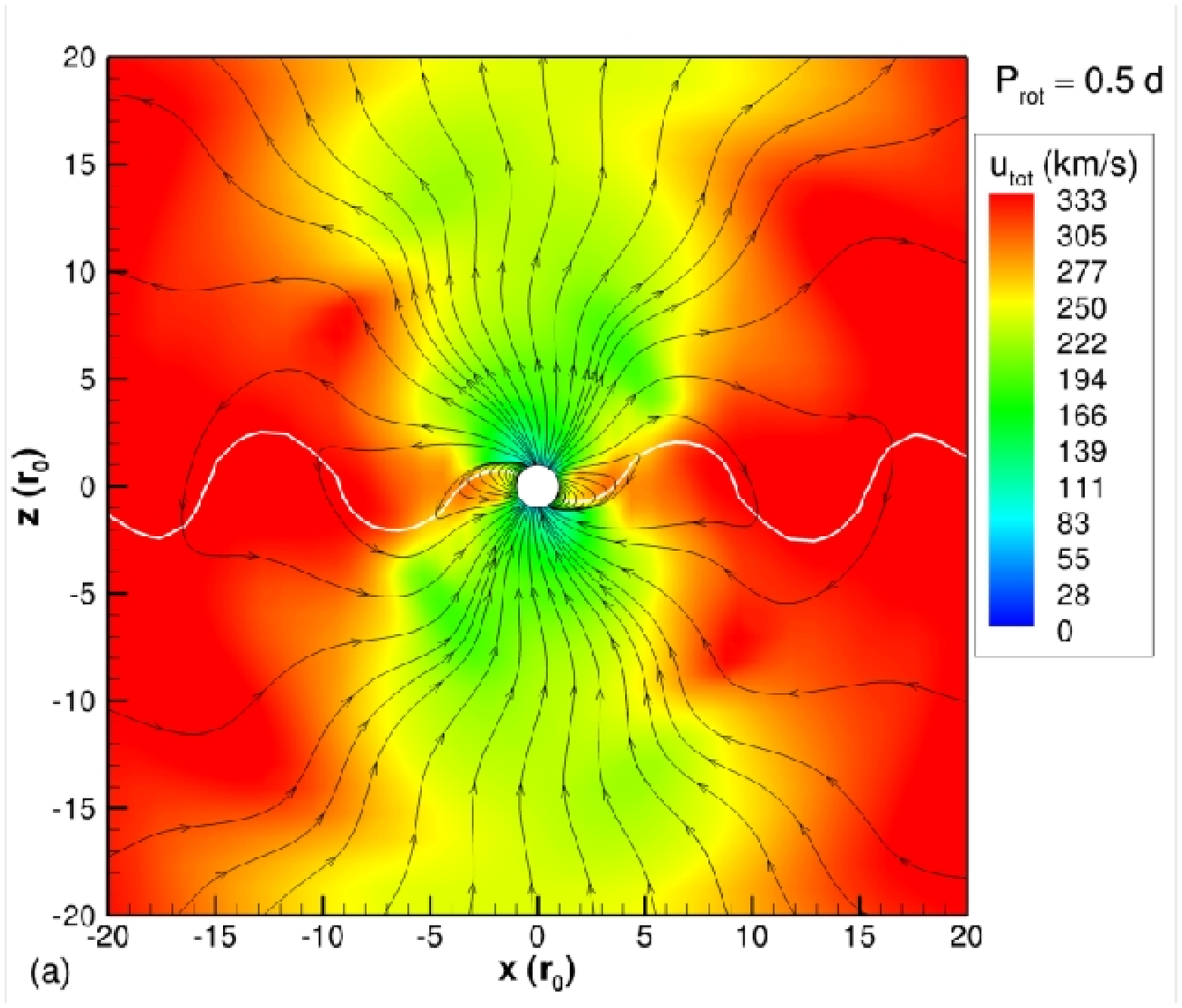}\\
  \includegraphics[height=7cm]{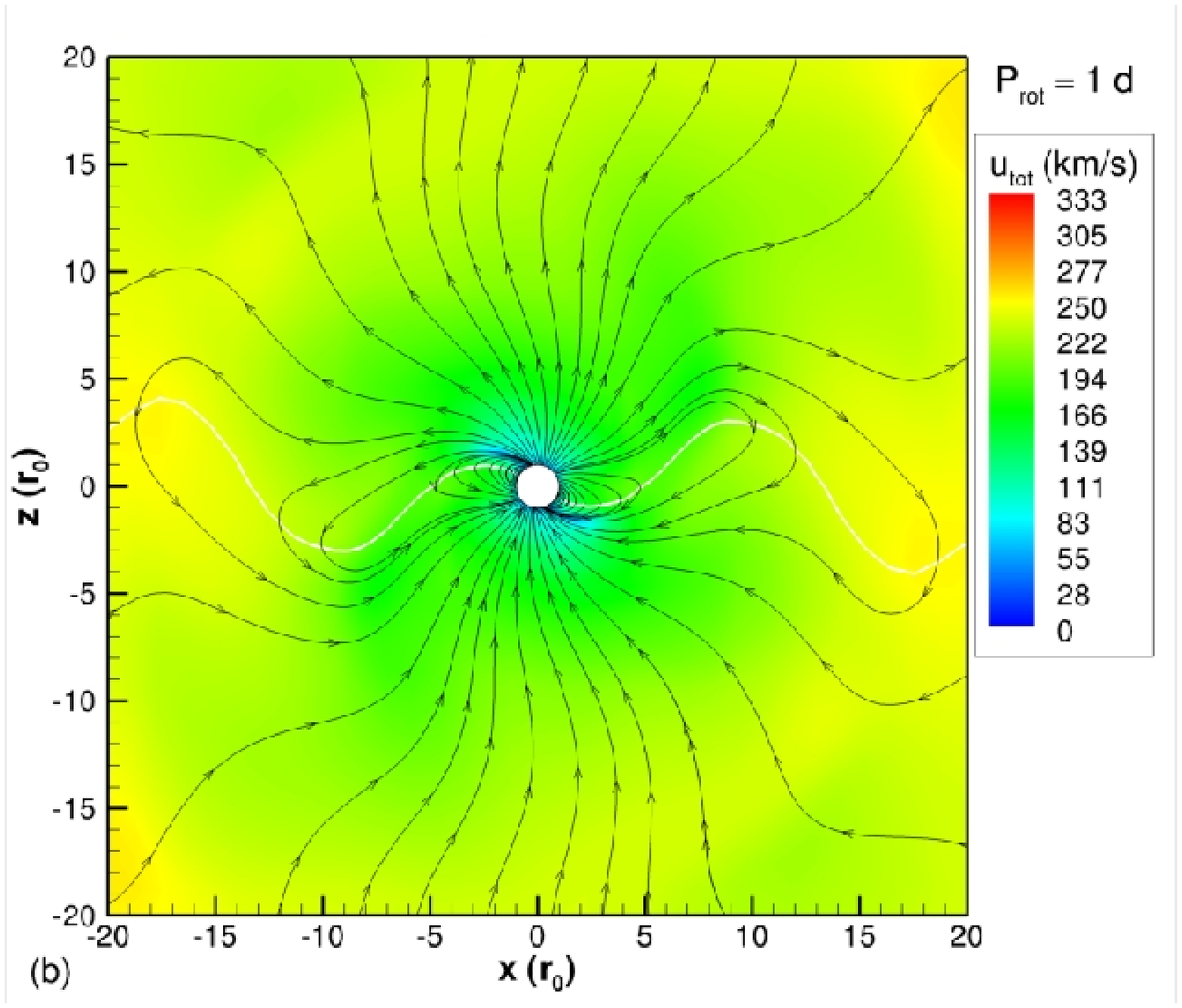}\\
  \includegraphics[height=7cm]{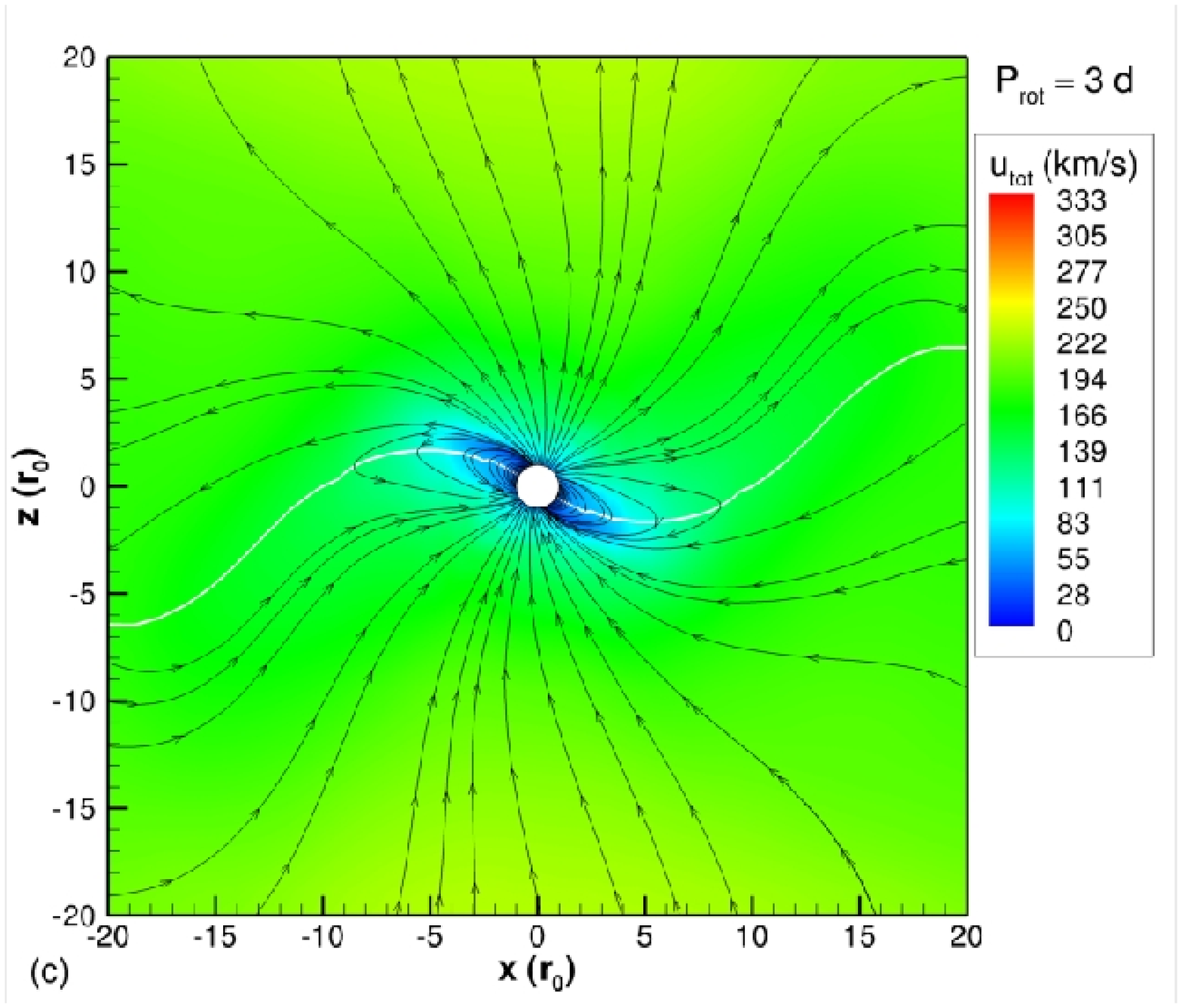}
  \caption{Meridional cuts of total wind velocities $u_{\rm tot}$ at the same stellar rotational phase for (a) $P_0=0.5$~d (T09), (b) $P_0=1$~d (T04), and (c) $P_0=3$~d (T08). Black and white lines have the same meaning as before. 
\label{fig.Prot}}
\end{figure}

Here we explore the rotational effects on the dynamics of the wind, keeping $\theta_t=30^{\rm o}$. We select cases T04, T08, and T09 to perform this comparison, where the stellar rotational periods are $1$, $3$, and $0.5$~d, respectively. These periods of rotation are in the lower range of observed periods for WTTSs  \citep[e.g.,][]{1993A&A...272..176B, 1996AJ....111..283C, 2002A&A...396..513H, 2006ApJ...646..297R}, as to explore the maximum effects on the wind. Longer stellar rotational periods imply a wind that is less disturbed by the precession of the magnetic field. Figure~\ref{fig.Prot} shows the meridional cuts of total wind velocities for $P_0=0.5$~d (Fig.~\ref{fig.Prot}a), $P_0=1$~d (Fig.~\ref{fig.Prot}b), and $P_0=3$~d (Fig.~\ref{fig.Prot}c) at the same stellar rotational phase. The wind is more accelerated for lower $P_0$ (i.e., larger rotational velocities), as a result of the coupling of magnetic fields and rotation \citep[e.g.,][]{1967ApJ...148..217W, 1976ApJ...210..498B}. At the equatorial plane, $u_{\rm tot}=360$~km~s$^{-1}$ for case T09, $u_{\rm tot}=250$~km~s$^{-1}$ for case T04, and $u_{\rm tot}=180$~km~s$^{-1}$ for case T08.

\subsection{The effect of a different $\beta_0$ on the wind}
\begin{figure}
  \includegraphics[height=7cm]{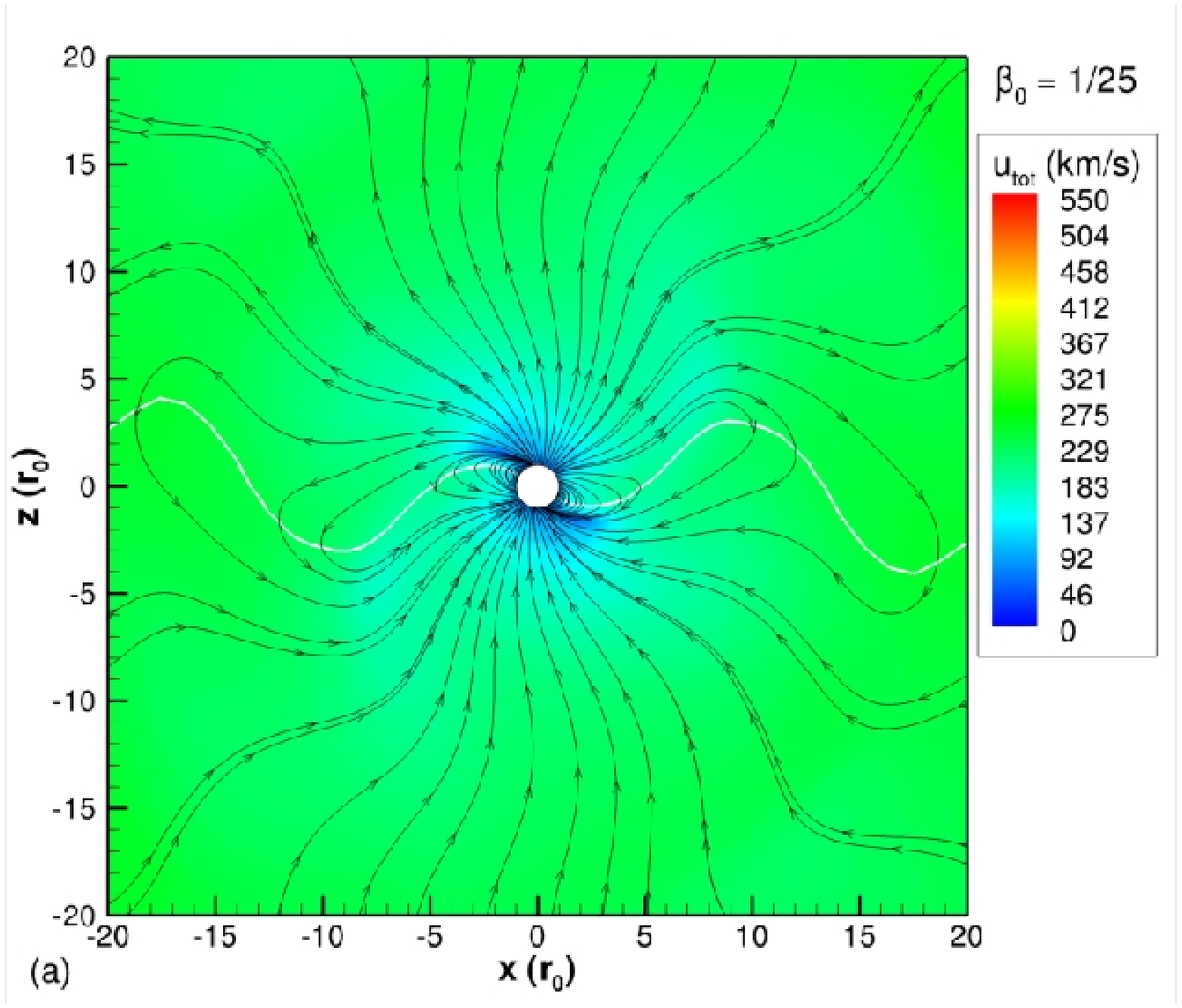}\\
  \includegraphics[height=7cm]{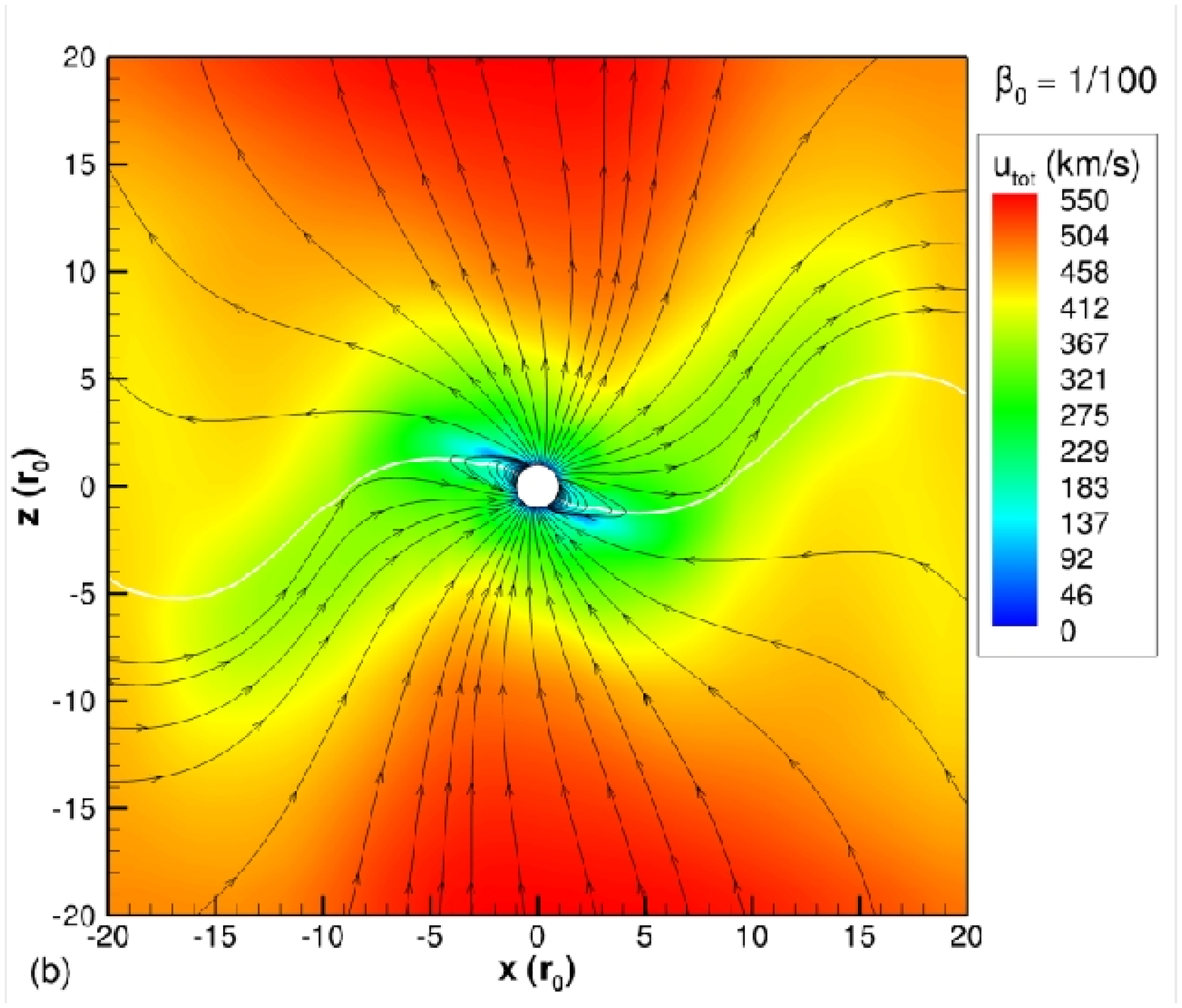}
  \caption{Comparison between simulations with different $\beta_0$: (a) $\beta_0=1/25$ (T04) and (b) $\beta_0=1/100$ (T10). Plots show meridional cuts of the total wind velocity $u_{\rm tot}$, magnetic field lines (black lines) and the surface $B_r=0$ (white line). \label{fig.beta}}
\end{figure}

In \citet{paper2,paper1}, we have shown that the ratio between the thermal and magnetic energy densities at the base of the wind ($\beta_0$) is a decisive factor in defining the magnetic configuration of the wind, as well as its velocity distribution. To study how $\beta_0$ influences the wind profile in the case of an oblique magnetic geometry, we compare simulations T04 and T10, which have the same model parameters, except for the density at the base of the wind (and thus, different $\beta_0$). Figure~\ref{fig.beta} shows meridional cuts of the total velocity of the wind plotted for both cases, as well as the magnetic field lines (black lines), and the surface $B_r=0$ (white line). Both panels show a snapshot at the same rotational phase of the star. As in the aligned case \citep{paper2,paper1}, the wind is more accelerated for low $\beta_0$, where the magnetic energy density at the base of the wind is more important than the thermal energy density. The ratio of open to closed magnetic field lines are larger for case T10, with lower $\beta_0$, showing that a faster wind is able to open the field lines more efficiently.  

\subsection{The ``wavelength'' of the magnetospheric oscillation}
The typical length $\lambda$ of the oscillation of the  isocontour of $B_r=0$ (white lines in Figs.~3, 7, 9 - 12) can be estimated as in \citet{2009ApJ...703....8L}
\begin{equation}\label{eq.lambda}
\lambda \simeq u_{\rm char} P_0  = 0.062~r_0 ~u_{\rm char}^{\rm (km/s)} P_0^{\rm(d)}  \, ,
\end{equation}
where $u_{\rm char}$ is the characteristic velocity of the plasma. 

For $P_0=1$~d, and different $\theta_t$ (cases T02 to T04), we showed that the velocity of the wind increases with $\theta_t$. Therefore, from Eq.~(\ref{eq.lambda}), it is immediate to conclude that $\lambda$ will be larger for larger $\theta_t$. In fact, our estimates for the inner portion of the grid show that $\lambda \simeq 13$, $14$, and $16~r_0$ for $\theta_t=10^{\rm o}$, $20^{\rm o}$, and $30^{\rm o}$, respectively, which are consistent to the values measured from the simulations, $\lambda_{\rm sim} \simeq 14$, $16$, and $17~r_0$. 

In the cases where $\gamma$ was compared (cases T04 and T07), we note that $\lambda$ is relatively larger for case T07 ($\gamma=1.1)$ than for case T04 ($\gamma=1.2)$. This is again due to the larger velocities of the wind for case T07. The value of $\lambda$ calculated by Eq.~(\ref{eq.lambda}) was $\simeq 16$ and $20~r_0$ for T04 and T07, respectively, while the measured values of $\lambda_{\rm sim}$ from the inner region of the simulations were $\simeq  17$ and $19~r_0$.

In the cases where $P_0$ was varied from $0.5$ to $3$~days (cases T04, T08, and T09), the velocity of the wind decreased with the increase of $P_0$. Both facts have a different effect on Eq.~(\ref{eq.lambda}). However, the increase in $P_0$ has proved to be more important in the increase of $\lambda$ than the effect provided by the increase in the wind velocity. We find $\lambda \simeq 11$, $16$, and $37~r_0$ for $P_0=0.5$~d (T09), $P_0=1$~d (T04), and $P_0=3$~d (T08), respectively, while the measured values were $\lambda_{\rm sim} \simeq 12$, $17$, and $36~r_0$.

Varying the ratio between the thermal and magnetic energy densities at the base of the wind (cases T04 and T10) has an important effect on the acceleration of the wind. Because case T10 presents a lower $\beta_0$ than case T04, T10 has a larger characteristic velocity of the plasma. Ultimately, this increases $\lambda$ from $\simeq 16~r_0$ (T04) to $ 30~r_0$ (T10). The measured values of $\lambda_{\rm sim}$ from the inner region of the simulations were $17$ and $31~r_0$.

\section{DISCUSSION OF STELLAR WIND RESULTS}\label{sec.discussion}
A MHD wind model provides solutions for the density, velocity, and temperature profiles, along with the magnetic field configuration of the wind. Because the solution of the MHD equations is complex, some models adopt several approximations. This is the case of the Weber-Davis model \citep{1967ApJ...148..217W}, first developed for the solar wind and later on adopted to describe winds of other stars \citep[e.g.,][among many other applications]{2005A&A...434.1191P, 2008MNRAS.389.1233L}. The simplifications involved in the Weber-Davis model are: the model is axisymmetric and stationary; it considers an open, radial magnetic field that acquires an azimuthal component due to the rotation of the star; the wind solution is  valid for the equatorial plane; it neglects meridional components of the magnetic and velocity fields. Because only the open magnetic field lines contribute to angular momentum loss, the Weber-Davis model is expected to overestimate angular momentum loss through a magnetized stellar wind, presenting shorter time-scales for stellar rotation brake than those models that consider the existence of closed field line regions \citep{1987MNRAS.226...57M}. Because of its uni-dimensional characteristic, the solution of the Weber-Davis model can be easily integrated. A detailed description of its solution is given in \citet{2005A&A...434.1191P}. 

Opposed to the Weber-Davis model, our model presents a multi-component corona, with the co-existence of open and closed field line regions. A latitudinally dependent velocity is observed, where the wind along the magnetic poles has a larger velocity than the wind along the equatorial regions. Details of the characteristics of the solutions of our wind model, in the context of aligned rotational axis and magnetic dipole moment, are presented in \citet{paper1}. Unfortunately, the three-dimensional nature of our model does not allow for analytical expressions of the solution of the variables of the wind.

Different wind scenarios on the framework of the Weber-Davis model were explored by \citet{2005A&A...434.1191P} and \citet{2007A&A...463...11H}. The first one investigated the characteristics of the stellar wind for a sample of stars hosting close-in giant planets, while the second one used empirical data to constrain theoretical wind scenarios. Despite the wind solutions from \citet{2005A&A...434.1191P} and \citet{2007A&A...463...11H} being focused mainly on cool main-sequence stars, we compare the overall trend of our model in respect to these works. Mass-loss rates on such models are similar to the solar wind value \citep[$10^{-14}~{\rm M}_\odot ~{\rm yr}^{-1}$][]{2005A&A...434.1191P} and do not exceed about $10$ times the solar value \citep{2007A&A...463...11H}. Wind terminal velocities found by \citet{2005A&A...434.1191P} range from $310$~km~s$^{-1}$ for an isothermal wind temperature of $T_0=5\times 10^5~$K to $760$~km~s$^{-1}$ for $T_0=2\times10^6~$K. Compared to our models, mass-loss rates are about $6$ orders of magnitude smaller, while velocities achieved are about the same order of magnitude. The difference in mass-loss rates is a consequence of the larger coronal densities adopted in our models. 

To our knowledge, there are no measurements of mass-loss rates and wind velocities for WTTSs to compare our results with. The detections of mass-loss rates are based on the early stage as a CTTS, when an accretion disk is still present \citep[e.g.,][]{1964ApJ...140.1409K, 2003ApJ...599L..41E, 2006ApJ...646..319E, 2007ApJ...657..897K, 2007ApJ...654L..91G}. Based on these detections, mass-loss rates are of the order of $10^{-10}$ to $10^{-7}~{\rm M}_\odot ~{\rm yr}^{-1}$ and wind terminal velocities $\simeq 400$~km~s$^{-1}$. The parameters in our models were chosen to have values compatible to these ones. \citet{paper2} considered models with the density at the base of the coronal wind spanning by two orders of magnitude, which resulted in mass-loss rates ranging between $\sim 10^{-9}$ and $8 \times 10^{-8}$~M$_\odot$~yr$^{-1}$. In the present paper, the main goal of \S\ref{sec.results} was to analyze the effects of the tilt angle on the wind. We therefore did not explore several values of base density and our models present mass-loss rates of about $ 9 \times 10^{-9}$~M$_\odot$~yr$^{-1}$. From Figures \ref{fig.radial-velocity}, \ref{fig.gamma}, \ref{fig.Prot}, and \ref{fig.beta} we note that the wind terminal velocities obtained are $\simeq 350$ to $500$~km~s$^{-1}$.

\citet{2007ApJ...657..897K} suggest that, if the winds of WTTSs are simply a scaled-up version of the solar wind, WTTS winds should then be stronger than those of CTTSs, because X-ray emission from WTTSs are stronger than the emission from CTTSs. However, the winds of CTTSs are believed to be powered by accretion, which could be the reason why the wind traced by HeI $\lambda$10830 is not detected in WTTSs \citep{2007ApJ...657..897K}. We would expect that when accretion ceases, the wind should become less strong. 

There is clearly a need of more observational constrains on the winds of WTTSs. In possession of that, we would be able to better constrain the parameters of our models.

\section{ON THE PLANET-WIND RECONNECTION} \label{sec.reconnection}
Finding planets around young stars is currently ongoing, with two recent detections of massive giant planets: one around a $5$~Myr-old star \citep{2010arXiv1006.3070L} and one around a $12~$Myr-old star \citep{2010arXiv1006.3314L}. The stellar wind is expected to directly influence the planet and its atmosphere, e.g., by changing the configuration of the planet's magnetosphere, producing nonthermal planetary magnetospheric radio emissions, etc. So far, the few theoretical works investigating the influence of the stellar wind on the magnetosphere of planets were based on simplified treatments of the stellar winds,  e.g., using the Parker wind model \citep{2007P&SS...55..618G,2007A&A...475..359G, 2010AJ....139...96L}, the Weber-Davis wind model \citep{2005A&A...434.1191P, 2006A&A...460..317P, 2007P&SS...55..589P}, assuming a solar-type stellar wind \citep{2004ApJ...602L..53I, 2004P&SS...52.1469F, 2005MNRAS.356.1053S, 2007P&SS...55..598Z}, or based on scalings for the mass-loss rates and wind terminal velocities \citep{2004A&A...425..753G, 2005A&A...437..717G}. The consideration of a realistic wind is crucial to determine how the interaction between the stellar wind and the magnetosphere of an extrasolar planet occur. 

Our 3D, time-dependent MHD simulations of stellar winds of WTTSs provide a powerful tool to investigate the planet-wind interaction, as it allows us to consider the effects of a more realistic wind  and obtain key insights on the detectability of radio emission from extrasolar planets. In this section, we estimate the reconnection rate and power released when reconnection between a close-in magnetized giant planet and the stellar wind takes place. This estimate is performed for four different wind simulations: T01, T02, T03, and T04, where the misalignment angle of the stellar rotation axis and its magnetic moment vector is $\theta_t=0^{\rm o}$, $10^{\rm o}$, $20^{\rm o}$, and $30^{\rm o}$, respectively.

\subsection{The Reconnection Rate}
In this section, we estimate the rate of reconnection between the stellar and planetary magnetic field lines. This is necessary to evaluate the planetary radio emission (\S\ref{subsec.radio}). 

The magnetic field of the stellar wind has three components $B_x$, $B_y$, and $B_z$. $B_x$ and $B_y$ are parallel to the stellar rotational equatorial plane. $B_z$ is perpendicular to the this plane. Considering a planet whose orbital plane coincides with the rotational equatorial plane of the star and considering that the planet's magnetic dipole moment is aligned in the $-z$-direction, then when the magnetic field of the stellar wind and the magnetic field of the planet are oriented anti-parallel to each other, magnetic field line reconnection can occur. This results in transfer of energy from the stellar wind to the planet's magnetosphere.

The reconnection rate is the amount of magnetic flux that reconnects per unit time per unit length of the reconnection line or, equivalently, the reconnection rate can be defined as the strength of the electric field parallel to the reconnection merging line \citep[e.g.,][]{2000mare.book.....P}. The reconnection line refers to the line where magnetic field lines reconnect. The rate at which reconnection between two different plasmas happens depends, among other things, on the velocity of the incident plasma on the reconnection site \citep{1973ApJ...180..247P}.  In the idealized case when the magnetic fields of two identical plasmas are exactly anti-parallel,  the reconnection rate (or the generated electric field at the reconnection site) can be estimated as \citep[e.g.,][]{2008JGRA..11307210B}
\begin{equation}\label{eq.recsimple}
E \simeq  \frac{v_{\rm in} B}{c} \, ,
\end{equation}
where $v_{\rm in} = C v_A$ is the inflow velocity of the plasma in the reconnection site, $v_A$ and $B$ are the Alv\'en speed and magnetic field of the ambient plasma, respectively, and $c$ is the speed of light. The factor $C=l/L$ is the dissipation region aspect ratio, i.e., a property of the geometry of the reconnection region which has a characteristic width $l$ and a characteristic length $L$. When reconnection occurs between two plasmas with different characteristics, Eq.~(\ref{eq.recsimple}) becomes more complex, taking into account the different magnetic field intensities and Alv\'en speeds of the two plasmas \citep{2007PhPl...14j2114C, 2008JGRA..11307210B}
\begin{equation}\label{eq.reccomplex}
E \simeq C \frac{2}{c} \left( \frac{B_{z,1}^3 B_{z,2}^3}{4 \pi (B_{z,2} \rho_1 + B_{z,1} \rho_2)(B_{z,1} + B_{z,2})} \right) ^{1/2}\,
\end{equation}
where the index ``1'' and ``2'' are used to distinguish both plasmas on the site of the interaction, $B_{z,1}$ and $B_{z,2}$ are oriented anti-parallel to each other, and $\rho$ is the mass density.

\subsection{Planetary Radio Emission}
The solar wind interaction with the magnetic planets of the Solar System (Earth, Jupiter, Saturn, Uranus, and Neptune) accelerates electrons that propagate along the planets magnetic field lines, producing electron cyclotron radiation at radio wavelengths \citep{1998JGR...10320159Z}. By analogy to the magnetic planets in the Solar System, predictions have been made that magnetized extra-solar planets should produce cyclotron maser emission \citep[e.g.,][]{1999JGR...10414025F, 2004P&SS...52.1469F, 2001Ap&SS.277..293Z, 2004ApJ...612..511L, 2004A&A...425..753G, 2005A&A...437..717G, 2007A&A...475..359G, 2007P&SS...55..618G, 2005MNRAS.356.1053S, 2007P&SS...55..598Z, 2008A&A...490..843J}.

The planetary radio emission depends on the planet's magnetic field intensity\footnote{ For predictions of planetary radio emission for non-magnetized planet, see \citet{2001Ap&SS.277..293Z, 2007P&SS...55..598Z, 2007A&A...475..359G}.} and on the stellar wind power: in general, it implies that the stronger the stellar wind is, the more radio-luminous should the planet be. So far, such radio signatures from stars hosting hot Jupiters have not yet been detected, and one possible reason for that may be due to the lack of instrumental sensitivity in the appropriate frequency range of the observations \citep{2000ApJ...545.1058B}. The theoretical estimates on the radio flux emitted by extrasolar planets carry along a big uncertainty due to the fact that the stellar wind properties are poorly constrained: \citet{1999JGR...10414025F} showed that a variation by a factor of $2$ in the wind velocity may change the level of radio power emission by a factor of $100$, with more recent works suggesting that the radio power emission is proportional to the incident wind power \citep[e.g.,][]{2005MNRAS.356.1053S, 2007P&SS...55..598Z, 2007A&A...475..359G}. Therefore, the potential observation of radio emission from extrasolar planets strongly depends on the nature of the stellar wind. 

Based on our simulated stellar winds (cases T01 to T04), we estimate the planet's radio power. The electric field generated in the interaction is calculated from Eq.~(\ref{eq.reccomplex}), where plasma $1$ refers to the characteristics of the planet's magnetosphere, while plasma $2$ refers to the local characteristics of the impacting stellar wind. Initially, we do not consider pile-up of the stellar wind magnetic field in the magnetosheath of the planet, but will do so in \S\ref{subsec.pile}. 

To derive the characteristics of the planet's magnetospheric plasma (plasma $1$), we consider a hot Jupiter with a dipolar magnetic field aligned in the $-z$-direction, and magnetic intensity at the equator  $B_p=50~$G. The density of the planetary plasma $\rho_1$ is taken to be negligible such that $\rho_2 B_{z,1} \gg \rho_1 B_{z,2}$ in Eq.~(\ref{eq.reccomplex})\footnote{In our simulations, for $r\lesssim 12~r_0$, the ratio $ |B_{z,2}/B_{z,1}|$ ranges from $0.2$ to $9.3$. For example, for case T02 at $r\simeq 5~r_0 \simeq 0.05~$AU, $B_{z,1}\simeq 6.1$~G and $B_{z,2}\simeq -3.4$~G. As $B_{z,1}$ and $ B_{z,2}$ have approximately the same order of magnitude, this condition implies that $\rho_1 \ll \rho_2 \sim 10^{-13}$~g~cm$^{-3}$. }. We assume that the planet has the same radius as Jupiter $R_p=R_{\rm Jup}\sim 0.05~r_0$.

As we do not include the planet in our simulation, next, we analytically calculate the area of the planet that will interact with the stellar wind.

\subsubsection{The Size of the Planet's Magnetosphere}
\begin{figure}
  \includegraphics[height=7cm]{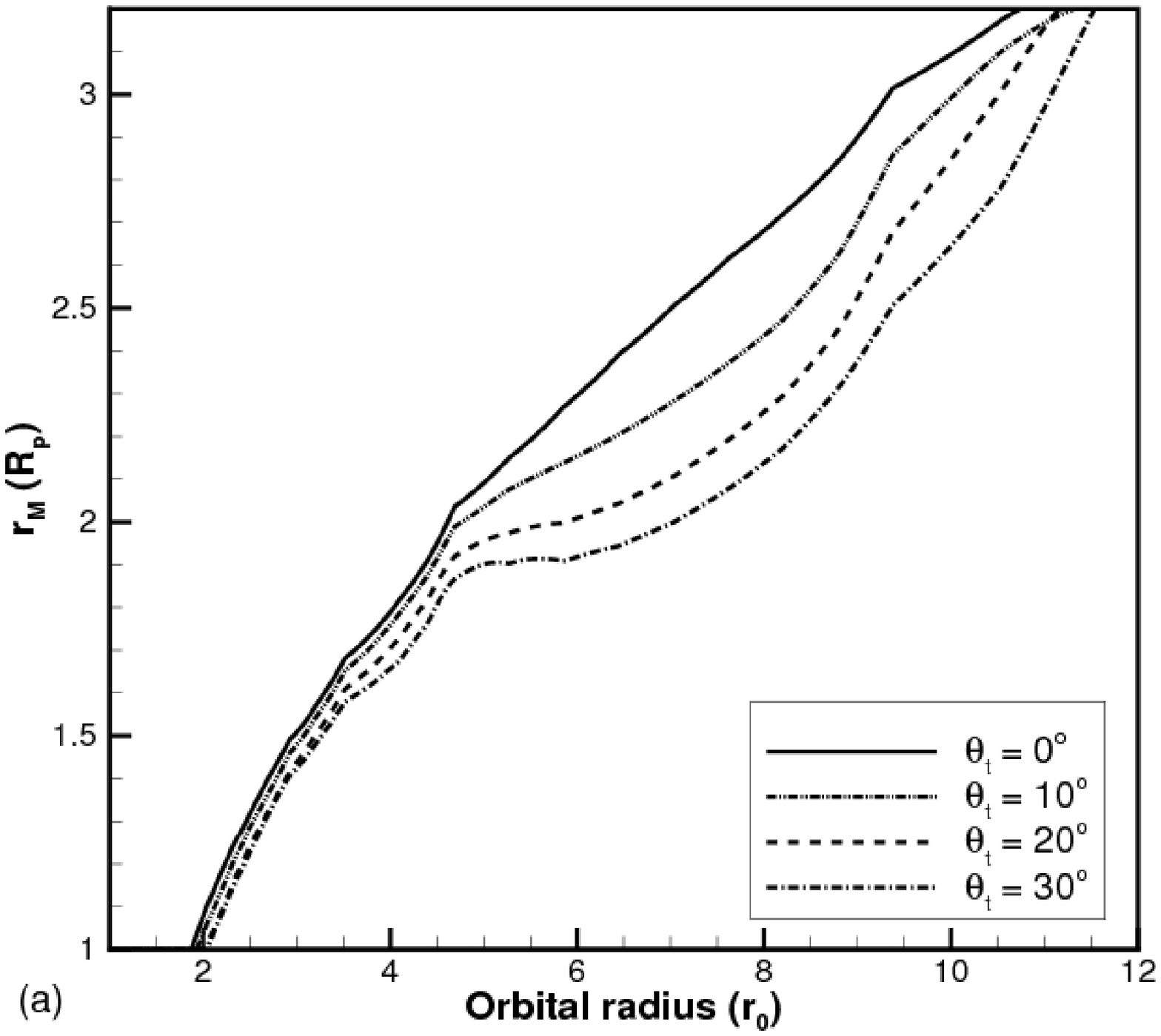}\\
  \includegraphics[height=7cm]{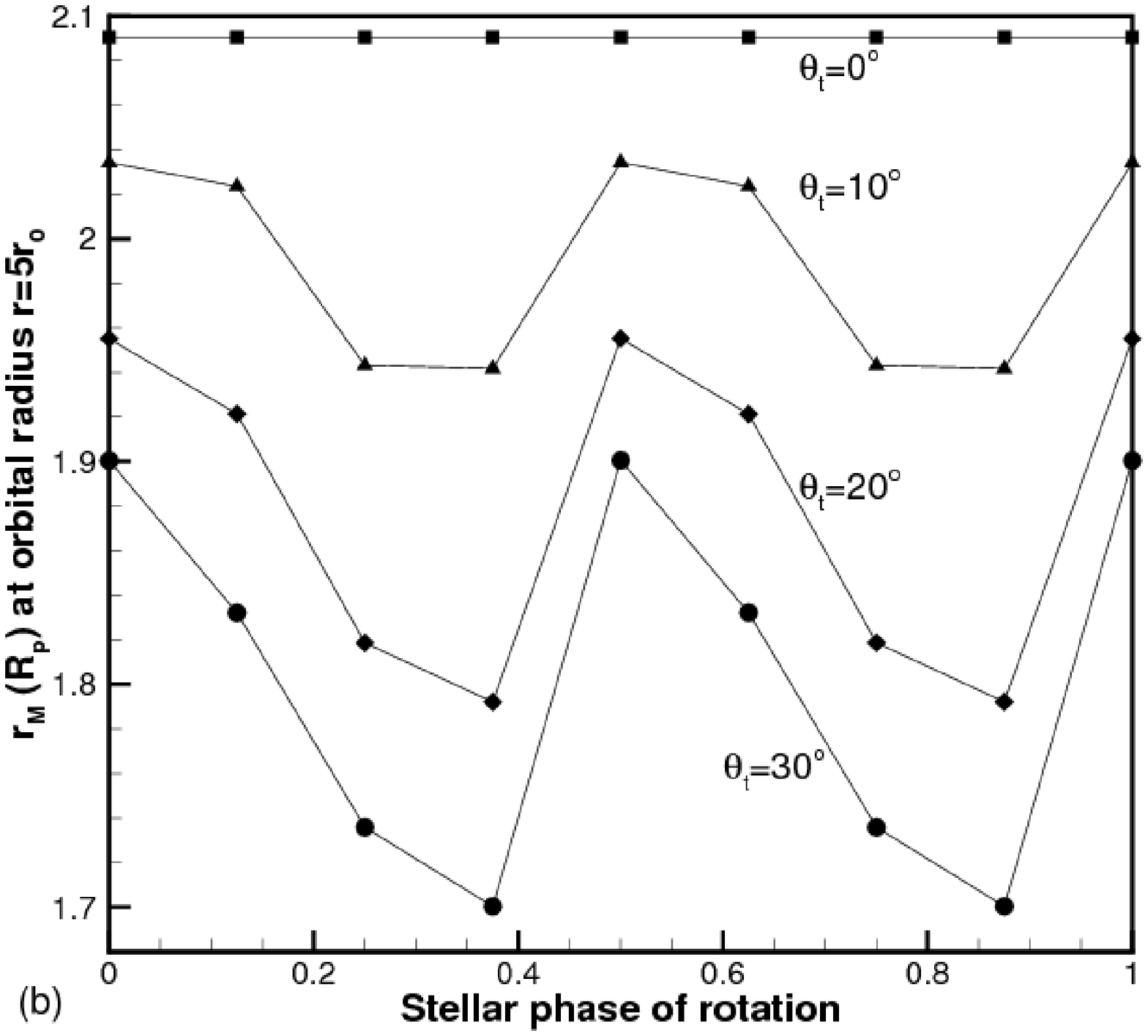}\\
  \caption{The magnetospheric radius $r_M$ [Eq.~(\ref{eq.rmagnetopause})] of a planet orbiting a star as given by simulations T01, T02, T03, and T04. (a) Considering a given phase of the stellar rotation ($t=240~$h), and planet located in the $x$-axis. (b) Considering an orbital radius of $r = 5~r_0 \simeq 0.05$~AU as function of stellar rotational phase (lines presented are merely symbol connectors and do not represent actual values of $r_M$).\label{fig.magnetosphere}}
\end{figure}

The interaction of the planet's magnetosphere with the wind takes place at a distance $r_M$ from the center of the planet, where there is balance between the wind total pressure and the magnetic pressure of the planet \citep{2008MNRAS.389.1233L,2008A&A...490..843J,paper2}
\begin{equation}\label{eq.equilibrium}
\frac{\rho_2 u_2^2}{2} + \frac{B_{\parallel,2}^2}{4\pi}= \frac{B_{z,1}^2}{4\pi}\, ,
\end{equation}
where $B_{z,1}$ is the $z$-component of the magnetic field of the planet at the equatorial plane 
\begin{equation}\label{eq.Bzeq}
B_{z,1}= \frac{B_p R_p^3}{r_M^3} \,  ,
\end{equation}
$B_{\parallel,2}$ is the parallel component of the stellar magnetic field to the boundary layer and $u_2=(u_\varphi-u_K)$ is the relative velocity between the wind azimuthal velocity $u_\varphi$ and the circular Keplerian velocity of the planet $u_K = (GM_\star/r)^{1/2}$. Substituting Eq.~(\ref{eq.Bzeq}) in (\ref{eq.equilibrium}), we have
\begin{equation}\label{eq.rmagnetopause}
\frac{r_M}{R_p} =  \left[ \frac{B_p^2/2\pi}{ (\rho_2 u_2^2 + B_{\parallel,2}^2/2\pi)} \right]^{1/6} \, .
\end{equation}
The size of the planet's magnetosphere $r_M$ depends on the local characteristics of the stellar wind, on the orbital radius of the planet (through $u_K$), and on the planetary magnetic field. Figure~\ref{fig.magnetosphere}a presents the size of the planet's magnetosphere if the stellar wind is given by cases T01 (aligned case, $\theta_t=0^{\rm o}$), T02 ($\theta_t=10^{\rm o}$), T03 ($\theta_t=20^{\rm o}$), and T04 ($\theta_t=30^{\rm o}$), calculated at $t=240~$h, for a range of planetary orbital radius up to $12~r_0 \sim 0.11~$AU. We note that the planet's magnetosphere becomes large as the misalignment angle $\theta_t$ is smaller (i.e., the highest values of $r_M$ are found for the aligned case). This is a result of a lower wind total pressure $\frac12(\rho_2 u_2^2 + B_{\parallel,2}^2/2\pi)$ as $\theta_t$ gets smaller. For the range of orbital radii analyzed here, both kinetic and magnetic terms of the wind total pressure make significant contributions, except for small orbital radii ($r \lesssim 3~r_0$), where the magnetic term dominates. We see that for $r \lesssim 2~r_0$, the magnetosphere of the planet has vanished, due to a large wind total pressure. As for the misaligned cases the wind impacting on the planet changes characteristics according to the stellar phase (see for instance Fig.~\ref{fig.evolution}), it is expected that the radius of the planet's magnetosphere will vary if the stellar rotational period is different from the orbital period of the planet. Considering a planet located at an orbital radius of $r = 5~r_0 \simeq 0.05$~AU in a circular orbit, for case T02, for instance, the variation in $r_M$ from its maximum possible value to its minimum possible value is around $5\%$, while for cases T03 and T04, this variation is around $8\%$ and $11\%$. Figure~\ref{fig.magnetosphere}b shows $r_M$ as a function of the stellar phase of rotation for a planet located at $r = 5~r_0 \simeq 0.05$~AU.

Knowing the value of $r_M$, we can then calculate from Eq.~(\ref{eq.Bzeq}) the value of the  magnetospheric magnetic field of the planet $B_{z,1}$ that will interact with the stellar wind. Because the size of the planet's magnetosphere can increase/decrease depending on the incident wind, $B_{z,1}$ will present variations along the planetary orbit. The values of $B_{z,1}$ can range between $6.1$ and $7.0$~G for case T02, $6.8$ and $8.9$~G for case T03, and $7.4$ and $10.4$~G for case T04, for a planet at an orbital radius $r = 5~r_0 \simeq 0.05$~AU. Thus, for instance, for $\theta_t=30^{\rm o}$, along the planetary orbit, the interacting planetary magnetic field varies up to a factor of $1.4$. The more misaligned is the stellar rotation axis in relation to the stellar magnetic moment vector, more variation is expected in the magnetospheric radius of the planet and on the interacting planetary magnetic field $B_{z,1}$.

\subsubsection{Estimate of the Planetary Radio Emission}\label{subsec.radio}
The power released with the reconnection event $P_{\rm rec} $ can be decomposed into a power released from the dissipation of kinetic energy carried by the stellar wind $P_k$ and a power released from the magnetic energy of the wind $P_B$ \citep[e.g.,][]{1999JGR...10414025F,2001Ap&SS.277..293Z}
\begin{equation}\label{eq.pwrrec}
P_{\rm rec} = a P_k + b P_B\, ,
\end{equation}
where $a$ and $b$ are efficiency ratios and we assumed that $P_{\rm rec}$ depends linearly on $P_k$ and $P_B$. Observations of the Solar System can be either explained by \{$a = 1\times 10^{-5}$, $b = 0$\} or \{$a=0$, $b=2\times 10^{-3}$\} \citep{2007P&SS...55..598Z}. In fact, \citet{2007P&SS...55..598Z} argues that it is not possible to decide which incident power actually drives the radio power observed from the magnetic planets of the Solar System. If both incident powers contribute to the radio emission, this implies that the coefficients $a$ and $b$ need to satisfy the relation $a/(1\times 10^{-5}) + b/(2\times 10^{-3}) =1$, as to match the observed radio power. However, we do not know if $a$ and $b$ should remain the same in other planetary systems. As we lack a better guess, we will adopt $a$ and $b$ as in the Solar System, i.e., either $P_{\rm rec} = 1\times 10^{-5} P_k$ or $P_{\rm rec} = 2\times 10^{-3} P_B$.

The magnetic power $P_B$ can be estimated as the Poynting flux of the stellar wind impacting on the planetary magnetospheric cross-section $S$ \citep{2007P&SS...55..598Z}
\begin{equation}\label{eq.pwrB}
P_B = \int c \frac{{\bf E}\times {\bf B}}{4\pi} \cdot {\rm d}{\bf S} \simeq c\frac{E B_{z,2}}{4\pi} \pi r_M^2\, ,
\end{equation}
where the electric field $E$ is given by Eq.~(\ref{eq.reccomplex}). The constant $C$ in Eq.~(\ref{eq.reccomplex}) is assumed to be $C =l/L \sim 0.1$, as derived by different analytical and numerical methods \citep[for a discussion, see][]{2008JGRA..11307210B}. This implies that we are assuming that the reconnection happens in a region with a characteristic width $l = 0.1 L$, where $L$ is the radius of the planet (the characteristic length). The kinetic power $P_k$ is 
\begin{equation}\label{eq.pwrk}
P_k = \int p_{\rm ram} {\bf u} \cdot {\rm d}{\bf S} \simeq  \rho_2 u_2^3 \pi r_M^2\, ,
\end{equation}
where $p_{\rm ram} = \rho_2 u_2^2 $ is the wind kinetic ram pressure. 

\begin{figure}
  \includegraphics[height=7cm]{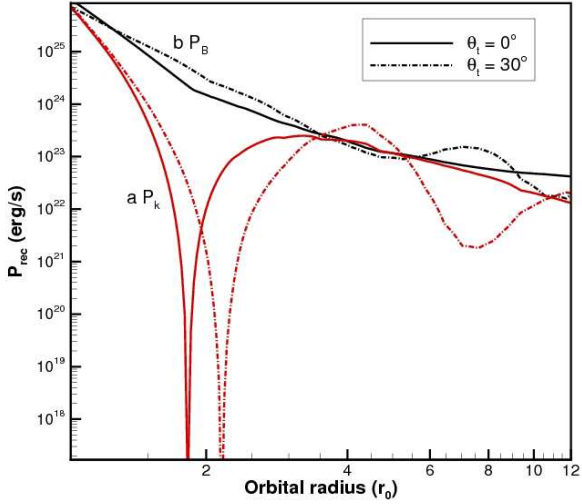}
  \caption{Estimated dissipated power $P_{\rm rec}$ due to the interaction of a hot-Jupiter with the stellar wind for cases T01 (aligned, solid lines) and T04 ($\theta_t=30^{\rm o}$, dot-dashed lines) as a function of orbital radius of the planet, considered to lie in the $x$-axis. Red curves consider the power released from the incident kinetic power ($P_{\rm rec} = a P_k$), while black curves are from the incident magnetic power ($P_{\rm rec} = b P_B$). The curves refer to the solutions for the stellar wind at $t=240$~h.  \label{fig.rec-power}}
\end{figure}

According to our stellar wind models, throughout the region of investigation ($r\lesssim 12~r_0\simeq 0.11$~AU), we find that $P_k> P_B$ \citep[the same behavior is found in Figure 11 of][]{2007P&SS...55..598Z}, except near the orbital radius where the planet co-rotates with the stellar wind and, thus, $u_2=(u_\varphi-u_K)\simeq 0$ and $P_k\simeq 0$. Figure~\ref{fig.rec-power} shows the estimated powers that are released for the wind cases T01 ($\theta_t=0^{\rm o}$, solid lines) and T04 ($\theta_t=30^{\rm o}$, dot-dashed lines) as a function of planetary orbital radius for a single rotation phase at $t=240~$h.

\begin{figure}
  \includegraphics[height=7cm]{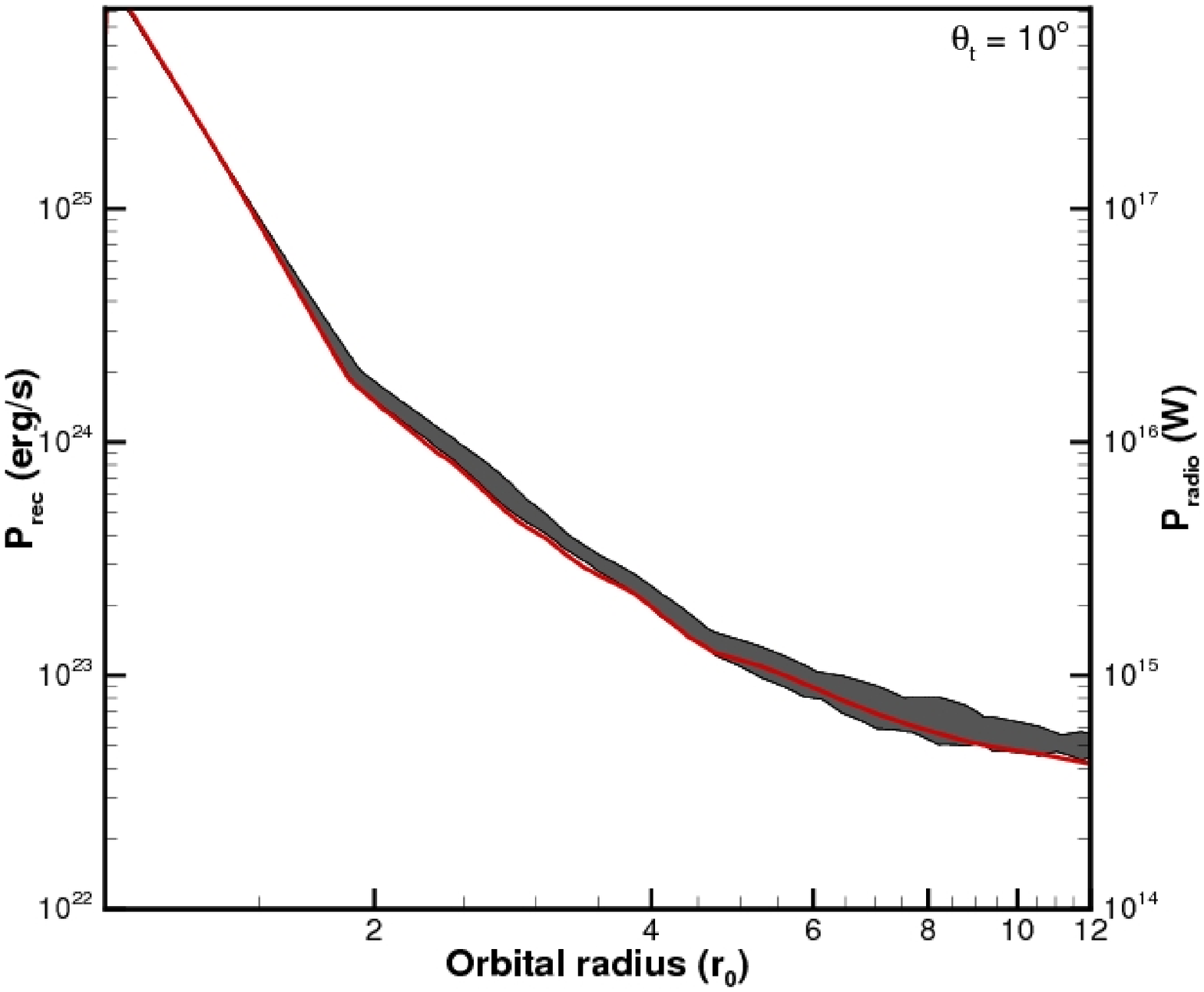}\\
  \includegraphics[height=7cm]{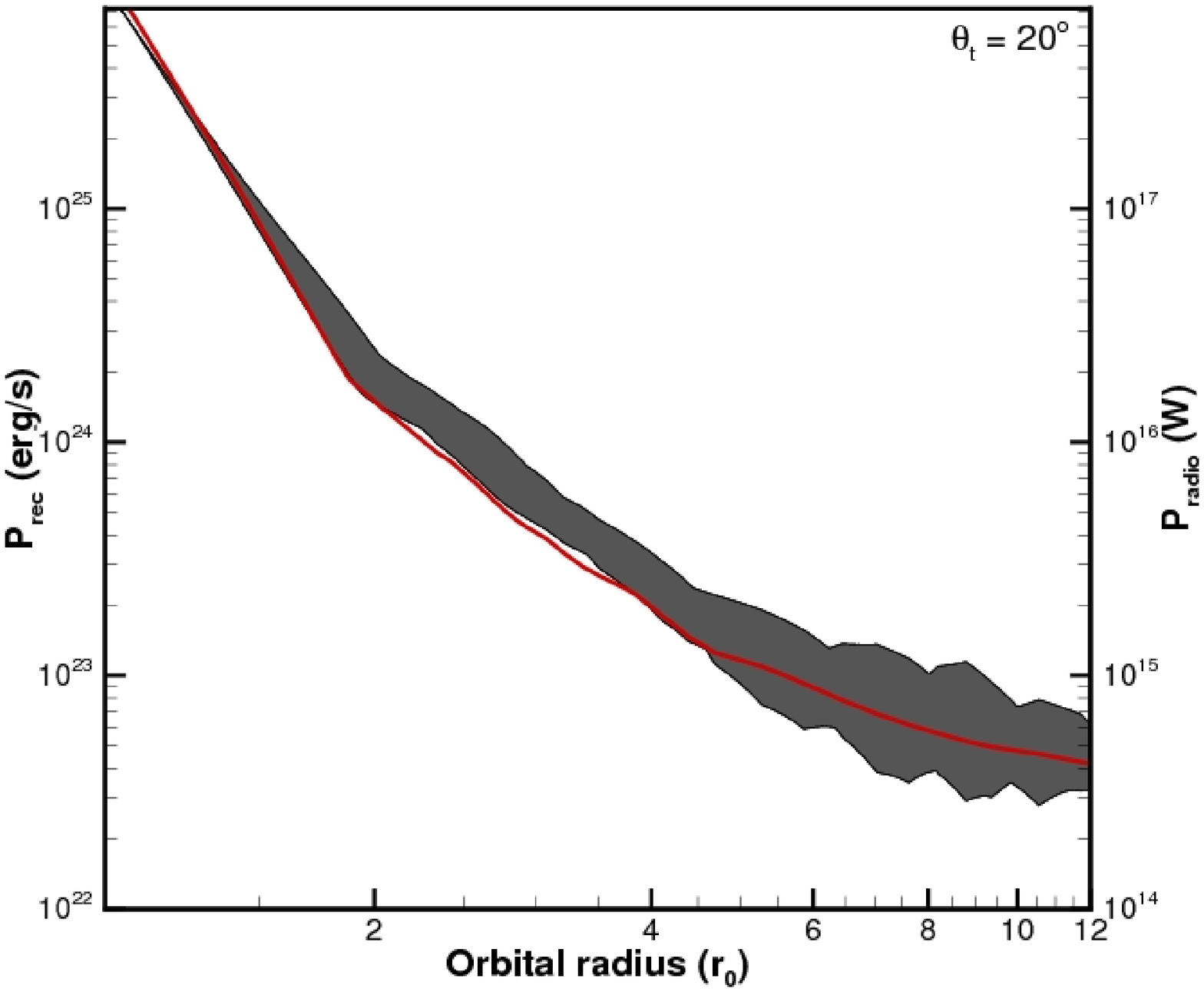}\\
  \includegraphics[height=7cm]{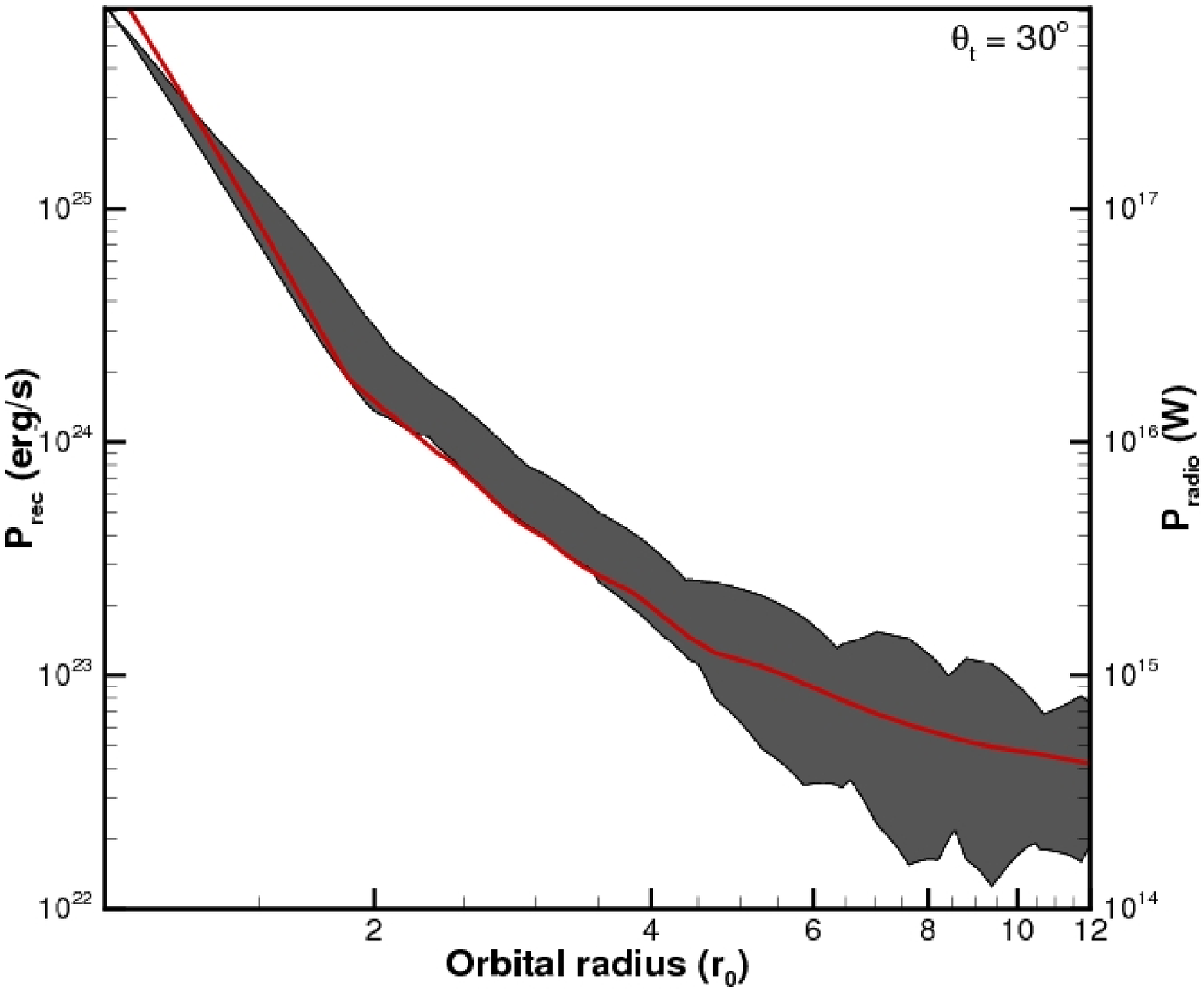}
  \caption{Estimated dissipated power $P_{\rm rec}= b P_B$ due to the interaction of a hot-Jupiter with the stellar wind for cases (a) T02 ($\theta_t=10^{\rm o}$), (b) T03 ($\theta_t=20^{\rm o}$), and (c) T04 ($\theta_t=30^{\rm o}$). The shaded area lies between maximum and minimum power that can be released in the interaction, according to the rotational phase of the star. The red curve is for the aligned case T01 ($\theta_t=0^{\rm o}$). The emitted radio power, assuming a conversion given by Eq.~(\ref{eq.pwrrad}), is shown in the vertical axes on the right of each plot. \label{fig.power}}
\end{figure}

The consideration of several phases of rotation of the star is taken in Figure~\ref{fig.power}, which shows the estimated power that is released for the wind cases T02 ($\theta_t=10^{\rm o}$, Fig.~\ref{fig.power}a), T03 ($\theta_t=20^{\rm o}$, Fig.~\ref{fig.power}b), and T04 ($\theta_t=30^{\rm o}$, Fig.~\ref{fig.power}c). Figure~\ref{fig.power} presents the power released from the incident magnetic power of the wind ($P_{\rm rec} = b P_B$). The shaded area lies between maximum and minimum power that can be released in the interaction, depending on the characteristics of the incident wind (i.e., phase of rotation of the star). We see that the maximum emitted power gets progressively high as $\theta_t$ increases, while the minimum released power gets progressively low as $\theta_t$ increases. This causes the shaded area to be larger for case T04 than for case T02 or T03. For a planet at orbital radius $r=5~r_0\simeq 0.05$~AU, the ratio between maximum and minimum released power due to a variation in the incident wind is a factor of $1.3$ for case T02 ($\theta_t=10^{\rm o}$), $2.2$ for case T03 ($\theta_t=20^{\rm o}$), and $3.7$ for case T04 ($\theta_t=30^{\rm o}$). Figure~\ref{fig.power} also shows the released power for the aligned case (red solid line), showing that an inclination between the axis of rotation of the star and the surface magnetic moment vector can contribute for an increase in the emitted power, depending on the phase of rotation of the star. 

Part of this released energy can be used to accelerate electrons, generating radio emission \citep{2008A&A...490..843J}
\begin{equation}\label{eq.pwrrad}
P_{\rm radio} =\eta P_{\rm rec}  \, .
\end{equation}
The efficiency $\eta$ in the conversion of $P_{\rm rec}$ into radio emission $P_{\rm radio}$ depends on the details of the physical processes that generate the radio emission (e.g., on the cyclotron-maser instability). Assuming $\eta=10\%$ \citep{2008A&A...490..843J}, the radio power emitted from a Jupiter-like planet orbiting at a distance $r=5~r_0 \simeq0.05$~AU is $P_{\rm radio}\sim 10^{15}$~W. The radio power is also shown in Fig.~\ref{fig.power} (vertical axes on the right). A time-dependent radio emission has also been estimated by \citet{2010MNRAS.tmp..735F}. For the magnetic planets of the Solar System $P_{\rm radio}\sim 10^{6.5}$~W (Neptune) to $\sim 10^{10.5}$~W (Jupiter), which means that, for our assumed giant planet orbiting our fictitious star, the radio power released is almost $\sim 5$ orders of magnitude larger than for Jupiter. This result suggests that stellar winds from pre-main sequence young stars have the potential to generate stronger planetary radio emission than the solar wind. Our results are in accordance to previous works developed on the framework of stellar winds of stars at the early main-sequence phase \citep{2005MNRAS.356.1053S, 2005A&A...437..717G}.

Table~\ref{tab.results} presents a summary of the results obtained for cases T01 to T04. The properties of the wind and of the reconnection site are described at $r=5~r_0$.

\subsubsection{Pile-up of the Stellar Wind Magnetic Field}\label{subsec.pile}
Around the Earth, the magnetic field of the solar wind piles-up in the magnetosheath. This causes the magnetic field strength to enhance in the reconnection site, and ultimately, increases the electric field $E$. By analogy, here we consider what would happen if magnetic field pile-up is considered in the planet's magnetosheath.

If we consider that the stellar wind is supersonic, a bow shock forms and the wind is deflected around the planet magnetosphere. For cases T01 to T04, the wind becomes supersonic at $r \sim 4~r0$ at the rotational equatorial plane. Downstream the shock, the field strength $B_{\parallel,2}$, the velocity $u_2$, and the density $\rho_2$ that appear on Eqs.~(\ref{eq.reccomplex}), (\ref{eq.rmagnetopause}), and (\ref{eq.pwrB}) will be given by shock conditions instead of arising directly from our stellar wind model. In this case, we use the Rankine-Hugoniot jump conditions to determine the magnetic field intensity $B_{\parallel,2}^{(s)}$, density $\rho_2^{(s)}$, and velocity $u_2^{(s)}$ on the magnetosheath \citep{opher-book-chapter}. Hence, except for Eqs.~(\ref{eq.reccomplex}), (\ref{eq.rmagnetopause}), and (\ref{eq.pwrB}), the equations presented in this section remain the same.

Assuming a perpendicular shock with strength $\delta=\rho_2^{(s)}/\rho_2=4$ (i.e., the density in the magnetosheath is four times the value of the stellar wind density), the Rankine-Hugoniot jump conditions state that the magnetic field at the magnetosheath is $\sim 4$ times higher than the local value of the wind magnetic field (i.e., $B_{\parallel,2}^{(s)} \sim 4B_{\parallel,2}$), while the velocity drops by a factor $\sim 4$ (i.e., $u_2^{(s)} \sim u_2/4$). As a consequence, the size $r_M^{(s)}$ of the planet's magnetosphere [Eq.~(\ref{eq.rmagnetopause})] diminishes with respect to $r_M$ when pile-up of the wind field lines was not considered. Within $r\lesssim 12~r_0$, the ratio $r_M^{(s)}/r_M\gtrsim 0.68$ for case T01, and becomes slightly more significant for higher $\theta_t$, presenting $r_M^{(s)}/r_M\gtrsim 0.63$ for case T04. Despite the decrease in $r_M^{(s)}$, the increase in the reconnection rate $E^{(s)}$ [Eq.~(\ref{eq.reccomplex})] is such that the dissipated power $P_{\rm rec}^{(s)}=b P_B^{(s)}$ [Eq.~(\ref{eq.pwrB})] increases when compared to the case when pile-up of the wind field lines was not considered. For case T01, $P_{\rm rec}^{(s)}$ increases $4.8$ -- $6.8$ times the values of $P_{\rm rec}$ presented in Fig.~\ref{fig.power}, depending on the orbital radius of the planet. For case T02, this increase ranges between $4.5$ -- $7.2$, $4.5$ -- $8.4$ for case T03, and $4.5$ -- $10.5$ for case T04, where these ranges depend now on both the location of the planet, as well as on the rotational phase of the star. This shows that for a supersonic stellar wind, the consideration of a perpendicular shock can increase the dissipated power due to the interaction of a hot-Jupiter with the stellar wind.

\subsection{On the Detectability of Planetary Radio Emission}
The detection of planetary radio emission depends on several factors, such as, on the distance $d$ to the extra-solar system, on whether the conical beam of the cyclotron emission is directed towards us, and on the emission bandwidth $\Delta f$ \citep{1999JGR...10414025F}. The radio flux that we detect on Earth is given by
\begin{equation}\label{eq.fluxrad}
\Phi_{\rm radio} = \frac{P_{\rm radio}}{d^2 w {\Delta f}}\, .
\end{equation}
where $w$ is the solid angle of the conical emission beam. The stellar wind ultimately controls the incident power on the planet, while the planet's characteristics control the frequency of the cyclotron emission $f_c$, and thus the emission bandwidth assumed to be $\Delta f = 0.5 f_c$ \citep{1999JGR...10414025F}.\footnote{Other authors \citep[e.g.,][]{2007A&A...475..359G} adopt a higher value of $\Delta f \simeq 0.9 - 1.0 f_c$. If this is the case, the emission bandwidth we obtain will have to be multiplied by a factor of $1.8$ -- $2.0$, and the resultant radio flux divided by the same factor. } For our fictitious planet, the assumed magnetic field at the pole is $100$~G (maximum field strength), which emits at (maximum) $f_c = 2.8 B = 280$~MHz ($B$ given in G and $f_c$ in MHz), with a bandwidth of $\Delta f = 140$~MHz. If our star is at a distance $d\sim10$~pc, using our estimated $P_{\rm radio}\sim 10^{15}$~W obtained in \S\ref{subsec.radio}, we find that the radio flux detected at Earth would be $\Phi_{\rm radio} \sim 7.5/w$~mJy. For a spherical emission, $w=4\pi$, and the detected flux is $\Phi_{\rm radio} \sim 0.6$~mJy, while for a hollow-cone beamed emission with a conical aperture of $45^{\rm o}$, $w\sim1.8$~sr, and $\Phi_{\rm radio} \sim 4$~mJy. 

One of the possibilities of the yet unsuccessful radio detections may be due to a probably small planetary field, which produces cyclotron emission in a low-frequency range, where instrumental sensitivity is still poor \citep{2000ApJ...545.1058B}. Low-frequency detectors, such as LOFAR, might be able to detect emission of few mJy at a frequency range of $10$ to $240$~MHz in the future \citep{2004P&SS...52.1469F, 2007A&A...475..359G}. Our hypothetical planet, for instance, could be observable by LOFAR.

Throughout \S\ref{sec.reconnection}, we have assumed a planet with equatorial magnetic field of $50$~G, which is about $6$ times larger compared to Jupiter's magnetic field of $8$~G at the equator \citep[or $\sim 16$~G at the pole, ][]{1998JGR...10311929C}. If we assume a planetary magnetic field intensity as Jupiter's, i.e., $B_p=8$~G at the equator, the power $P_{\rm rec}=bP_B$ released from the reconnection between planetary and stellar wind field lines at $r=5~r_0\simeq 0.05~$AU will be a factor of $2.7$ to $3.4$ smaller than the estimates performed in \S\ref{subsec.radio} for $B_p=50~$G. Also, because the planetary magnetic field is smaller, the magnetosphere of the planet will face a strong wind pressure and will vanish for $r\lesssim 4~r_0$ (in this case, $r_M =R_p$). For a value of $P_{\rm rec} \simeq 3\times 10^{22}$~erg~s$^{-1}$, and adopting the same efficiency $\eta =10\%$ for the conversion of the released energy into radio power, the radio flux of the planet arriving at Earth (adopting $d=10$~pc) would be $\Phi_{\rm radio}=14/w~$mJy, at a bandwidth of ${\Delta f}=22.4$~MHz, for a given solid angle $w$ of the conical emission beam. Comparing to $B_p=50$~G, the drop on the radio power caused by a smaller planetary magnetic field ($B_p=8$~G) is more than compensated by a smaller ${\Delta f}$ leading to a radio flux that is about twice the value obtained for $B_p=50$~G.

\section{THE INFLUENCE OF THE WIND ON PLANET MIGRATION}\label{sec.migration}
In \citet{paper2}, we investigated the action of magnetic torques from the stellar wind acting on the planet and whether such torques were able to significantly remove orbital angular momentum of the planet, causing planetary migration. The idea is that the wind exerts a pressure $p_{\rm tot } = \frac12(\rho_2 u_2^2 + B_{\parallel,2}^2/2\pi)$ on the area of the planet's magnetosphere $A_{\rm eff} = \pi r_M^2$. This force $p_{\rm tot} A_{\rm eff}$ will produce a torque $\frac{d L_p}{dt}$ at the planetary orbital radius $r$ \citep{2008MNRAS.389.1233L}
\begin{equation}\label{eq.torque.planet2}
\left| \frac{d L_p}{dt} \right| \simeq \ (p_{\rm tot} A_{\rm eff}) r \, 
\end{equation}
where $L_p =M_p v_K r$ is the orbital angular momentum of the planet, and $M_p$ is the mass of the planet. A change in the planet's angular momentum also leads to
\begin{equation}\label{eq.torque.planet}
\left| \frac{d L_p}{dt} \right| \simeq \frac12 M_p v_K \frac{d r}{dt} \simeq \frac12 M_p v_K \frac{r}{\tau_w}\, ,
\end{equation}
where $\tau_w$ is the time-scale for an appreciable radial motion of the planet \citep{1996Natur.380..606L}. From Eqs.~(\ref{eq.torque.planet2}) and (\ref{eq.torque.planet}), we can estimate such time-scale 
\begin{equation}\label{eq.time-scale}
{\tau_w} \simeq \frac12  \frac{M_p v_K}{p_{\rm tot} A_{\rm eff}} \, .
\end{equation}

\begin{figure}
  \includegraphics[height=7cm]{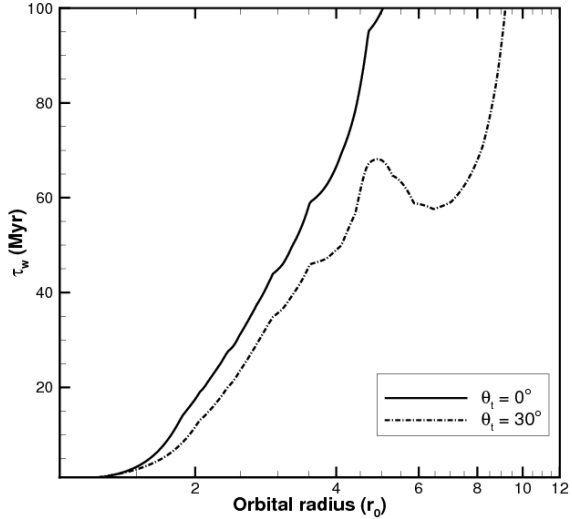}
  \caption{Time-scale $\tau_w$ for an appreciable radial motion of the planet. Solid line is for case T01 (aligned case) and dot-dashed line for case T04 ($\theta_t=30^{\rm o}$). The planet is assumed to lie in the $x$-axis. The curves refer to the solutions for the stellar wind at $t=240$~h. \label{fig.mig}}
\end{figure}

Within the range of parameters adopted in the simulations performed in \citet{paper2}, we showed that the stellar winds of WTTSs were not expected to have strong influence on the migration of close-in giant planets. The winds analyzed in that case assumed that the axis of rotation of the star and the stellar magnetic dipole moment were aligned. One aspect that was not investigated in \citet{paper2} is the effect of a tilted magnetic moment with respect to the rotation axis. Using the results of the simulations presented in the present paper, we thus compare the time-scales $\tau_w$ obtained for the aligned case (T01) and the misaligned case T04 ($\theta_t=30^{\rm o}$). We consider a planet with the same mass and radius as Jupiter, and the magnetic field at the equator of $B_p=50$~G. This comparison is shown in Figure~\ref{fig.mig}, where we note that an inclination of the stellar magnetic field acts to reduce $\tau_w$ when compared to the aligned case. The cases with intermediate tilt angles investigated (i.e., $\theta_t=10^{\rm o}$, $20^{\rm o}$) results in time-scales $\tau_w (r)$ that lie between the solid line for case T01 and the dot-dashed line for case T04. Figure~\ref{fig.mig} illustrates one single rotational phase of the star. For other phases of rotation of the star, we observe the same behavior: $\tau_w$ calculated for the misaligned cases is smaller than for the aligned case. 

The last column of Table~\ref{tab.results} presents the time-scale $\tau_w$ calculated for cases T01 to T04 for a planet at $r=5~r_0$. Compared to the aligned case, where $\tau_w \simeq 100$~Myr, case T04 ($\theta_t = 30^{\rm o}$), for example, shows considerably smaller time-scales ranging from $\tau_w \sim 40$ to $70$~Myr. We expect that larger misalignment angles $\theta_t>30^{\rm o}$ or other effects, such as an increase in the wind coronal base density or magnetic field intensity \citep[as discussed in][]{paper2}, could reduce $\tau_w$. The time-scales derived here seem to be larger (and therefore less important) than those estimated by other processes, such as from the interaction of the protoplanet with the disk wherein it was formed \citep{2006RPPh...69..119P}. However, as suggested by \citet{2010A&A...512A..77L}, the removal of planetary orbital angular momentum by the stellar wind may be important for synchronizing stellar rotation with the orbital motion of their planets during the pre-main sequence phase.

\section{CONCLUSION}\label{sec.conclusions}
We have presented simulations of magnetized stellar winds where the surface stellar magnetic moment is tilted with respect to the axis of rotation of the star. Such configuration requires a fully 3D approach, as the system does not present axisymmetry. By adopting a dipolar surface distribution of magnetic flux, we showed that the interaction of magnetic field lines and the wind leads to a periodic final solution, with the same rotational period as the star. The final magnetic field configuration of the stellar magnetosphere presents an oscillatory pattern. 

By varying several parameters of the simulations, we explored the effects of the misalignment angle $\theta_t$, stellar period of rotation $P_0$, heating index $\gamma$, and plasma-$\beta$ parameter at the magnetic pole of the star $\beta_0$ in the final periodic solution of our simulations. We showed that the increase in $\theta_t$ or the decrease in $P_0$ lead to a more accelerated wind. The same is true if $\gamma$ or $\beta_0$ are decreased, as already demonstrated in the axisymmetric cases of our previous paper \citep{paper2}.

We estimated the power released in the interaction between a close-in giant planet and the stellar wind. If the planet and wind are magnetized, the interaction results in reconnection process, which releases energy that can be used to accelerate electrons. These electrons propagate along the planet's magnetic field, producing cyclotron radiation at radio wavelengths. This calculation is motivated by radio emission observed in the magnetic planets of the Solar System (Earth, Neptune, Saturn, Uranus, and Jupiter). We  showed that the intensity of the radio emission varies, as the wind impacting on the planet changes according to the stellar phase of rotation. If radio emission from a planet orbiting a star with misaligned rotation axis and magnetic axis is to be detected, we showed here that the radio power will present a larger temporal variation for higher $\theta_t$. 

Our estimates show that the radio power emitted by the fictitious extra-solar planet orbiting our star at $\sim 0.05~$AU can be $5$ orders of magnitude larger than the non-thermal radio power emitted by Jupiter. This suggests that the stellar wind from a young star has the potential to generate strong planetary radio emission, which could be detected by LOFAR.

As a final point, we answered the question posed in \citet{paper2} about whether winds from misaligned stellar magnetospheres could cause a significant effect on planetary migration. In \citet{paper2}, only the case of winds from stars where the rotation axis and the surface dipolar magnetic moment were aligned was considered. Compared to the aligned case, we showed here that the time-scale for an appreciable radial motion of the planet is shorter for larger misalignment angles.

\acknowledgments
The authors would like to thank E.~Shkolnik for useful comments. We also appreciate the comments and suggestions from the anonymous referee that greatly improved the manuscript. AAV acknowledges support from FAPESP (04-13846-6), CAPES (BEX4686/06-3), and an STFC grant. MO acknowledges the support by National Science Foundation CAREER Grant ATM-0747654. VJ-P thanks CNPq (305905/2007-4). The simulations presented here were performed at the Columbia supercomputer, at NASA Ames Research Center. 


\clearpage

\begin{landscape}
\begin{deluxetable}{c c c c c c c c c c c} 
\tablewidth{0pt}
\tablecaption{Summary of the local characteristics of the wind and planet at the reconnection site for a planet at orbital radius $r=5~r_0\simeq 0.05$~AU. The columns represent, respectively: the simulation name, the local wind density $\rho_2$, the relative velocity $u_2=(u_\varphi-u_K)$, the mass-loss rate per unit solid angle $\rho u_r r^2$, the local $z$-component of the wind $B_{z,2}$ and of the magnetosphere of the planet $B_{z,1}$, the $E$ electric field (or reconnection rate) at the reconnection site, the size of the planetary magnetosphere $r_M$, the powers released in the interaction adopting $P_{\rm rec}=aP_k$ and $P_{\rm rec}=bP_B$, and time-scale for planet migration $\tau_w$ (\S\ref{sec.migration}). Range of numbers are due to different phases of rotation of the star. \label{tab.results}}  
\tablehead{
\colhead{Name}	&	\colhead{$	\rho_2	$}	&	\colhead{$	u_2	$}	&	\colhead{$	\rho u_r r^2	$}	&	\colhead{$	-B_{z,2}	$}	&	\colhead{$	B_{z,1}	$}	&	\colhead{$	E	$}	&	\colhead{$	r_M	$}	&	\colhead{$	P_{\rm rec}=aP_k	$}	&	\colhead{$	P_{\rm rec}=bP_B	$}	&	\colhead{$	\tau_w	$}\\	
\colhead{}	&	\colhead{$(10^{-13}~{\rm g~cm^{-3}})$}	&\colhead{(km~s$^{-1}$)}&\colhead{$(10^{-9} {\rm M}_\odot~{\rm yr}^{-1})$}&\colhead{(G)}&\colhead{(G)}&\colhead{($10^{-5}$~stV~cm$^{-1}$)}&\colhead{$(R_p)$}&\colhead{($10^{23}$~erg~s$^{-1}$)}&\colhead{($10^{23}$~erg~s$^{-1}$)}&	\colhead{(Myr)}		
}
\startdata
T01	&	$	1.13	$	&	$	52	$	&	$	9.45	$	&	$	3.30	$	&	$	5.63	$	&	$	1.08	$	&	$	2.09	$	&	$	1.16	$	&	$	1.16	$	&	$	99	$	\\
T02	&	$	0.93-1.10	$	&	$	47-56	$	&	$	8.44-9.66	$	&	$	3.33-3.78	$	&	$	6.07-7.00	$	&	$	1.08-1.34	$	&	$	1.95-2.03	$	&	$	0.61-1.26	$	&	$	1.09-1.42	$	&	$	74-89	$	\\
T03	&	$	0.62- 1.07	$	&	$	32-59	$	&	$	6.59-10.4	$	&	$	3.25-4.50	$	&	$	6.80-8.88	$	&	$	1.05-1.91	$	&	$	1.79-1.96	$	&	$	0.11-1.18	$	&	$	0.92-2.05	$	&	$	54-76	$	\\
T04	&	$	0.39-1.07	$	&	$	14-64	$	&	$	5.03-10.9	$	&	$	2.84-4.62	$	&	$	7.37-10.4	$	&	$	0.91-2.36	$	&	$	1.70-1.90	$	&	$	0.0055-1.67	$	&	$	0.63-2.35	$	&	$	43-68	$	
\enddata
\end{deluxetable}
\clearpage
\end{landscape}
\end{document}